\documentclass[%
 reprint,
 amsmath,amssymb,
 aps,
 prd,
]{revtex4-2}

\usepackage{graphicx}
\usepackage{dcolumn}
\usepackage{bm}

\begin{document}

\title{Hunting Galactic Axion Dark Matter with Gravitationally Lensed Fast Radio Bursts}

\author{Ran Gao$^{1}$, Zhengxiang Li$^{1,2}$, Kai Liao$^{3}$, He Gao$^{4,1,2}$, Bing Zhang$^{5,6}$, Zong-Hong Zhu$^{1,2,3}$}

\affiliation{$^1$Institute for Frontiers in Astronomy and Astrophysics, Beijing Normal University, Beijing 102206, China;\\
$^2$Department of Astronomy, Beijing Normal University, Beijing 100875, China;\\
$^3$Department of Astronomy, School of Physics and Technology, Wuhan university, Wuhan 430072, China;\\
$^4$Purple Mountain Observatory, Chinese Academy of Sciences, Nanjing, 210023, People's Republic of China\\
$^5$Nevada Center for Astrophysics, University of Nevada, Las Vegas, NV 89154, USA;\\
$^6$Department of Physics and Astronomy, University of Nevada, Las Vegas, NV 89154, USA}

\date{\today}

\begin{abstract}
Ultralight axion or axionlike particles are one of the most promising candidates for dark matter because they are a well-motivated solution for the theoretical strong $CP$ problem and observational issues on small scales, i.e. the core-cusp problem and the satellite problem. A tiny coupling of axions and photons induces birefringence. We propose the differential birefringence measurements of multiple images of gravitationally lensed fast radio burst (FRB) systems as probes of the Galactic axion dark matter (ADM) background. In addition to general advantages of lensing systems, i.e. alleviating systematics and intrinsic astrophysical dependencies, precise measurements of lensing time delay and polarization angle in gravitationally lensed FRB systems make them a more robust and powerful probe. We show that, with a single lensed FRB system (which may be detected in large numbers in the SKA era), the axion-photon coupling under the ADM background could be constrained to be $g_{a\gamma} < 7.3 \times 10^{-11}~ \mathrm{GeV^{-1}}$ for an axion mass $m_a\sim10^{-20}~\mathrm{eV}$. This will be of great significance in achieving synergistic searches of the Galactic ADM with other astrophysical probes and laboratorial experiments.

\end{abstract}
\maketitle

\section{Introduction} \label{sec:intro}
Solid and abundant evidence for the existence of dark matter (DM) has been found over a wide range of astrophysical and cosmological scales~\citep{2000RPPh...63..793B,2005PhR...405..279B}. The fundamental nature of this component has been one of the most pressing open questions in astronomy and physics for decades. In the context of microscopic particle physics, a particularly intriguing class of potential DM candidates is supplied by light bosonic degrees of freedom (d.o.f), of which axions and axionlike particles are prototypical~\citep{2009NJPh...11j5008D,2016PhR...643....1M,2015ARNPS..65..485G} (below we will not distinguish these two concepts and call both of them axions for short in the context of dark matter). They can act as fuzzy dark matter and provide a natural solution to the intense challenges of observations on small scales, such as the core-cusp problem and the satellite problem~\citep{2015PNAS..11212249W,2000PhRvL..85.1158H,2017PhRvD..95d3541H,2014NatPh..10..496S}. In addition to the being well-motivated d.o.f from the standpoint of UV theory~\citep{2006JHEP...05..078C,2010PhRvD..81l3530A}, the canonical QCD axion is also extremely well-motivated as a solution to the strong $CP$ problem~\citep{1977PhRvL..38.1440P,1978PhRvL..40..223W,1978PhRvL..40..279W}. Therefore, the axion dark matter (ADM), an ideal and very promising candidate, is nowadays in the limelight, and massive efforts are underway to search for their signatures.

In the past decades, a plethora of works have attempted to test the ultralight ADM by using the two distinct predictions of interaction between axion and photon, $\mathcal{L_{\mathit{a\gamma}}}=-g_{\mathit{a\gamma}}a F_{\mu\nu}\widetilde{F}^{\mu\nu}/4$. One is the interconversion between axions and photons in the presence of magnetic fields, such as ``light shining through a wall" experiments~\citep{2010PhLB..689..149E,2013JInst...8.9001B}, the axion helioscope~\citep{2014JInst...9.5002A,2017NatPh..13..584A,2019JCAP...06..047A}, astrophysical observations of quasars or active galactic nucleus (AGN)~\citep{2012JCAP...07..041P,2017ApJ...847..101B,2020JCAP...04..055A} and SN1987A~\citep{2015JCAP...02..006P}. However, there is no signal of the conversion detected so far~\citep{2013PhRvD..87c5027M,2014JCAP...09..026A,2017PhRvD..96e1701K} since these astrophysical constraints heavily depend on the uncertainty of cosmic magnetic fields, including their strength and structure. Therefore, several previous works also attempted to design experimental instruments in laboratory for detecting ADM, including cavities~\citep{2018PhRvL.121p1301O}, wire arrays~\citep{1994PhRvD..50.4744S}, dielectric plates~\citep{2013JCAP...04..016H,2019EPJC...79..186B}, and interferometers~\citep{2019PhRvL.123k1301N} (see~\citep{2015ARNPS..65..485G,2021RvMP...93a5004S,2022arXiv220314923A} for a review). The other is the photon birefringence in the presence of axion background. This effect is caused by the modification of the photon dispersion relation and the rotation of the linear polarization plane due to the oscillating ADM background~\citep{1990PhRvD..41.1231C,1992PhLB..289...67H,1998PhRvL..81.3067C,1999PhRvL..83.1506L,2018PhRvD..98c5021D},including the $E$-mode polarization of the cosmic microwave background~\citep{2018arXiv181107873S,2019PhRvD.100a5040F}, multiepoch observations of inherently linearly polarized synchrotron emission from AGN~\citep{2019JCAP...02..059I}, linearly polarized pulsar light~\citep{2020PhRvD.101f3012L,2023PhRvL.130l1401L}, and linear polarization at certain wavelength caused by the scattering of light of the parent star in a protoplanetary disk~\citep{2019PhRvL.122s1101F}. However, broadly speaking, almost all these tests or constraints suffer from the uncertainties of instrumental offsets and intrinsic brightness or polarization of astrophysical sources.

On the other hand, differential birefringence measurements of multiple images of gravitationally lensed polarized sources have been proposed as robust probes of ADM since they can significantly alleviate systematics and astrophysical dependencies~\citep{2021PhRvL.126s1102B}. However, it should be noted that, in such a case with a persistent source quasar, the differential birefringence angle toward two gravitationally lensed images is usually measured at the same time and thus depends only on the properties of the axion field in the emitting region. That is, the traditional lensed quasar systems only probe the ADM in the host galaxy of the source. However, almost all other astrophysical methods or experiments probe ADM in the local or Galactic region. Therefore, any new probe for detecting the Galactic ADM background is of great importance for complementarity and cross-check. Moreover, the precision of time delay measurement and the inference of no changes in the polarization angle between the polarized components of the two images from no intensity variations of the two lensed images might weaken the confidence level of these constraints.  

In this paper, we propose gravitaionally lensed fast radio burst (FRB) systems as robust probes of Galactic ADM. Within this kind of systems, sources are FRBs, which are bright radio transients with millisecond duration first discovered in 2007~\citep{2007Sci...318..777L}. Several observational properties, including extragalactic origin~\citep{2017Natur.541...58C}, high event rate~\citep{2013Sci...341...53T,2016MNRAS.460L..30C,2021ApJ...909L...8N}, short duration, make FRBs have been proposed as powerful probes for astrophysical and cosmological studies (see~\citep{2019ARA&A..57..417C,2022arXiv221203972Z} for a review). In particular, as a typical transient with very short duration, gravitationally lensed FRB systems have overwhelming advantages in fundamental physics tests and cosmological applications, such as searching for compact dark matter~\citep{2016PhRvL.117i1301M,2018A&A...614A..50W,2020MNRAS.496..564K,2020ApJ...896L..11L,2022ApJ...928..124Z,2022PhRvD.105j3528K}, probing the proper motion of the FRB source~\citep{2017ApJ...847...19D}, testing the validity of of general relativity~\citep{2022MNRAS.516.1977G}, and measuring the expansion rate and space-time curvature of the universe~\citep{2018NatCo...9.3833L,2021A&A...645A..44W} (see~\citep{2019RPPh...82l6901O,2022ChPhL..39k9801L} for a review). 
Here, we establish gravitationally lensed FRB systems (lensed by an intervening galaxy) as ideal and powerful tool for hunting Galactic ADM. Specifically, most of FRBs are highly linearly polarized and, for bright bursts with high signal-to-noise ratio, polarization angles of them can be measured in great accuracy using facilities with high sensitivity. More importantly, as a source with millisecond duration, time delay between lensed images also can be measured with unprecedented precision. And thus two signals emitting from the source at the same time would be clearly separated and reported as two lensed images in the observation region. Ultimately, differential birefringence angle measurements of these time-separated images decode the time variation of the oscillating axion field background in the Milky Way.

\section{Method} \label{sec:meth}
In this section, we present a brief introduction to the birefringence effect and the lensing effect.
\subsection{birefringence effect}
It is well known that the axion-photon coupling gives rise to modifications to electrodynamics in an axion field background~\citep{1987PhRvL..58.1799W}. We consider the relevant Lagrangian density as
\begin{equation}
    \mathcal{L}=-\frac{1}{4} F_{\mu \nu} F^{\mu \nu}-\frac{1}{2} \partial_{\mu} a \partial^{\mu} a+\frac{g_{a \gamma}}{4} a F_{\mu \nu} \tilde{F}^{\mu \nu}-\frac{1}{2} m_{a}^{2} a^{2}
\end{equation}
where $F_{\mu \nu}$ and $\tilde{F}^{\mu \nu}$ denote the electromagnetic field strength tensor and its dual, respectively. $g_{a\gamma}$ is the coupling constant between the axion and photon field. $m_{a}$ is the axion mass.
The equation of motion for $a$ is given by the Klein-Gordon equation, and a simple solution to it is the coherently oscillating axion field when we ignore the backaction term
\begin{equation}\label{eq2}
    a\left(t, x^{i}\right)=\frac{\sqrt{2 \rho_{a}\left(x^{i}\right)}}{m_{a}} \sin \left[m_{a} t+\delta\left(x^{i}\right)\right],
\end{equation}
where $x^{i}$ is the three-dimensional spatial coordinates, $\rho_{a}$ is the energy density of the axion field, $\delta$ is the phase.
Here we use the natural unit system, so $m_{a} t$ is dimensionless.
When $\delta$ is approximately constant within the size patch of the de Broglie wavelength $\lambda_\mathrm{dB}$, the inhomogeneity of the axion field could be characterized by the spatial dependence of $\rho_{a}$ and $\delta$.

The parity-violating coupling term $aF\tilde{F}$ leads to birefringence. When temporal and spatial variations of the axion background field are much smaller than the frequency of the photons propagating in the background, satisfying $10^{-16}\left(m_{a} / 10^{-22} \mathrm{eV}\right)(\mathrm{GHz} / \nu) \ll 1$, the polarization angle rotation effect is independent of the frequency of the light and the rotation angle is given as
\begin{align}\label{eq3}
    \Delta\theta_{a} &= \frac{g_{a\gamma}}{2}\Delta a(t_\mathrm{obs},x^i_\mathrm{obs};t_\mathrm{em},x^i_\mathrm{em})\\ \nonumber
    &= \frac{g_{a\gamma}}{2}\int_C ds~n^{\mu} \partial_{\mu} a\\ \nonumber
    &= \frac{g_{a\gamma}}{2}[a(t_\mathrm{obs},x^i_\mathrm{obs})-a(t_\mathrm{em},x^i_\mathrm{em})],
\end{align}
where the axion field at the point of observation and at the point of photon emission are denoted by the subscripts ``obs" and ``em", respectively. $C$ is the path of the photon in spacetime from the point of emission to the point of observation, and $n^{\mu}$ is the null tangent vector to $C$. It should be strongly emphasized that the net polarization angle rotation is independent of the details  of the axion field configuration along the photon path at all points between emission and observation. That is, the net polarization angle rotation depends only on the initial and final axion field values~\citep{1992PhLB..289...67H,2019PhRvD.100a5040F,2021PhRvD.103h1306S,2021PhRvL.126s1102B}.

\subsection{FRB lensing effect}
Most FRBs are natural and ideal sources with linear polarization. For an observed FRB, the measured polarization angle consists of two main components: $\theta(\nu)=\theta_0+\mathrm{RM}(c/\nu)^2$. The first component $\theta_0$ is the polarization angle of the source and it is frequency-independent. The second one is the additional chromatic birefringence caused by Faraday rotation effect when radio signals pass through magnetized plasma. The frequency dependencies of the Faraday rotation are captured by polarization measurements over large bandwidths and can be robustly modeled by fitting Stokes parameters. Furthermore, the Faraday rotation-corrected polarization angle $\theta_0$ could be simultaneously determined from the Stokes parameters fitting~\citep{2012MNRAS.421.3300O,2017MNRAS.469.4034O}. For this Faraday rotation-corrected and frequency-free term, it can be further decomposed as the following three compositions: $\theta_0=\theta_{\text{FRB}}+\Delta\theta_a+\delta\theta$. $\theta_{\text{FRB}}$ is its intrinsic polarization angle, $\Delta\theta_a$ is the achromatic birefringence induced by  the coupling between the photon and the axion field, and $\delta\theta$ is the systematic calibration offset and random error in observation.
Obviously, for observations of an FRB along a single line of sight, it is difficult or unfeasible to directly derive $\Delta\theta_a$ without knowing $\theta_{\text{FRB}}$ and $\delta\theta$. Fortunately, gravitational lensing of polarized sources have been proposed as probes with a unique advantage in overcoming this difficulty and alleviating the unknown $\theta_{\text{FRB}}$ and $\delta\theta$~\citep{2021PhRvL.126s1102B}. For a gravitationally lensed FRB system, millisecond duration signals simultaneously emitting from the source can be observed as time-separated or lensed images due to gravitational time delay. Differential birefringence of the time-separated images provide information of the time variation of the axion field around the observer, i.e. on Earth or in Milky Way. In a lensed FRB system, the polarization angles for the two images of A and B ($\theta_{0,\mathrm{A}}$ and $\theta_{0,\mathrm{B}}$) are $\theta_{0,\mathrm{A}}=\theta_{\text{FRB}}+\Delta\theta_{a,\mathrm{A}}+\delta\theta_\mathrm{A}$, $\theta_{0,\mathrm{B}}=\theta_{\text{FRB}}+\Delta\theta_{a,\mathrm{B}}+\delta\theta_\mathrm{B}$, respectively.
For FRB-like sources with very short durations, pulses emitting from it at the same time would separately arrive at our galaxy with a gravitational lensing time delay $\Delta t =\left|t_\mathrm{A}-t_\mathrm{B}\right|$.
$t_\mathrm{A}$ and $t_\mathrm{B}$ are the observation times of the two images.
The initial polarization angles, $\theta_{\text{FRB}}$, for the two images are the same since they simultaneously emitted from the source. In addition, if we consider that there is no systematic deviation between $\delta\theta_\mathrm{A}$ and $\delta\theta_\mathrm{B}$, then the difference between the two polarization angles is $\Delta \theta_{a, \text {lens}} \equiv \theta_{0,\mathrm{A}}-\theta_{0,\mathrm{B}} = \Delta\theta_{a,\mathrm{A}}-\Delta\theta_{a,\mathrm{B}}$, and is not dependent on $\theta_{\text{FRB}}$ and $\delta\theta$.

For a gravitationally lensed FRB system, signals for both the two images A and B simultaneously emit from the source. Therefore, their $a(t_\mathrm{em},x^i_\mathrm{em})$ is the same. It has been shown that gravitational lensing does not affect the birefringence angle~\citep{2021PhRvD.103h1306S}. Consequently, the differential birefringence angle is only dependent on the axion field in the region of observation. Combining equations (\ref{eq2}) and (\ref{eq3}), we can obtain
\begin{equation} \label{eq4}
     \Delta \theta_{a, \text {lens}}=K \sin \left[\frac{m_{a} \Delta t}{2}\right] \sin \left(m_{a} t_{\mathrm{obs}}+\delta_{\mathrm{obs}}-\pi / 2\right), 
\end{equation}
with $K$ in normalized units being
\begin{equation} \label{eq5}
    K=1.225^{\circ}\left[\frac{\rho_{a, \mathrm{obs}}}{0.3~\mathrm{GeV}~\mathrm{cm}^{-3}}\right]^{\frac{1}{2}} \frac{g_{a \gamma}}{10^{-12} ~\mathrm{GeV}^{-1}}\left[\frac{m_{a}}{10^{-22}~\mathrm{eV}}\right]^{-1}.
\end{equation}
Here, we use the dark matter density in our galaxy $\rho_{a, \mathrm{A}} = \rho_{a, \mathrm{B}} \equiv \rho_{a, \mathrm{obs}}$ as the energy density of axion field in the observation region, $t_\mathrm{obs} = \frac{1}{2}(t_\mathrm{A}+t_\mathrm{B})$ is the average observation time.
The phase factor $\delta_\mathrm{obs}$ remains constant at the observer.
For a single observation, since the phase of $\sin \left(m_{a} t_{\mathrm{obs}}+\delta_{\mathrm{obs}}-\pi / 2\right)$ in Eq. (\ref{eq4}) is uncertain, we use the root mean square of the oscillating axion field with a random phase, $1/\sqrt{2}$, in the following calculation.

For gravitationally lensed repeating FRB systems, it is viable to accurately measure multiple $ \Delta \theta_{a, \text {lens}}$ over a time sequence. This merit gives rise to several advantages.
First, we can use statistical methods to significantly reduce the error $\delta\theta$ with a large number of bursts from a repeating source.
Moreover, because of the persistence of repeating FRBs, we can perform further periodic analysis to determine the range of $m_a$.
Last but not least, for each two gravitational lensing delayed bursts or images, the fact that their initial polarization angles are almost the same and the precise measurements of observation times yield reliable constraints on the axion field in the observation region or Milky Way.

\section{SIMULATIONS AND RESULTS} \label{sec:sim}
\subsection{Single lensed FRB case}
Combining recent analysis of the FRB20201124A polarization~\citet{2022RAA....22l4003J} (a total of 536 FRB bursts with signal-to-noise ratio greater than 50, $S/N>50$, in 4 days and most of them are highly linearly polarized bursts) and the calibration of 3C286 polarization (with an error of $0.3^\circ$)~\citet{2022Natur.601...49C}, it indicates that the FAST telescope can measure the polarization angle of high $S/N$ bursts of FRBs with an conservative accuracy of $\sim1^\circ$. Meanwhile, the measured error for the polarization angle difference between two images of a lensed FRB system also is $\sim1^\circ$. For a very tiny polarization angle difference due to the ultralight axion field birefringence, it would be completely overwhelmed by the measure error. In this context, we can only derive an upper limit on the birefringence effect, typically, $\Delta \theta_{a, \text {lens}} < 1^\circ$. 
In addition, gravitational lensing of an FRB by a lens of mass $M_\mathrm{L}$ induces two images separated by a typical time delay $\sim$few $\times$ $(M_\mathrm{L}/30 M_{\odot})$ msec. Therefore, for a galaxy lens with a dark matter halo mass $\sim 10^{12}M_{\odot}h^{-1}$ ($h$ is the Hubble constant in units of $100\ \mathrm{km}\ \mathrm{s}^{-1}\ \mathrm{Mpc}^{-1}$), the typical time delay is $\sim \mathcal{O}(10)$ days. For $\rho_{a, \mathrm{obs}}$ in Equation (\ref{eq5}), we take the dark matter energy density in Milky Way, $\rho_{a, \mathrm{obs}} = 0.3~\mathrm{GeV}~\mathrm{cm}^{-3}$~\citep{1986PhRvD..33..889T,2021arXiv211110615L,2022JCAP...06..014C,2022PhRvD.106d2011F}. With these typical value determined, we substitute them into Eq. \ref{eq4} and yield constraints on the Galactic axion field from single lensed FRB system. Results are shown in Figure \ref{fig:ma_gay} and denoted by the thick blue line, i.e. $g_{a\gamma} < 7.3 \times 10^{-11}~\mathrm{GeV^{-1}}$ for an axion mass $m_a\sim10^{-20}~\mathrm{eV}$

For a repeating FRB source, once it is strongly lensed by an intervening galaxy, a series of image multiplets from the same source will exhibit a fixed pattern in their mutual time delays, appearing over and over again as we detect the repeating bursts. This property can help us effectively identify whether a repeating FRB has been lensed~\citep{2018NatCo...9.3833L}. 
We usually observe hundreds of bursts ($N\sim\mathcal O(100)$) in an active episode and measure their polarization angles. The precision of differential birefringence measurements would be statistically increase by a factor of $\sqrt{N}$. Assuming there are $N\sim\mathcal O(100)$ pairs of images with high enough SNR, the measured upper limit would approximately be $\Delta \theta_{a, \text {lens}} < 0.1^\circ$. Corresponding results are shown in Figure \ref{fig:ma_gay} and labeled by the thick red line.
However, it should be noted that, as the precision increases, systematic errors may become dominant and results derived from $\Delta \theta_{a, \text {lens}} < 0.1^\circ$ would be overestimated.

In addition to estimating the polarization angle due to the axion-photon interaction, we should also clarify the range of the axion mass $m_a$ for a single observation. As the axion mass increases, its period becomes shorter. In general, the difference of the axion field for images A and B would be significant when the oscillating period $T_a=2\pi/m_a$ is comparable with the lensing time delay $\Delta t$. 
That is, the differential birefringence would be negligible when $T_a\gg \Delta t$ since the values of the field for images A and B are very close to each other. On the other hand, differential birefringence is also negligible when $T_a \ll t_{\mathrm{sampling}}$, since this effect is averaged out. For a lensed FRB system with a typical time delay of $\mathcal{O}(10)$ days, the corresponding $m_a \sim 10^{-21}$ eV. For FRBs observed by FAST, effective polarization information can be observed in the time domain as long as the processing time resolution, $t_{\mathrm{sampling}}$, is shorter than the field oscillation period. Therefore, for lensed FRB systems with $t_{\mathrm{sampling}} \sim \mathcal{O}(1)~\mathrm{ms}$, they can provide the axion field information for axion mass as heavy as $m_a\sim4.1\times10^{-12}~\mathrm{eV}$. This range is almost 6 orders of magnitude higher than the one derived from strongly lensed quasar~\citep{2021PhRvL.126s1102B}.

\subsection{Statistical sample}
For a statistical sample of $N$ lensed FRB system, the uncertainty of the measurement for the differential birefringence angle $\Delta \theta_{a, \text {lens}}$ is approximately reduced by a factor of $1/\sqrt{N}$. For each lens system, the corresponding time delays $\Delta t$ and observation times $t_\mathrm{obs}$ are independent, so the estimation is done by statistical averaging.
The average value of $\sin \left(m_{a} t_{\mathrm{obs}}+\delta_{\mathrm{obs}}-\pi / 2\right)$ is $ \langle | \sin (m_{a} t_{\mathrm{obs}}+\delta_{\mathrm{obs}}-\pi / 2)|\rangle = 2/\pi$.
Assuming that $\Delta t$ satisfies a uniform distribution $\Delta t\in[0,\Delta t_\mathrm{max}]$ in which $\Delta t_\mathrm{max}$ is the maximum time delay in the lensed FRB systems, $\left \langle \left|\sin \left[\frac{m_a \Delta t}{2} \right]\right| \right\rangle$ is given by the following equation 
\begin{equation}\label{eq6}
\left \langle \left|\sin \left[\frac{m_a \Delta t}{2} \right]\right| \right\rangle = \frac{1}{\Delta t_\mathrm{max}} \int_{0}^{\Delta t_\mathrm{max}} \left|\sin \left[\frac{m_a \Delta t}{2} \right]\right| d(\Delta t).
\end{equation}

\begin{figure*}[htbp]
    \centering
    \includegraphics[width=2.0\columnwidth]{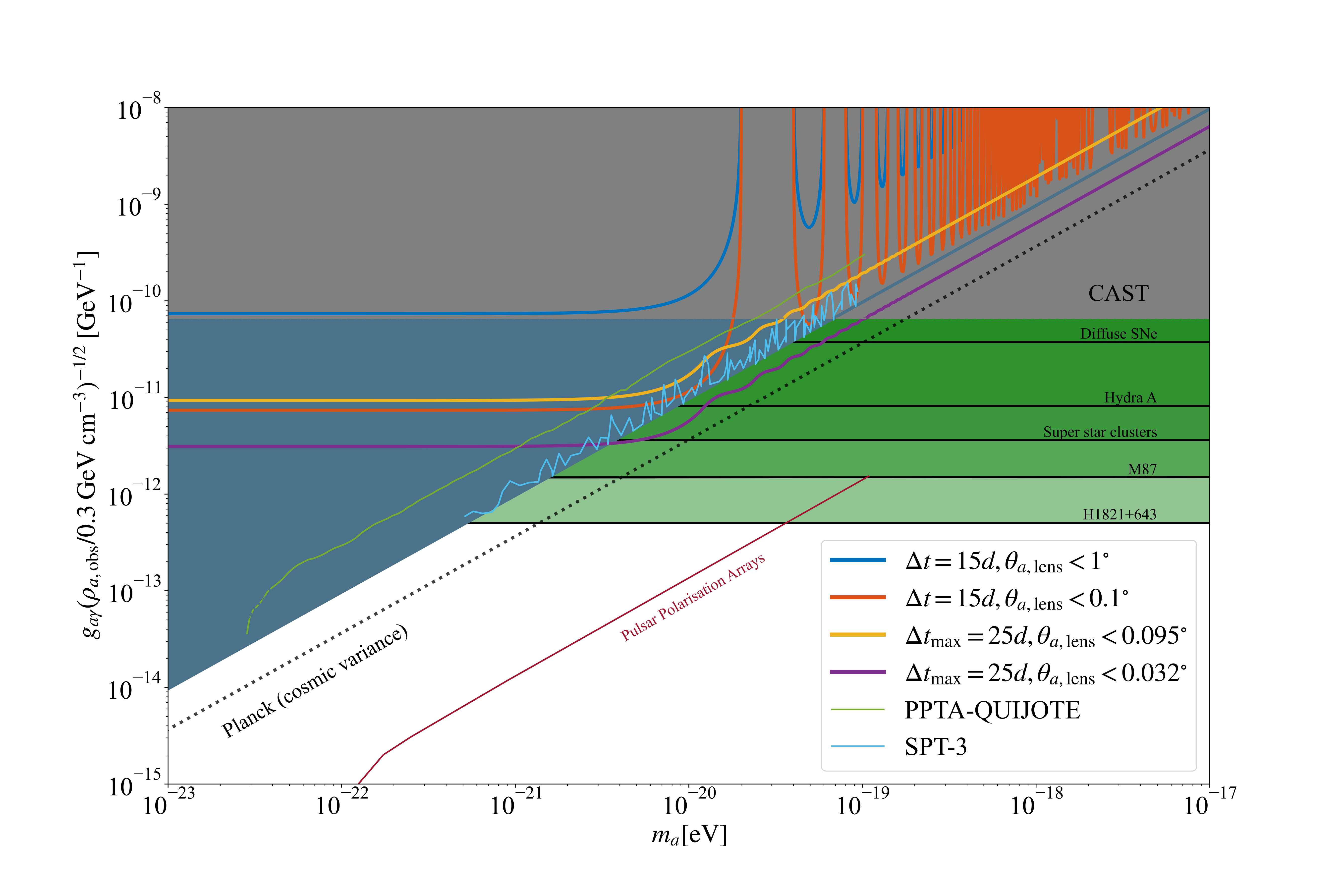}
    \caption{The thick blue and red lines denote constraints on $g_{a\gamma}$ from a single lensed FRB system. The yellow and purple thick lines represent constraints on $g_{a\gamma}$ from 10 lensed FRB systems. The parameters used in these calculations are listed in the box in the lower right corner. For the sake of comparison, we also show results from the CAST \citep{2017NatPh..13..584A} (gray), CMB polarization measurements~\citep{2020A&A...641A...6P} (dark blue, black), and some other astrophysical methods \citep{2022JCAP...06..014C,2022PhRvD.106d2011F,2022PhRvD.105f3028C,2013ApJ...772...44W,2020PhRvL.125z1102D,2017JCAP...12..036M,2022MNRAS.510.1264S,2023PhRvL.130l1401L}. It should be noted that not all currently available constraints are included in this plot to avoid over-crowdedness. Please refer to the following link\footnote{https://raw.githubusercontent.com/cajohare/AxionLimits/master/plots/plots\_png/AxionPhoton\_Ultralight\_with\_Projections.png} for a more complete presentation of existing constraints. }
    \label{fig:ma_gay}
\end{figure*}

Subsequently, $\langle|\Delta \theta_{a, \text {lens}}|\rangle$ is
\begin{equation} \label{eq7}
    \begin{aligned}
        \langle|\Delta \theta_{a, \text {lens}}|\rangle 
        & = \frac{2K}{\pi\Delta t_\mathrm{max}}\int_{0}^{\Delta t_\mathrm{max}} \left | \sin \left[\frac{m_a \Delta t}{2} \right]\right| d(\Delta t) \\
        & = \frac{4K}{m_a\pi\Delta t_\mathrm{max}}(2n+1-\cos\eta).
    \end{aligned}
\end{equation}
Here $m_a \Delta t_\mathrm{max} / 2$ is  $m_a \Delta t_\mathrm{max} / 2 = n\pi + \eta $, where $n \in \mathbb{N}$ and $\eta < \pi$.
Both the yellow and purple thick lines in Figure \ref{fig:ma_gay} are the expected upper limits derived from Equation \ref{eq7}, assuming 10 gravitationally lensed FRBs. The angle of polarization calculated from the yellow line is $ 0.3^\circ / \sqrt{10} \simeq 0.095^\circ $, corresponding to the polarization angle limit of an on-off FRB source. Meanwhile, the angle of polarization calculated by the purple line is $0.1^\circ/\sqrt{10} \simeq 0.032^\circ$, which is the angle of polarization corresponding to the statistically estimated repeating FRB.

\section{Conclusions and Discussions} \label{sec:con}
In this paper, we have proposed and established differential birefringence from gravitationally lensed FRB systems as a robust probe of Galactic axion dark matter background. With the state-of-the-art observations and data reduction techniques, we obtained that the axion-photon coupling under the ADM background could be constrained to be $g_{a\gamma} < 7.3 \times 10^{-11}~\mathrm{GeV^{-1}}$ for an axion mass $m_a\sim10^{-20}~\mathrm{eV}$ from a single lensed FRB system. It is most notably that, in addition to merits of traditional lensed quasar systems, lensed FRBs are ideal and promising probes to detect local axion field near Earth or in Milky Way thanks to the unique observational properties of FRB sources, such as millisecond duration and highly polarized emission. Constraints from this probe will be complementary and provide cross-check with other astrophysical or laboratorial searches. This will be of great value in exploring the nature of local axion field around Earth or in Milky Way. For instance, preliminary analysis suggests that constraints on $g_{a\gamma}$ from a single lensed repeating FRB would be comparable with results from the currently existing CMB polarization measurements~\citep{2022PhRvD.106d2011F}. Meanwhile, this probe will complement the parameter space by the CMB polarization measurements (several orders of magnitude for the mass range) and overcome the cosmic variance. This suggests that gravitationally lensed FRBs are an effective way to search for Galactic ultra-light axions. Moreover, as shown in Fig.~\ref{fig:ma_gay}, constraints or detection of axions from differential birefringence measurements in lensed FRB systems provide significantly stronger constraints compared to those obtained from the solar axion experiments, such as CAST~\citep{2017NatPh..13..584A}.

In addition, Eq. \ref{eq4} shows that the polarization angle difference is expected to oscillate with a period $T_a=2\pi/m_a$. Considering the observational characteristics of repeating FRBs, it is likely to report a large number of bursts in a few days, and it is foreseen that good limits can be obtained for periods less than $\sim 10$ days $(m_a>4.7 \times 10^{-21} \mathrm{eV})$. Note that it is more difficult to impose long period limits, since repeating FRBs usually do not remain active for a very long time. For the size of the angle $\Delta \theta_{a, \text {lens}}$ in this work, it is only a relatively simple estimate. The actual situation may be more complicated. Since the observation times of the two images are different, this may introduce additional systematic errors which are needed to be carefully considered. 

It is worth looking forward to the revolutionary development of radio observation equipment in the future. Once the SKA starts operating, it is expected to observe hundreds of FRBs every day. For FRBs with $z>1$, the probability of being lensed is $10^{-4}$. Thus, in the future, the SKA is expected to discover $>10$ strongly lensed FRBs per year. According to conservative estimates based on past work, within a decade of SKA operation, it is expected to detect $\mathcal{O}(10)$ strongly lensed FRB events~\citep{2018NatCo...9.3833L}, and the SKA can achieve high-precision polarization and position observations for these events. Overall, the SKA is expected to play a crucial role in our quest of understanding the nature of ultralight axions, or even the nature of dark matter. 

\begin{acknowledgments}
We would like to thank Jinchen Jiang for helpful discussions. This work was supported by the National Key Research and Development Program of China Grant No. 2021YFC2203001, the National Natural Science Foundation of China under Grants Nos. 12322301, 12222302, 12275021, 12021003, and 11920101003, National SKA Program of China (2022SKA0130100), the science research grants from the China Manned Space Project with No. CMS-CSST-2021-B11, the Strategic Priority Research Program of the Chinese Academy of Sciences, Grant No. XDB23040100, and the Interdiscipline Research Funds of Beijing Normal University.
\end{acknowledgments}

\bibliography{lensFRB}

\begin{thebibliography}{80}%
\makeatletter
\providecommand \@ifxundefined [1]{%
 \@ifx{#1\undefined}
}%
\providecommand \@ifnum [1]{%
 \ifnum #1\expandafter \@firstoftwo
 \else \expandafter \@secondoftwo
 \fi
}%
\providecommand \@ifx [1]{%
 \ifx #1\expandafter \@firstoftwo
 \else \expandafter \@secondoftwo
 \fi
}%
\providecommand \natexlab [1]{#1}%
\providecommand \enquote  [1]{``#1''}%
\providecommand \bibnamefont  [1]{#1}%
\providecommand \bibfnamefont [1]{#1}%
\providecommand \citenamefont [1]{#1}%
\providecommand \href@noop [0]{\@secondoftwo}%
\providecommand \href [0]{\begingroup \@sanitize@url \@href}%
\providecommand \@href[1]{\@@startlink{#1}\@@href}%
\providecommand \@@href[1]{\endgroup#1\@@endlink}%
\providecommand \@sanitize@url [0]{\catcode `\\12\catcode `\$12\catcode `\&12\catcode `\#12\catcode `\^12\catcode `\_12\catcode `\%12\relax}%
\providecommand \@@startlink[1]{}%
\providecommand \@@endlink[0]{}%
\providecommand \url  [0]{\begingroup\@sanitize@url \@url }%
\providecommand \@url [1]{\endgroup\@href {#1}{\urlprefix }}%
\providecommand \urlprefix  [0]{URL }%
\providecommand \Eprint [0]{\href }%
\providecommand \doibase [0]{https://doi.org/}%
\providecommand \selectlanguage [0]{\@gobble}%
\providecommand \bibinfo  [0]{\@secondoftwo}%
\providecommand \bibfield  [0]{\@secondoftwo}%
\providecommand \translation [1]{[#1]}%
\providecommand \BibitemOpen [0]{}%
\providecommand \bibitemStop [0]{}%
\providecommand \bibitemNoStop [0]{.\EOS\space}%
\providecommand \EOS [0]{\spacefactor3000\relax}%
\providecommand \BibitemShut  [1]{\csname bibitem#1\endcsname}%
\let\auto@bib@innerbib\@empty
\bibitem [{\citenamefont {{Bergstr{\"o}m}}(2000)}]{2000RPPh...63..793B}%
  \BibitemOpen
  \bibfield  {author} {\bibinfo {author} {\bibfnamefont {L.}~\bibnamefont {{Bergstr{\"o}m}}},\ }\bibfield  {title} {\bibinfo {title} {{Non-baryonic dark matter: observational evidence and detection methods}},\ }\href {https://doi.org/10.1088/0034-4885/63/5/2r3} {\bibfield  {journal} {\bibinfo  {journal} {Reports on Progress in Physics}\ }\textbf {\bibinfo {volume} {63}},\ \bibinfo {pages} {793} (\bibinfo {year} {2000})},\ \Eprint {https://arxiv.org/abs/hep-ph/0002126} {arXiv:hep-ph/0002126 [hep-ph]} \BibitemShut {NoStop}%
\bibitem [{\citenamefont {{Bertone}}\ \emph {et~al.}(2005)\citenamefont {{Bertone}}, \citenamefont {{Hooper}},\ and\ \citenamefont {{Silk}}}]{2005PhR...405..279B}%
  \BibitemOpen
  \bibfield  {author} {\bibinfo {author} {\bibfnamefont {G.}~\bibnamefont {{Bertone}}}, \bibinfo {author} {\bibfnamefont {D.}~\bibnamefont {{Hooper}}},\ and\ \bibinfo {author} {\bibfnamefont {J.}~\bibnamefont {{Silk}}},\ }\bibfield  {title} {\bibinfo {title} {{Particle dark matter: evidence, candidates and constraints}},\ }\href {https://doi.org/10.1016/j.physrep.2004.08.031} {\bibfield  {journal} {\bibinfo  {journal} {Physics Reports}\ }\textbf {\bibinfo {volume} {405}},\ \bibinfo {pages} {279} (\bibinfo {year} {2005})},\ \Eprint {https://arxiv.org/abs/hep-ph/0404175} {arXiv:hep-ph/0404175 [hep-ph]} \BibitemShut {NoStop}%
\bibitem [{\citenamefont {{Duffy}}\ and\ \citenamefont {{van Bibber}}(2009)}]{2009NJPh...11j5008D}%
  \BibitemOpen
  \bibfield  {author} {\bibinfo {author} {\bibfnamefont {L.~D.}\ \bibnamefont {{Duffy}}}\ and\ \bibinfo {author} {\bibfnamefont {K.}~\bibnamefont {{van Bibber}}},\ }\bibfield  {title} {\bibinfo {title} {{Axions as dark matter particles}},\ }\href {https://doi.org/10.1088/1367-2630/11/10/105008} {\bibfield  {journal} {\bibinfo  {journal} {New Journal of Physics}\ }\textbf {\bibinfo {volume} {11}},\ \bibinfo {eid} {105008} (\bibinfo {year} {2009})},\ \Eprint {https://arxiv.org/abs/0904.3346} {arXiv:0904.3346 [hep-ph]} \BibitemShut {NoStop}%
\bibitem [{\citenamefont {{Marsh}}(2016)}]{2016PhR...643....1M}%
  \BibitemOpen
  \bibfield  {author} {\bibinfo {author} {\bibfnamefont {D.~J.~E.}\ \bibnamefont {{Marsh}}},\ }\bibfield  {title} {\bibinfo {title} {{Axion cosmology}},\ }\href {https://doi.org/10.1016/j.physrep.2016.06.005} {\bibfield  {journal} {\bibinfo  {journal} {Physics Reports}\ }\textbf {\bibinfo {volume} {643}},\ \bibinfo {pages} {1} (\bibinfo {year} {2016})},\ \Eprint {https://arxiv.org/abs/1510.07633} {arXiv:1510.07633 [astro-ph.CO]} \BibitemShut {NoStop}%
\bibitem [{\citenamefont {{Graham}}\ \emph {et~al.}(2015)\citenamefont {{Graham}}, \citenamefont {{Irastorza}}, \citenamefont {{Lamoreaux}}, \citenamefont {{Lindner}},\ and\ \citenamefont {{van Bibber}}}]{2015ARNPS..65..485G}%
  \BibitemOpen
  \bibfield  {author} {\bibinfo {author} {\bibfnamefont {P.~W.}\ \bibnamefont {{Graham}}}, \bibinfo {author} {\bibfnamefont {I.~G.}\ \bibnamefont {{Irastorza}}}, \bibinfo {author} {\bibfnamefont {S.~K.}\ \bibnamefont {{Lamoreaux}}}, \bibinfo {author} {\bibfnamefont {A.}~\bibnamefont {{Lindner}}},\ and\ \bibinfo {author} {\bibfnamefont {K.~A.}\ \bibnamefont {{van Bibber}}},\ }\bibfield  {title} {\bibinfo {title} {{Experimental Searches for the Axion and Axion-Like Particles}},\ }\href {https://doi.org/10.1146/annurev-nucl-102014-022120} {\bibfield  {journal} {\bibinfo  {journal} {Annual Review of Nuclear and Particle Science}\ }\textbf {\bibinfo {volume} {65}},\ \bibinfo {pages} {485} (\bibinfo {year} {2015})},\ \Eprint {https://arxiv.org/abs/1602.00039} {arXiv:1602.00039 [hep-ex]} \BibitemShut {NoStop}%
\bibitem [{\citenamefont {{Weinberg}}\ \emph {et~al.}(2015)\citenamefont {{Weinberg}}, \citenamefont {{Bullock}}, \citenamefont {{Governato}}, \citenamefont {{Kuzio de Naray}},\ and\ \citenamefont {{Peter}}}]{2015PNAS..11212249W}%
  \BibitemOpen
  \bibfield  {author} {\bibinfo {author} {\bibfnamefont {D.~H.}\ \bibnamefont {{Weinberg}}}, \bibinfo {author} {\bibfnamefont {J.~S.}\ \bibnamefont {{Bullock}}}, \bibinfo {author} {\bibfnamefont {F.}~\bibnamefont {{Governato}}}, \bibinfo {author} {\bibfnamefont {R.}~\bibnamefont {{Kuzio de Naray}}},\ and\ \bibinfo {author} {\bibfnamefont {A.~H.~G.}\ \bibnamefont {{Peter}}},\ }\bibfield  {title} {\bibinfo {title} {{Cold dark matter: Controversies on small scales}},\ }\href {https://doi.org/10.1073/pnas.1308716112} {\bibfield  {journal} {\bibinfo  {journal} {Proceedings of the National Academy of Science}\ }\textbf {\bibinfo {volume} {112}},\ \bibinfo {pages} {12249} (\bibinfo {year} {2015})},\ \Eprint {https://arxiv.org/abs/1306.0913} {arXiv:1306.0913 [astro-ph.CO]} \BibitemShut {NoStop}%
\bibitem [{\citenamefont {{Hu}}\ \emph {et~al.}(2000)\citenamefont {{Hu}}, \citenamefont {{Barkana}},\ and\ \citenamefont {{Gruzinov}}}]{2000PhRvL..85.1158H}%
  \BibitemOpen
  \bibfield  {author} {\bibinfo {author} {\bibfnamefont {W.}~\bibnamefont {{Hu}}}, \bibinfo {author} {\bibfnamefont {R.}~\bibnamefont {{Barkana}}},\ and\ \bibinfo {author} {\bibfnamefont {A.}~\bibnamefont {{Gruzinov}}},\ }\bibfield  {title} {\bibinfo {title} {{Fuzzy Cold Dark Matter: The Wave Properties of Ultralight Particles}},\ }\href {https://doi.org/10.1103/PhysRevLett.85.1158} {\bibfield  {journal} {\bibinfo  {journal} {\prl}\ }\textbf {\bibinfo {volume} {85}},\ \bibinfo {pages} {1158} (\bibinfo {year} {2000})},\ \Eprint {https://arxiv.org/abs/astro-ph/0003365} {arXiv:astro-ph/0003365 [astro-ph]} \BibitemShut {NoStop}%
\bibitem [{\citenamefont {{Hui}}\ \emph {et~al.}(2017)\citenamefont {{Hui}}, \citenamefont {{Ostriker}}, \citenamefont {{Tremaine}},\ and\ \citenamefont {{Witten}}}]{2017PhRvD..95d3541H}%
  \BibitemOpen
  \bibfield  {author} {\bibinfo {author} {\bibfnamefont {L.}~\bibnamefont {{Hui}}}, \bibinfo {author} {\bibfnamefont {J.~P.}\ \bibnamefont {{Ostriker}}}, \bibinfo {author} {\bibfnamefont {S.}~\bibnamefont {{Tremaine}}},\ and\ \bibinfo {author} {\bibfnamefont {E.}~\bibnamefont {{Witten}}},\ }\bibfield  {title} {\bibinfo {title} {{Ultralight scalars as cosmological dark matter}},\ }\href {https://doi.org/10.1103/PhysRevD.95.043541} {\bibfield  {journal} {\bibinfo  {journal} {\prd}\ }\textbf {\bibinfo {volume} {95}},\ \bibinfo {eid} {043541} (\bibinfo {year} {2017})},\ \Eprint {https://arxiv.org/abs/1610.08297} {arXiv:1610.08297 [astro-ph.CO]} \BibitemShut {NoStop}%
\bibitem [{\citenamefont {{Schive}}\ \emph {et~al.}(2014)\citenamefont {{Schive}}, \citenamefont {{Chiueh}},\ and\ \citenamefont {{Broadhurst}}}]{2014NatPh..10..496S}%
  \BibitemOpen
  \bibfield  {author} {\bibinfo {author} {\bibfnamefont {H.-Y.}\ \bibnamefont {{Schive}}}, \bibinfo {author} {\bibfnamefont {T.}~\bibnamefont {{Chiueh}}},\ and\ \bibinfo {author} {\bibfnamefont {T.}~\bibnamefont {{Broadhurst}}},\ }\bibfield  {title} {\bibinfo {title} {{Cosmic structure as the quantum interference of a coherent dark wave}},\ }\href {https://doi.org/10.1038/nphys2996} {\bibfield  {journal} {\bibinfo  {journal} {Nature Physics}\ }\textbf {\bibinfo {volume} {10}},\ \bibinfo {pages} {496} (\bibinfo {year} {2014})},\ \Eprint {https://arxiv.org/abs/1406.6586} {arXiv:1406.6586 [astro-ph.GA]} \BibitemShut {NoStop}%
\bibitem [{\citenamefont {{Conlon}}(2006)}]{2006JHEP...05..078C}%
  \BibitemOpen
  \bibfield  {author} {\bibinfo {author} {\bibfnamefont {J.~P.}\ \bibnamefont {{Conlon}}},\ }\bibfield  {title} {\bibinfo {title} {{The QCD axion and moduli stabilisation}},\ }\href {https://doi.org/10.1088/1126-6708/2006/05/078} {\bibfield  {journal} {\bibinfo  {journal} {Journal of High Energy Physics}\ }\textbf {\bibinfo {volume} {2006}},\ \bibinfo {eid} {078} (\bibinfo {year} {2006})},\ \Eprint {https://arxiv.org/abs/hep-th/0602233} {arXiv:hep-th/0602233 [hep-th]} \BibitemShut {NoStop}%
\bibitem [{\citenamefont {{Arvanitaki}}\ \emph {et~al.}(2010)\citenamefont {{Arvanitaki}}, \citenamefont {{Dimopoulos}}, \citenamefont {{Dubovsky}}, \citenamefont {{Kaloper}},\ and\ \citenamefont {{March-Russell}}}]{2010PhRvD..81l3530A}%
  \BibitemOpen
  \bibfield  {author} {\bibinfo {author} {\bibfnamefont {A.}~\bibnamefont {{Arvanitaki}}}, \bibinfo {author} {\bibfnamefont {S.}~\bibnamefont {{Dimopoulos}}}, \bibinfo {author} {\bibfnamefont {S.}~\bibnamefont {{Dubovsky}}}, \bibinfo {author} {\bibfnamefont {N.}~\bibnamefont {{Kaloper}}},\ and\ \bibinfo {author} {\bibfnamefont {J.}~\bibnamefont {{March-Russell}}},\ }\bibfield  {title} {\bibinfo {title} {{String axiverse}},\ }\href {https://doi.org/10.1103/PhysRevD.81.123530} {\bibfield  {journal} {\bibinfo  {journal} {\prd}\ }\textbf {\bibinfo {volume} {81}},\ \bibinfo {eid} {123530} (\bibinfo {year} {2010})},\ \Eprint {https://arxiv.org/abs/0905.4720} {arXiv:0905.4720 [hep-th]} \BibitemShut {NoStop}%
\bibitem [{\citenamefont {{Peccei}}\ and\ \citenamefont {{Quinn}}(1977)}]{1977PhRvL..38.1440P}%
  \BibitemOpen
  \bibfield  {author} {\bibinfo {author} {\bibfnamefont {R.~D.}\ \bibnamefont {{Peccei}}}\ and\ \bibinfo {author} {\bibfnamefont {H.~R.}\ \bibnamefont {{Quinn}}},\ }\bibfield  {title} {\bibinfo {title} {{CP conservation in the presence of pseudoparticles}},\ }\href {https://doi.org/10.1103/PhysRevLett.38.1440} {\bibfield  {journal} {\bibinfo  {journal} {\prl}\ }\textbf {\bibinfo {volume} {38}},\ \bibinfo {pages} {1440} (\bibinfo {year} {1977})}\BibitemShut {NoStop}%
\bibitem [{\citenamefont {{Weinberg}}(1978)}]{1978PhRvL..40..223W}%
  \BibitemOpen
  \bibfield  {author} {\bibinfo {author} {\bibfnamefont {S.}~\bibnamefont {{Weinberg}}},\ }\bibfield  {title} {\bibinfo {title} {{A new light boson?}},\ }\href {https://doi.org/10.1103/PhysRevLett.40.223} {\bibfield  {journal} {\bibinfo  {journal} {\prl}\ }\textbf {\bibinfo {volume} {40}},\ \bibinfo {pages} {223} (\bibinfo {year} {1978})}\BibitemShut {NoStop}%
\bibitem [{\citenamefont {{Wilczek}}(1978)}]{1978PhRvL..40..279W}%
  \BibitemOpen
  \bibfield  {author} {\bibinfo {author} {\bibfnamefont {F.}~\bibnamefont {{Wilczek}}},\ }\bibfield  {title} {\bibinfo {title} {{Problem of strong P and T invariance in the presence of instantons}},\ }\href {https://doi.org/10.1103/PhysRevLett.40.279} {\bibfield  {journal} {\bibinfo  {journal} {\prl}\ }\textbf {\bibinfo {volume} {40}},\ \bibinfo {pages} {279} (\bibinfo {year} {1978})}\BibitemShut {NoStop}%
\bibitem [{\citenamefont {{Ehret}}\ \emph {et~al.}(2010)\citenamefont {{Ehret}}, \citenamefont {{Frede}}, \citenamefont {{Ghazaryan}}, \citenamefont {{Hildebrandt}}, \citenamefont {{Knabbe}}, \citenamefont {{Kracht}}, \citenamefont {{Lindner}}, \citenamefont {{List}}, \citenamefont {{Meier}}, \citenamefont {{Meyer}}, \citenamefont {{Notz}}, \citenamefont {{Redondo}}, \citenamefont {{Ringwald}}, \citenamefont {{Wiedemann}},\ and\ \citenamefont {{Willke}}}]{2010PhLB..689..149E}%
  \BibitemOpen
  \bibfield  {author} {\bibinfo {author} {\bibfnamefont {K.}~\bibnamefont {{Ehret}}}, \bibinfo {author} {\bibfnamefont {M.}~\bibnamefont {{Frede}}}, \bibinfo {author} {\bibfnamefont {S.}~\bibnamefont {{Ghazaryan}}}, \bibinfo {author} {\bibfnamefont {M.}~\bibnamefont {{Hildebrandt}}}, \bibinfo {author} {\bibfnamefont {E.-A.}\ \bibnamefont {{Knabbe}}}, \bibinfo {author} {\bibfnamefont {D.}~\bibnamefont {{Kracht}}}, \bibinfo {author} {\bibfnamefont {A.}~\bibnamefont {{Lindner}}}, \bibinfo {author} {\bibfnamefont {J.}~\bibnamefont {{List}}}, \bibinfo {author} {\bibfnamefont {T.}~\bibnamefont {{Meier}}}, \bibinfo {author} {\bibfnamefont {N.}~\bibnamefont {{Meyer}}}, \bibinfo {author} {\bibfnamefont {D.}~\bibnamefont {{Notz}}}, \bibinfo {author} {\bibfnamefont {J.}~\bibnamefont {{Redondo}}}, \bibinfo {author} {\bibfnamefont {A.}~\bibnamefont {{Ringwald}}}, \bibinfo {author} {\bibfnamefont {G.}~\bibnamefont {{Wiedemann}}},\ and\ \bibinfo {author} {\bibfnamefont {B.}~\bibnamefont {{Willke}}},\ }\bibfield  {title}
  {\bibinfo {title} {{New ALPS results on hidden-sector lightweights}},\ }\href {https://doi.org/10.1016/j.physletb.2010.04.066} {\bibfield  {journal} {\bibinfo  {journal} {Physics Letters B}\ }\textbf {\bibinfo {volume} {689}},\ \bibinfo {pages} {149} (\bibinfo {year} {2010})},\ \Eprint {https://arxiv.org/abs/1004.1313} {arXiv:1004.1313 [hep-ex]} \BibitemShut {NoStop}%
\bibitem [{\citenamefont {{B{\"a}hre}}\ \emph {et~al.}(2013)\citenamefont {{B{\"a}hre}}, \citenamefont {{D{\"o}brich}}, \citenamefont {{Dreyling-Eschweiler}}, \citenamefont {{Ghazaryan}}, \citenamefont {{Hodajerdi}}, \citenamefont {{Horns}}, \citenamefont {{Januschek}}, \citenamefont {{Knabbe}}, \citenamefont {{Lindner}}, \citenamefont {{Notz}}, \citenamefont {{Ringwald}}, \citenamefont {{von Seggern}}, \citenamefont {{Stromhagen}}, \citenamefont {{Trines}},\ and\ \citenamefont {{Willke}}}]{2013JInst...8.9001B}%
  \BibitemOpen
  \bibfield  {author} {\bibinfo {author} {\bibfnamefont {R.}~\bibnamefont {{B{\"a}hre}}}, \bibinfo {author} {\bibfnamefont {B.}~\bibnamefont {{D{\"o}brich}}}, \bibinfo {author} {\bibfnamefont {J.}~\bibnamefont {{Dreyling-Eschweiler}}}, \bibinfo {author} {\bibfnamefont {S.}~\bibnamefont {{Ghazaryan}}}, \bibinfo {author} {\bibfnamefont {R.}~\bibnamefont {{Hodajerdi}}}, \bibinfo {author} {\bibfnamefont {D.}~\bibnamefont {{Horns}}}, \bibinfo {author} {\bibfnamefont {F.}~\bibnamefont {{Januschek}}}, \bibinfo {author} {\bibfnamefont {E.~A.}\ \bibnamefont {{Knabbe}}}, \bibinfo {author} {\bibfnamefont {A.}~\bibnamefont {{Lindner}}}, \bibinfo {author} {\bibfnamefont {D.}~\bibnamefont {{Notz}}}, \bibinfo {author} {\bibfnamefont {A.}~\bibnamefont {{Ringwald}}}, \bibinfo {author} {\bibfnamefont {J.~E.}\ \bibnamefont {{von Seggern}}}, \bibinfo {author} {\bibfnamefont {R.}~\bibnamefont {{Stromhagen}}}, \bibinfo {author} {\bibfnamefont {D.}~\bibnamefont {{Trines}}},\ and\ \bibinfo {author} {\bibfnamefont {B.}~\bibnamefont
  {{Willke}}},\ }\bibfield  {title} {\bibinfo {title} {{Any light particle search II {\textemdash} Technical Design Report}},\ }\href {https://doi.org/10.1088/1748-0221/8/09/T09001} {\bibfield  {journal} {\bibinfo  {journal} {Journal of Instrumentation}\ }\textbf {\bibinfo {volume} {8}}\bibfield  {number} {\bibinfo  {number} { (9)},\ \bibinfo {eid} {T09001}},\ }\Eprint {https://arxiv.org/abs/1302.5647} {arXiv:1302.5647 [physics.ins-det]} \BibitemShut {NoStop}%
\bibitem [{\citenamefont {{Armengaud}}\ \emph {et~al.}(2014)\citenamefont {{Armengaud}}, \citenamefont {{Avignone}}, \citenamefont {{Betz}}, \citenamefont {{Brax}}, \citenamefont {{Brun}}, \citenamefont {{Cantatore}}, \citenamefont {{Carmona}}, \citenamefont {{Carosi}}, \citenamefont {{Caspers}}, \citenamefont {{Caspi}}, \citenamefont {{Cetin}}, \citenamefont {{Chelouche}}, \citenamefont {{Christensen}}, \citenamefont {{Dael}}, \citenamefont {{Dafni}}, \citenamefont {{Davenport}}, \citenamefont {{Derbin}}, \citenamefont {{Desch}}, \citenamefont {{Diago}}, \citenamefont {{D{\"o}brich}}, \citenamefont {{Dratchnev}}, \citenamefont {{Dudarev}}, \citenamefont {{Eleftheriadis}}, \citenamefont {{Fanourakis}}, \citenamefont {{Ferrer-Ribas}}, \citenamefont {{Gal{\'a}n}}, \citenamefont {{Garc{\'\i}a}}, \citenamefont {{Garza}}, \citenamefont {{Geralis}}, \citenamefont {{Gimeno}}, \citenamefont {{Giomataris}}, \citenamefont {{Gninenko}}, \citenamefont {{G{\'o}mez}}, \citenamefont {{Gonz{\'a}lez-D{\'\i}az}}, \citenamefont
  {{Guendelman}}, \citenamefont {{Hailey}}, \citenamefont {{Hiramatsu}}, \citenamefont {{Hoffmann}}, \citenamefont {{Horns}}, \citenamefont {{Iguaz}}, \citenamefont {{Irastorza}}, \citenamefont {{Isern}}, \citenamefont {{Imai}}, \citenamefont {{Jakobsen}}, \citenamefont {{Jaeckel}}, \citenamefont {{Jakov{\v{c}}i{\'c}}}, \citenamefont {{Kaminski}}, \citenamefont {{Kawasaki}}, \citenamefont {{Karuza}}, \citenamefont {{Kr{\v{c}}mar}}, \citenamefont {{Kousouris}}, \citenamefont {{Krieger}}, \citenamefont {{Laki{\'c}}}, \citenamefont {{Limousin}}, \citenamefont {{Lindner}}, \citenamefont {{Liolios}}, \citenamefont {{Luz{\'o}n}}, \citenamefont {{Matsuki}}, \citenamefont {{Muratova}}, \citenamefont {{Nones}}, \citenamefont {{Ortega}}, \citenamefont {{Papaevangelou}}, \citenamefont {{Pivovaroff}}, \citenamefont {{Raffelt}}, \citenamefont {{Redondo}}, \citenamefont {{Ringwald}}, \citenamefont {{Russenschuck}}, \citenamefont {{Ruz}}, \citenamefont {{Saikawa}}, \citenamefont {{Savvidis}}, \citenamefont {{Sekiguchi}},
  \citenamefont {{Semertzidis}}, \citenamefont {{Shilon}}, \citenamefont {{Sikivie}}, \citenamefont {{Silva}}, \citenamefont {{ten Kate}}, \citenamefont {{Tomas}}, \citenamefont {{Troitsky}}, \citenamefont {{Vafeiadis}}, \citenamefont {{van Bibber}}, \citenamefont {{Vedrine}}, \citenamefont {{Villar}}, \citenamefont {{Vogel}}, \citenamefont {{Walckiers}}, \citenamefont {{Weltman}}, \citenamefont {{Wester}}, \citenamefont {{Yildiz}},\ and\ \citenamefont {{Zioutas}}}]{2014JInst...9.5002A}%
  \BibitemOpen
  \bibfield  {author} {\bibinfo {author} {\bibfnamefont {E.}~\bibnamefont {{Armengaud}}}, \bibinfo {author} {\bibfnamefont {F.~T.}\ \bibnamefont {{Avignone}}}, \bibinfo {author} {\bibfnamefont {M.}~\bibnamefont {{Betz}}}, \bibinfo {author} {\bibfnamefont {P.}~\bibnamefont {{Brax}}}, \bibinfo {author} {\bibfnamefont {P.}~\bibnamefont {{Brun}}}, \bibinfo {author} {\bibfnamefont {G.}~\bibnamefont {{Cantatore}}}, \bibinfo {author} {\bibfnamefont {J.~M.}\ \bibnamefont {{Carmona}}}, \bibinfo {author} {\bibfnamefont {G.~P.}\ \bibnamefont {{Carosi}}}, \bibinfo {author} {\bibfnamefont {F.}~\bibnamefont {{Caspers}}}, \bibinfo {author} {\bibfnamefont {S.}~\bibnamefont {{Caspi}}}, \bibinfo {author} {\bibfnamefont {S.~A.}\ \bibnamefont {{Cetin}}}, \bibinfo {author} {\bibfnamefont {D.}~\bibnamefont {{Chelouche}}}, \bibinfo {author} {\bibfnamefont {F.~E.}\ \bibnamefont {{Christensen}}}, \bibinfo {author} {\bibfnamefont {A.}~\bibnamefont {{Dael}}}, \bibinfo {author} {\bibfnamefont {T.}~\bibnamefont {{Dafni}}}, \bibinfo
  {author} {\bibfnamefont {M.}~\bibnamefont {{Davenport}}}, \bibinfo {author} {\bibfnamefont {A.~V.}\ \bibnamefont {{Derbin}}}, \bibinfo {author} {\bibfnamefont {K.}~\bibnamefont {{Desch}}}, \bibinfo {author} {\bibfnamefont {A.}~\bibnamefont {{Diago}}}, \bibinfo {author} {\bibfnamefont {B.}~\bibnamefont {{D{\"o}brich}}}, \bibinfo {author} {\bibfnamefont {I.}~\bibnamefont {{Dratchnev}}}, \bibinfo {author} {\bibfnamefont {A.}~\bibnamefont {{Dudarev}}}, \bibinfo {author} {\bibfnamefont {C.}~\bibnamefont {{Eleftheriadis}}}, \bibinfo {author} {\bibfnamefont {G.}~\bibnamefont {{Fanourakis}}}, \bibinfo {author} {\bibfnamefont {E.}~\bibnamefont {{Ferrer-Ribas}}}, \bibinfo {author} {\bibfnamefont {J.}~\bibnamefont {{Gal{\'a}n}}}, \bibinfo {author} {\bibfnamefont {J.~A.}\ \bibnamefont {{Garc{\'\i}a}}}, \bibinfo {author} {\bibfnamefont {J.~G.}\ \bibnamefont {{Garza}}}, \bibinfo {author} {\bibfnamefont {T.}~\bibnamefont {{Geralis}}}, \bibinfo {author} {\bibfnamefont {B.}~\bibnamefont {{Gimeno}}}, \bibinfo {author}
  {\bibfnamefont {I.}~\bibnamefont {{Giomataris}}}, \bibinfo {author} {\bibfnamefont {S.}~\bibnamefont {{Gninenko}}}, \bibinfo {author} {\bibfnamefont {H.}~\bibnamefont {{G{\'o}mez}}}, \bibinfo {author} {\bibfnamefont {D.}~\bibnamefont {{Gonz{\'a}lez-D{\'\i}az}}}, \bibinfo {author} {\bibfnamefont {E.}~\bibnamefont {{Guendelman}}}, \bibinfo {author} {\bibfnamefont {C.~J.}\ \bibnamefont {{Hailey}}}, \bibinfo {author} {\bibfnamefont {T.}~\bibnamefont {{Hiramatsu}}}, \bibinfo {author} {\bibfnamefont {D.~H.~H.}\ \bibnamefont {{Hoffmann}}}, \bibinfo {author} {\bibfnamefont {D.}~\bibnamefont {{Horns}}}, \bibinfo {author} {\bibfnamefont {F.~J.}\ \bibnamefont {{Iguaz}}}, \bibinfo {author} {\bibfnamefont {I.~G.}\ \bibnamefont {{Irastorza}}}, \bibinfo {author} {\bibfnamefont {J.}~\bibnamefont {{Isern}}}, \bibinfo {author} {\bibfnamefont {K.}~\bibnamefont {{Imai}}}, \bibinfo {author} {\bibfnamefont {A.~C.}\ \bibnamefont {{Jakobsen}}}, \bibinfo {author} {\bibfnamefont {J.}~\bibnamefont {{Jaeckel}}}, \bibinfo {author}
  {\bibfnamefont {K.}~\bibnamefont {{Jakov{\v{c}}i{\'c}}}}, \bibinfo {author} {\bibfnamefont {J.}~\bibnamefont {{Kaminski}}}, \bibinfo {author} {\bibfnamefont {M.}~\bibnamefont {{Kawasaki}}}, \bibinfo {author} {\bibfnamefont {M.}~\bibnamefont {{Karuza}}}, \bibinfo {author} {\bibfnamefont {M.}~\bibnamefont {{Kr{\v{c}}mar}}}, \bibinfo {author} {\bibfnamefont {K.}~\bibnamefont {{Kousouris}}}, \bibinfo {author} {\bibfnamefont {C.}~\bibnamefont {{Krieger}}}, \bibinfo {author} {\bibfnamefont {B.}~\bibnamefont {{Laki{\'c}}}}, \bibinfo {author} {\bibfnamefont {O.}~\bibnamefont {{Limousin}}}, \bibinfo {author} {\bibfnamefont {A.}~\bibnamefont {{Lindner}}}, \bibinfo {author} {\bibfnamefont {A.}~\bibnamefont {{Liolios}}}, \bibinfo {author} {\bibfnamefont {G.}~\bibnamefont {{Luz{\'o}n}}}, \bibinfo {author} {\bibfnamefont {S.}~\bibnamefont {{Matsuki}}}, \bibinfo {author} {\bibfnamefont {V.~N.}\ \bibnamefont {{Muratova}}}, \bibinfo {author} {\bibfnamefont {C.}~\bibnamefont {{Nones}}}, \bibinfo {author} {\bibfnamefont
  {I.}~\bibnamefont {{Ortega}}}, \bibinfo {author} {\bibfnamefont {T.}~\bibnamefont {{Papaevangelou}}}, \bibinfo {author} {\bibfnamefont {M.~J.}\ \bibnamefont {{Pivovaroff}}}, \bibinfo {author} {\bibfnamefont {G.}~\bibnamefont {{Raffelt}}}, \bibinfo {author} {\bibfnamefont {J.}~\bibnamefont {{Redondo}}}, \bibinfo {author} {\bibfnamefont {A.}~\bibnamefont {{Ringwald}}}, \bibinfo {author} {\bibfnamefont {S.}~\bibnamefont {{Russenschuck}}}, \bibinfo {author} {\bibfnamefont {J.}~\bibnamefont {{Ruz}}}, \bibinfo {author} {\bibfnamefont {K.}~\bibnamefont {{Saikawa}}}, \bibinfo {author} {\bibfnamefont {I.}~\bibnamefont {{Savvidis}}}, \bibinfo {author} {\bibfnamefont {T.}~\bibnamefont {{Sekiguchi}}}, \bibinfo {author} {\bibfnamefont {Y.~K.}\ \bibnamefont {{Semertzidis}}}, \bibinfo {author} {\bibfnamefont {I.}~\bibnamefont {{Shilon}}}, \bibinfo {author} {\bibfnamefont {P.}~\bibnamefont {{Sikivie}}}, \bibinfo {author} {\bibfnamefont {H.}~\bibnamefont {{Silva}}}, \bibinfo {author} {\bibfnamefont {H.}~\bibnamefont {{ten
  Kate}}}, \bibinfo {author} {\bibfnamefont {A.}~\bibnamefont {{Tomas}}}, \bibinfo {author} {\bibfnamefont {S.}~\bibnamefont {{Troitsky}}}, \bibinfo {author} {\bibfnamefont {T.}~\bibnamefont {{Vafeiadis}}}, \bibinfo {author} {\bibfnamefont {K.}~\bibnamefont {{van Bibber}}}, \bibinfo {author} {\bibfnamefont {P.}~\bibnamefont {{Vedrine}}}, \bibinfo {author} {\bibfnamefont {J.~A.}\ \bibnamefont {{Villar}}}, \bibinfo {author} {\bibfnamefont {J.~K.}\ \bibnamefont {{Vogel}}}, \bibinfo {author} {\bibfnamefont {L.}~\bibnamefont {{Walckiers}}}, \bibinfo {author} {\bibfnamefont {A.}~\bibnamefont {{Weltman}}}, \bibinfo {author} {\bibfnamefont {W.}~\bibnamefont {{Wester}}}, \bibinfo {author} {\bibfnamefont {S.~C.}\ \bibnamefont {{Yildiz}}},\ and\ \bibinfo {author} {\bibfnamefont {K.}~\bibnamefont {{Zioutas}}},\ }\bibfield  {title} {\bibinfo {title} {{Conceptual design of the International Axion Observatory (IAXO)}},\ }\href {https://doi.org/10.1088/1748-0221/9/05/T05002} {\bibfield  {journal} {\bibinfo  {journal}
  {Journal of Instrumentation}\ }\textbf {\bibinfo {volume} {9}}\bibfield  {number} {\bibinfo  {number} { (5)},\ \bibinfo {eid} {T05002}},\ }\Eprint {https://arxiv.org/abs/1401.3233} {arXiv:1401.3233 [physics.ins-det]} \BibitemShut {NoStop}%
\bibitem [{\citenamefont {{Anastassopoulos}}\ \emph {et~al.}(2017)\citenamefont {{Anastassopoulos}}, \citenamefont {{Aune}}, \citenamefont {{Barth}}, \citenamefont {{Belov}}, \citenamefont {{Br{\"a}uninger}}, \citenamefont {{Cantatore}}, \citenamefont {{Carmona}}, \citenamefont {{Castel}}, \citenamefont {{Cetin}}, \citenamefont {{Christensen}}, \citenamefont {{Collar}}, \citenamefont {{Dafni}}, \citenamefont {{Davenport}}, \citenamefont {{Decker}}, \citenamefont {{Dermenev}}, \citenamefont {{Desch}}, \citenamefont {{Eleftheriadis}}, \citenamefont {{Fanourakis}}, \citenamefont {{Ferrer-Ribas}}, \citenamefont {{Fischer}}, \citenamefont {{Garc{\'\i}a}}, \citenamefont {{Gardikiotis}}, \citenamefont {{Garza}}, \citenamefont {{Gazis}}, \citenamefont {{Geralis}}, \citenamefont {{Giomataris}}, \citenamefont {{Gninenko}}, \citenamefont {{Hailey}}, \citenamefont {{Hasinoff}}, \citenamefont {{Hoffmann}}, \citenamefont {{Iguaz}}, \citenamefont {{Irastorza}}, \citenamefont {{Jakobsen}}, \citenamefont {{Jacoby}},
  \citenamefont {{Jakov{\v{c}}i{\'c}}}, \citenamefont {{Kaminski}}, \citenamefont {{Karuza}}, \citenamefont {{Kralj}}, \citenamefont {{Kr{\v{c}}mar}}, \citenamefont {{Kostoglou}}, \citenamefont {{Krieger}}, \citenamefont {{Laki{\'c}}}, \citenamefont {{Laurent}}, \citenamefont {{Liolios}}, \citenamefont {{Ljubi{\v{c}}i{\'c}}}, \citenamefont {{Luz{\'o}n}}, \citenamefont {{Maroudas}}, \citenamefont {{Miceli}}, \citenamefont {{Neff}}, \citenamefont {{Ortega}}, \citenamefont {{Papaevangelou}}, \citenamefont {{Paraschou}}, \citenamefont {{Pivovaroff}}, \citenamefont {{Raffelt}}, \citenamefont {{Rosu}}, \citenamefont {{Ruz}}, \citenamefont {{Ch{\'o}liz}}, \citenamefont {{Savvidis}}, \citenamefont {{Schmidt}}, \citenamefont {{Semertzidis}}, \citenamefont {{Solanki}}, \citenamefont {{Stewart}}, \citenamefont {{Vafeiadis}}, \citenamefont {{Vogel}}, \citenamefont {{Yildiz}},\ and\ \citenamefont {{Zioutas}}}]{2017NatPh..13..584A}%
  \BibitemOpen
  \bibfield  {author} {\bibinfo {author} {\bibfnamefont {V.}~\bibnamefont {{Anastassopoulos}}}, \bibinfo {author} {\bibfnamefont {S.}~\bibnamefont {{Aune}}}, \bibinfo {author} {\bibfnamefont {K.}~\bibnamefont {{Barth}}}, \bibinfo {author} {\bibfnamefont {A.}~\bibnamefont {{Belov}}}, \bibinfo {author} {\bibfnamefont {H.}~\bibnamefont {{Br{\"a}uninger}}}, \bibinfo {author} {\bibfnamefont {G.}~\bibnamefont {{Cantatore}}}, \bibinfo {author} {\bibfnamefont {J.~M.}\ \bibnamefont {{Carmona}}}, \bibinfo {author} {\bibfnamefont {J.~F.}\ \bibnamefont {{Castel}}}, \bibinfo {author} {\bibfnamefont {S.~A.}\ \bibnamefont {{Cetin}}}, \bibinfo {author} {\bibfnamefont {F.}~\bibnamefont {{Christensen}}}, \bibinfo {author} {\bibfnamefont {J.~I.}\ \bibnamefont {{Collar}}}, \bibinfo {author} {\bibfnamefont {T.}~\bibnamefont {{Dafni}}}, \bibinfo {author} {\bibfnamefont {M.}~\bibnamefont {{Davenport}}}, \bibinfo {author} {\bibfnamefont {T.~A.}\ \bibnamefont {{Decker}}}, \bibinfo {author} {\bibfnamefont {A.}~\bibnamefont
  {{Dermenev}}}, \bibinfo {author} {\bibfnamefont {K.}~\bibnamefont {{Desch}}}, \bibinfo {author} {\bibfnamefont {C.}~\bibnamefont {{Eleftheriadis}}}, \bibinfo {author} {\bibfnamefont {G.}~\bibnamefont {{Fanourakis}}}, \bibinfo {author} {\bibfnamefont {E.}~\bibnamefont {{Ferrer-Ribas}}}, \bibinfo {author} {\bibfnamefont {H.}~\bibnamefont {{Fischer}}}, \bibinfo {author} {\bibfnamefont {J.~A.}\ \bibnamefont {{Garc{\'\i}a}}}, \bibinfo {author} {\bibfnamefont {A.}~\bibnamefont {{Gardikiotis}}}, \bibinfo {author} {\bibfnamefont {J.~G.}\ \bibnamefont {{Garza}}}, \bibinfo {author} {\bibfnamefont {E.~N.}\ \bibnamefont {{Gazis}}}, \bibinfo {author} {\bibfnamefont {T.}~\bibnamefont {{Geralis}}}, \bibinfo {author} {\bibfnamefont {I.}~\bibnamefont {{Giomataris}}}, \bibinfo {author} {\bibfnamefont {S.}~\bibnamefont {{Gninenko}}}, \bibinfo {author} {\bibfnamefont {C.~J.}\ \bibnamefont {{Hailey}}}, \bibinfo {author} {\bibfnamefont {M.~D.}\ \bibnamefont {{Hasinoff}}}, \bibinfo {author} {\bibfnamefont {D.~H.~H.}\ \bibnamefont
  {{Hoffmann}}}, \bibinfo {author} {\bibfnamefont {F.~J.}\ \bibnamefont {{Iguaz}}}, \bibinfo {author} {\bibfnamefont {I.~G.}\ \bibnamefont {{Irastorza}}}, \bibinfo {author} {\bibfnamefont {A.}~\bibnamefont {{Jakobsen}}}, \bibinfo {author} {\bibfnamefont {J.}~\bibnamefont {{Jacoby}}}, \bibinfo {author} {\bibfnamefont {K.}~\bibnamefont {{Jakov{\v{c}}i{\'c}}}}, \bibinfo {author} {\bibfnamefont {J.}~\bibnamefont {{Kaminski}}}, \bibinfo {author} {\bibfnamefont {M.}~\bibnamefont {{Karuza}}}, \bibinfo {author} {\bibfnamefont {N.}~\bibnamefont {{Kralj}}}, \bibinfo {author} {\bibfnamefont {M.}~\bibnamefont {{Kr{\v{c}}mar}}}, \bibinfo {author} {\bibfnamefont {S.}~\bibnamefont {{Kostoglou}}}, \bibinfo {author} {\bibfnamefont {C.}~\bibnamefont {{Krieger}}}, \bibinfo {author} {\bibfnamefont {B.}~\bibnamefont {{Laki{\'c}}}}, \bibinfo {author} {\bibfnamefont {J.~M.}\ \bibnamefont {{Laurent}}}, \bibinfo {author} {\bibfnamefont {A.}~\bibnamefont {{Liolios}}}, \bibinfo {author} {\bibfnamefont {A.}~\bibnamefont
  {{Ljubi{\v{c}}i{\'c}}}}, \bibinfo {author} {\bibfnamefont {G.}~\bibnamefont {{Luz{\'o}n}}}, \bibinfo {author} {\bibfnamefont {M.}~\bibnamefont {{Maroudas}}}, \bibinfo {author} {\bibfnamefont {L.}~\bibnamefont {{Miceli}}}, \bibinfo {author} {\bibfnamefont {S.}~\bibnamefont {{Neff}}}, \bibinfo {author} {\bibfnamefont {I.}~\bibnamefont {{Ortega}}}, \bibinfo {author} {\bibfnamefont {T.}~\bibnamefont {{Papaevangelou}}}, \bibinfo {author} {\bibfnamefont {K.}~\bibnamefont {{Paraschou}}}, \bibinfo {author} {\bibfnamefont {M.~J.}\ \bibnamefont {{Pivovaroff}}}, \bibinfo {author} {\bibfnamefont {G.}~\bibnamefont {{Raffelt}}}, \bibinfo {author} {\bibfnamefont {M.}~\bibnamefont {{Rosu}}}, \bibinfo {author} {\bibfnamefont {J.}~\bibnamefont {{Ruz}}}, \bibinfo {author} {\bibfnamefont {E.~R.}\ \bibnamefont {{Ch{\'o}liz}}}, \bibinfo {author} {\bibfnamefont {I.}~\bibnamefont {{Savvidis}}}, \bibinfo {author} {\bibfnamefont {S.}~\bibnamefont {{Schmidt}}}, \bibinfo {author} {\bibfnamefont {Y.~K.}\ \bibnamefont {{Semertzidis}}},
  \bibinfo {author} {\bibfnamefont {S.~K.}\ \bibnamefont {{Solanki}}}, \bibinfo {author} {\bibfnamefont {L.}~\bibnamefont {{Stewart}}}, \bibinfo {author} {\bibfnamefont {T.}~\bibnamefont {{Vafeiadis}}}, \bibinfo {author} {\bibfnamefont {J.~K.}\ \bibnamefont {{Vogel}}}, \bibinfo {author} {\bibfnamefont {S.~C.}\ \bibnamefont {{Yildiz}}},\ and\ \bibinfo {author} {\bibfnamefont {K.}~\bibnamefont {{Zioutas}}},\ }\bibfield  {title} {\bibinfo {title} {{New CAST limit on the axion-photon interaction}},\ }\href {https://doi.org/10.1038/nphys4109} {\bibfield  {journal} {\bibinfo  {journal} {Nature Physics}\ }\textbf {\bibinfo {volume} {13}},\ \bibinfo {pages} {584} (\bibinfo {year} {2017})},\ \Eprint {https://arxiv.org/abs/1705.02290} {arXiv:1705.02290 [hep-ex]} \BibitemShut {NoStop}%
\bibitem [{\citenamefont {{Armengaud}}\ \emph {et~al.}(2019)\citenamefont {{Armengaud}}, \citenamefont {{Atti{\'e}}}, \citenamefont {{Basso}}, \citenamefont {{Brun}}, \citenamefont {{Bykovskiy}}, \citenamefont {{Carmona}}, \citenamefont {{Castel}}, \citenamefont {{Cebri{\'a}n}}, \citenamefont {{Cicoli}}, \citenamefont {{Civitani}}, \citenamefont {{Cogollos}}, \citenamefont {{Conlon}}, \citenamefont {{Costa}}, \citenamefont {{Dafni}}, \citenamefont {{Daido}}, \citenamefont {{Derbin}}, \citenamefont {{Descalle}}, \citenamefont {{Desch}}, \citenamefont {{Dratchnev}}, \citenamefont {{D{\"o}brich}}, \citenamefont {{Dudarev}}, \citenamefont {{Ferrer-Ribas}}, \citenamefont {{Fleck}}, \citenamefont {{Gal{\'a}n}}, \citenamefont {{Galanti}}, \citenamefont {{Garrido}}, \citenamefont {{Gascon}}, \citenamefont {{Gastaldo}}, \citenamefont {{Germani}}, \citenamefont {{Ghisellini}}, \citenamefont {{Giannotti}}, \citenamefont {{Giomataris}}, \citenamefont {{Gninenko}}, \citenamefont {{Golubev}}, \citenamefont {{Graciani}},
  \citenamefont {{Irastorza}}, \citenamefont {{Jakov{\v{c}}i{\'c}}}, \citenamefont {{Kaminski}}, \citenamefont {{Kr{\v{c}}mar}}, \citenamefont {{Krieger}}, \citenamefont {{Laki{\'c}}}, \citenamefont {{Lasserre}}, \citenamefont {{Laurent}}, \citenamefont {{Limousin}}, \citenamefont {{Lindner}}, \citenamefont {{Lomskaya}}, \citenamefont {{Lubsandorzhiev}}, \citenamefont {{Luz{\'o}n}}, \citenamefont {{Marsh}}, \citenamefont {{Margalejo}}, \citenamefont {{Mescia}}, \citenamefont {{Meyer}}, \citenamefont {{Miralda-Escud{\'e}}}, \citenamefont {{Mirallas}}, \citenamefont {{Muratova}}, \citenamefont {{Navick}}, \citenamefont {{Nones}}, \citenamefont {{Notari}}, \citenamefont {{Nozik}}, \citenamefont {{Ortiz de Sol{\'o}rzano}}, \citenamefont {{Pantuev}}, \citenamefont {{Papaevangelou}}, \citenamefont {{Pareschi}}, \citenamefont {{Perez}}, \citenamefont {{Picatoste}}, \citenamefont {{Pivovaroff}}, \citenamefont {{Redondo}}, \citenamefont {{Ringwald}}, \citenamefont {{Roncadelli}}, \citenamefont {{Ruiz-Ch{\'o}liz}},
  \citenamefont {{Ruz}}, \citenamefont {{Saikawa}}, \citenamefont {{Salvad{\'o}}}, \citenamefont {{Samperiz}}, \citenamefont {{Schiffer}}, \citenamefont {{Schmidt}}, \citenamefont {{Schneekloth}}, \citenamefont {{Schott}}, \citenamefont {{Silva}}, \citenamefont {{Tagliaferri}}, \citenamefont {{Takahashi}}, \citenamefont {{Tavecchio}}, \citenamefont {{ten Kate}}, \citenamefont {{Tkachev}}, \citenamefont {{Troitsky}}, \citenamefont {{Unzhakov}}, \citenamefont {{Vedrine}}, \citenamefont {{Vogel}}, \citenamefont {{Weinsheimer}}, \citenamefont {{Weltman}},\ and\ \citenamefont {{Yin}}}]{2019JCAP...06..047A}%
  \BibitemOpen
  \bibfield  {author} {\bibinfo {author} {\bibfnamefont {E.}~\bibnamefont {{Armengaud}}}, \bibinfo {author} {\bibfnamefont {D.}~\bibnamefont {{Atti{\'e}}}}, \bibinfo {author} {\bibfnamefont {S.}~\bibnamefont {{Basso}}}, \bibinfo {author} {\bibfnamefont {P.}~\bibnamefont {{Brun}}}, \bibinfo {author} {\bibfnamefont {N.}~\bibnamefont {{Bykovskiy}}}, \bibinfo {author} {\bibfnamefont {J.~M.}\ \bibnamefont {{Carmona}}}, \bibinfo {author} {\bibfnamefont {J.~F.}\ \bibnamefont {{Castel}}}, \bibinfo {author} {\bibfnamefont {S.}~\bibnamefont {{Cebri{\'a}n}}}, \bibinfo {author} {\bibfnamefont {M.}~\bibnamefont {{Cicoli}}}, \bibinfo {author} {\bibfnamefont {M.}~\bibnamefont {{Civitani}}}, \bibinfo {author} {\bibfnamefont {C.}~\bibnamefont {{Cogollos}}}, \bibinfo {author} {\bibfnamefont {J.~P.}\ \bibnamefont {{Conlon}}}, \bibinfo {author} {\bibfnamefont {D.}~\bibnamefont {{Costa}}}, \bibinfo {author} {\bibfnamefont {T.}~\bibnamefont {{Dafni}}}, \bibinfo {author} {\bibfnamefont {R.}~\bibnamefont {{Daido}}}, \bibinfo
  {author} {\bibfnamefont {A.~V.}\ \bibnamefont {{Derbin}}}, \bibinfo {author} {\bibfnamefont {M.~A.}\ \bibnamefont {{Descalle}}}, \bibinfo {author} {\bibfnamefont {K.}~\bibnamefont {{Desch}}}, \bibinfo {author} {\bibfnamefont {I.~S.}\ \bibnamefont {{Dratchnev}}}, \bibinfo {author} {\bibfnamefont {B.}~\bibnamefont {{D{\"o}brich}}}, \bibinfo {author} {\bibfnamefont {A.}~\bibnamefont {{Dudarev}}}, \bibinfo {author} {\bibfnamefont {E.}~\bibnamefont {{Ferrer-Ribas}}}, \bibinfo {author} {\bibfnamefont {I.}~\bibnamefont {{Fleck}}}, \bibinfo {author} {\bibfnamefont {J.}~\bibnamefont {{Gal{\'a}n}}}, \bibinfo {author} {\bibfnamefont {G.}~\bibnamefont {{Galanti}}}, \bibinfo {author} {\bibfnamefont {L.}~\bibnamefont {{Garrido}}}, \bibinfo {author} {\bibfnamefont {D.}~\bibnamefont {{Gascon}}}, \bibinfo {author} {\bibfnamefont {L.}~\bibnamefont {{Gastaldo}}}, \bibinfo {author} {\bibfnamefont {C.}~\bibnamefont {{Germani}}}, \bibinfo {author} {\bibfnamefont {G.}~\bibnamefont {{Ghisellini}}}, \bibinfo {author} {\bibfnamefont
  {M.}~\bibnamefont {{Giannotti}}}, \bibinfo {author} {\bibfnamefont {I.}~\bibnamefont {{Giomataris}}}, \bibinfo {author} {\bibfnamefont {S.}~\bibnamefont {{Gninenko}}}, \bibinfo {author} {\bibfnamefont {N.}~\bibnamefont {{Golubev}}}, \bibinfo {author} {\bibfnamefont {R.}~\bibnamefont {{Graciani}}}, \bibinfo {author} {\bibfnamefont {I.~G.}\ \bibnamefont {{Irastorza}}}, \bibinfo {author} {\bibfnamefont {K.}~\bibnamefont {{Jakov{\v{c}}i{\'c}}}}, \bibinfo {author} {\bibfnamefont {J.}~\bibnamefont {{Kaminski}}}, \bibinfo {author} {\bibfnamefont {M.}~\bibnamefont {{Kr{\v{c}}mar}}}, \bibinfo {author} {\bibfnamefont {C.}~\bibnamefont {{Krieger}}}, \bibinfo {author} {\bibfnamefont {B.}~\bibnamefont {{Laki{\'c}}}}, \bibinfo {author} {\bibfnamefont {T.}~\bibnamefont {{Lasserre}}}, \bibinfo {author} {\bibfnamefont {P.}~\bibnamefont {{Laurent}}}, \bibinfo {author} {\bibfnamefont {O.}~\bibnamefont {{Limousin}}}, \bibinfo {author} {\bibfnamefont {A.}~\bibnamefont {{Lindner}}}, \bibinfo {author} {\bibfnamefont
  {I.}~\bibnamefont {{Lomskaya}}}, \bibinfo {author} {\bibfnamefont {B.}~\bibnamefont {{Lubsandorzhiev}}}, \bibinfo {author} {\bibfnamefont {G.}~\bibnamefont {{Luz{\'o}n}}}, \bibinfo {author} {\bibfnamefont {M.~C.~D.}\ \bibnamefont {{Marsh}}}, \bibinfo {author} {\bibfnamefont {C.}~\bibnamefont {{Margalejo}}}, \bibinfo {author} {\bibfnamefont {F.}~\bibnamefont {{Mescia}}}, \bibinfo {author} {\bibfnamefont {M.}~\bibnamefont {{Meyer}}}, \bibinfo {author} {\bibfnamefont {J.}~\bibnamefont {{Miralda-Escud{\'e}}}}, \bibinfo {author} {\bibfnamefont {H.}~\bibnamefont {{Mirallas}}}, \bibinfo {author} {\bibfnamefont {V.~N.}\ \bibnamefont {{Muratova}}}, \bibinfo {author} {\bibfnamefont {X.~F.}\ \bibnamefont {{Navick}}}, \bibinfo {author} {\bibfnamefont {C.}~\bibnamefont {{Nones}}}, \bibinfo {author} {\bibfnamefont {A.}~\bibnamefont {{Notari}}}, \bibinfo {author} {\bibfnamefont {A.}~\bibnamefont {{Nozik}}}, \bibinfo {author} {\bibfnamefont {A.}~\bibnamefont {{Ortiz de Sol{\'o}rzano}}}, \bibinfo {author} {\bibfnamefont
  {V.}~\bibnamefont {{Pantuev}}}, \bibinfo {author} {\bibfnamefont {T.}~\bibnamefont {{Papaevangelou}}}, \bibinfo {author} {\bibfnamefont {G.}~\bibnamefont {{Pareschi}}}, \bibinfo {author} {\bibfnamefont {K.}~\bibnamefont {{Perez}}}, \bibinfo {author} {\bibfnamefont {E.}~\bibnamefont {{Picatoste}}}, \bibinfo {author} {\bibfnamefont {M.~J.}\ \bibnamefont {{Pivovaroff}}}, \bibinfo {author} {\bibfnamefont {J.}~\bibnamefont {{Redondo}}}, \bibinfo {author} {\bibfnamefont {A.}~\bibnamefont {{Ringwald}}}, \bibinfo {author} {\bibfnamefont {M.}~\bibnamefont {{Roncadelli}}}, \bibinfo {author} {\bibfnamefont {E.}~\bibnamefont {{Ruiz-Ch{\'o}liz}}}, \bibinfo {author} {\bibfnamefont {J.}~\bibnamefont {{Ruz}}}, \bibinfo {author} {\bibfnamefont {K.}~\bibnamefont {{Saikawa}}}, \bibinfo {author} {\bibfnamefont {J.}~\bibnamefont {{Salvad{\'o}}}}, \bibinfo {author} {\bibfnamefont {M.~P.}\ \bibnamefont {{Samperiz}}}, \bibinfo {author} {\bibfnamefont {T.}~\bibnamefont {{Schiffer}}}, \bibinfo {author} {\bibfnamefont
  {S.}~\bibnamefont {{Schmidt}}}, \bibinfo {author} {\bibfnamefont {U.}~\bibnamefont {{Schneekloth}}}, \bibinfo {author} {\bibfnamefont {M.}~\bibnamefont {{Schott}}}, \bibinfo {author} {\bibfnamefont {H.}~\bibnamefont {{Silva}}}, \bibinfo {author} {\bibfnamefont {G.}~\bibnamefont {{Tagliaferri}}}, \bibinfo {author} {\bibfnamefont {F.}~\bibnamefont {{Takahashi}}}, \bibinfo {author} {\bibfnamefont {F.}~\bibnamefont {{Tavecchio}}}, \bibinfo {author} {\bibfnamefont {H.}~\bibnamefont {{ten Kate}}}, \bibinfo {author} {\bibfnamefont {I.}~\bibnamefont {{Tkachev}}}, \bibinfo {author} {\bibfnamefont {S.}~\bibnamefont {{Troitsky}}}, \bibinfo {author} {\bibfnamefont {E.}~\bibnamefont {{Unzhakov}}}, \bibinfo {author} {\bibfnamefont {P.}~\bibnamefont {{Vedrine}}}, \bibinfo {author} {\bibfnamefont {J.~K.}\ \bibnamefont {{Vogel}}}, \bibinfo {author} {\bibfnamefont {C.}~\bibnamefont {{Weinsheimer}}}, \bibinfo {author} {\bibfnamefont {A.}~\bibnamefont {{Weltman}}},\ and\ \bibinfo {author} {\bibfnamefont {W.}~\bibnamefont
  {{Yin}}},\ }\bibfield  {title} {\bibinfo {title} {{Physics potential of the International Axion Observatory (IAXO)}},\ }\href {https://doi.org/10.1088/1475-7516/2019/06/047} {\bibfield  {journal} {\bibinfo  {journal} {Journal of Cosmology and Astroparticle Physics}\ }\textbf {\bibinfo {volume} {2019}}\bibfield  {number} {\bibinfo  {number} { (6)},\ \bibinfo {eid} {047}},\ }\Eprint {https://arxiv.org/abs/1904.09155} {arXiv:1904.09155 [hep-ph]} \BibitemShut {NoStop}%
\bibitem [{\citenamefont {{Payez}}\ \emph {et~al.}(2012)\citenamefont {{Payez}}, \citenamefont {{Cudell}},\ and\ \citenamefont {{Hutsem{\'e}kers}}}]{2012JCAP...07..041P}%
  \BibitemOpen
  \bibfield  {author} {\bibinfo {author} {\bibfnamefont {A.}~\bibnamefont {{Payez}}}, \bibinfo {author} {\bibfnamefont {J.~R.}\ \bibnamefont {{Cudell}}},\ and\ \bibinfo {author} {\bibfnamefont {D.}~\bibnamefont {{Hutsem{\'e}kers}}},\ }\bibfield  {title} {\bibinfo {title} {{New polarimetric constraints on axion-like particles}},\ }\href {https://doi.org/10.1088/1475-7516/2012/07/041} {\bibfield  {journal} {\bibinfo  {journal} {Journal of Cosmology and Astroparticle Physics}\ }\textbf {\bibinfo {volume} {2012}}\bibfield  {number} {\bibinfo  {number} { (7)},\ \bibinfo {eid} {041}},\ }\Eprint {https://arxiv.org/abs/1204.6187} {arXiv:1204.6187 [astro-ph.CO]} \BibitemShut {NoStop}%
\bibitem [{\citenamefont {{Berg}}\ \emph {et~al.}(2017)\citenamefont {{Berg}}, \citenamefont {{Conlon}}, \citenamefont {{Day}}, \citenamefont {{Jennings}}, \citenamefont {{Krippendorf}}, \citenamefont {{Powell}},\ and\ \citenamefont {{Rummel}}}]{2017ApJ...847..101B}%
  \BibitemOpen
  \bibfield  {author} {\bibinfo {author} {\bibfnamefont {M.}~\bibnamefont {{Berg}}}, \bibinfo {author} {\bibfnamefont {J.~P.}\ \bibnamefont {{Conlon}}}, \bibinfo {author} {\bibfnamefont {F.}~\bibnamefont {{Day}}}, \bibinfo {author} {\bibfnamefont {N.}~\bibnamefont {{Jennings}}}, \bibinfo {author} {\bibfnamefont {S.}~\bibnamefont {{Krippendorf}}}, \bibinfo {author} {\bibfnamefont {A.~J.}\ \bibnamefont {{Powell}}},\ and\ \bibinfo {author} {\bibfnamefont {M.}~\bibnamefont {{Rummel}}},\ }\bibfield  {title} {\bibinfo {title} {{Constraints on Axion-like Particles from X-Ray Observations of NGC1275}},\ }\href {https://doi.org/10.3847/1538-4357/aa8b16} {\bibfield  {journal} {\bibinfo  {journal} {\apj}\ }\textbf {\bibinfo {volume} {847}},\ \bibinfo {eid} {101} (\bibinfo {year} {2017})},\ \Eprint {https://arxiv.org/abs/1605.01043} {arXiv:1605.01043 [astro-ph.HE]} \BibitemShut {NoStop}%
\bibitem [{\citenamefont {{Ayad}}\ and\ \citenamefont {{Beck}}(2020)}]{2020JCAP...04..055A}%
  \BibitemOpen
  \bibfield  {author} {\bibinfo {author} {\bibfnamefont {A.}~\bibnamefont {{Ayad}}}\ and\ \bibinfo {author} {\bibfnamefont {G.}~\bibnamefont {{Beck}}},\ }\bibfield  {title} {\bibinfo {title} {{Probing a cosmic axion-like particle background within the jets of active galactic nuclei}},\ }\href {https://doi.org/10.1088/1475-7516/2020/04/055} {\bibfield  {journal} {\bibinfo  {journal} {Journal of Cosmology and Astroparticle Physics}\ }\textbf {\bibinfo {volume} {2020}}\bibfield  {number} {\bibinfo  {number} { (4)},\ \bibinfo {eid} {055}},\ }\Eprint {https://arxiv.org/abs/1911.10078} {arXiv:1911.10078 [astro-ph.HE]} \BibitemShut {NoStop}%
\bibitem [{\citenamefont {{Payez}}\ \emph {et~al.}(2015)\citenamefont {{Payez}}, \citenamefont {{Evoli}}, \citenamefont {{Fischer}}, \citenamefont {{Giannotti}}, \citenamefont {{Mirizzi}},\ and\ \citenamefont {{Ringwald}}}]{2015JCAP...02..006P}%
  \BibitemOpen
  \bibfield  {author} {\bibinfo {author} {\bibfnamefont {A.}~\bibnamefont {{Payez}}}, \bibinfo {author} {\bibfnamefont {C.}~\bibnamefont {{Evoli}}}, \bibinfo {author} {\bibfnamefont {T.}~\bibnamefont {{Fischer}}}, \bibinfo {author} {\bibfnamefont {M.}~\bibnamefont {{Giannotti}}}, \bibinfo {author} {\bibfnamefont {A.}~\bibnamefont {{Mirizzi}}},\ and\ \bibinfo {author} {\bibfnamefont {A.}~\bibnamefont {{Ringwald}}},\ }\bibfield  {title} {\bibinfo {title} {{Revisiting the SN1987A gamma-ray limit on ultralight axion-like particles}},\ }\href {https://doi.org/10.1088/1475-7516/2015/02/006} {\bibfield  {journal} {\bibinfo  {journal} {Journal of Cosmology and Astroparticle Physics}\ }\textbf {\bibinfo {volume} {2015}}\bibfield  {number} {\bibinfo  {number} { (2)},\ \bibinfo {pages} {006}},\ }\Eprint {https://arxiv.org/abs/1410.3747} {arXiv:1410.3747 [astro-ph.HE]} \BibitemShut {NoStop}%
\bibitem [{\citenamefont {{Meyer}}\ \emph {et~al.}(2013)\citenamefont {{Meyer}}, \citenamefont {{Horns}},\ and\ \citenamefont {{Raue}}}]{2013PhRvD..87c5027M}%
  \BibitemOpen
  \bibfield  {author} {\bibinfo {author} {\bibfnamefont {M.}~\bibnamefont {{Meyer}}}, \bibinfo {author} {\bibfnamefont {D.}~\bibnamefont {{Horns}}},\ and\ \bibinfo {author} {\bibfnamefont {M.}~\bibnamefont {{Raue}}},\ }\bibfield  {title} {\bibinfo {title} {{First lower limits on the photon-axion-like particle coupling from very high energy gamma-ray observations}},\ }\href {https://doi.org/10.1103/PhysRevD.87.035027} {\bibfield  {journal} {\bibinfo  {journal} {\prd}\ }\textbf {\bibinfo {volume} {87}},\ \bibinfo {eid} {035027} (\bibinfo {year} {2013})},\ \Eprint {https://arxiv.org/abs/1302.1208} {arXiv:1302.1208 [astro-ph.HE]} \BibitemShut {NoStop}%
\bibitem [{\citenamefont {{Angus}}\ \emph {et~al.}(2014)\citenamefont {{Angus}}, \citenamefont {{Conlon}}, \citenamefont {{Marsh}}, \citenamefont {{Powell}},\ and\ \citenamefont {{Witkowski}}}]{2014JCAP...09..026A}%
  \BibitemOpen
  \bibfield  {author} {\bibinfo {author} {\bibfnamefont {S.}~\bibnamefont {{Angus}}}, \bibinfo {author} {\bibfnamefont {J.~P.}\ \bibnamefont {{Conlon}}}, \bibinfo {author} {\bibfnamefont {M.~C.~D.}\ \bibnamefont {{Marsh}}}, \bibinfo {author} {\bibfnamefont {A.~J.}\ \bibnamefont {{Powell}}},\ and\ \bibinfo {author} {\bibfnamefont {L.~T.}\ \bibnamefont {{Witkowski}}},\ }\bibfield  {title} {\bibinfo {title} {{Soft X-ray excess in the Coma cluster from a Cosmic Axion Background}},\ }\href {https://doi.org/10.1088/1475-7516/2014/09/026} {\bibfield  {journal} {\bibinfo  {journal} {Journal of Cosmology and Astroparticle Physics}\ }\textbf {\bibinfo {volume} {2014}}\bibfield  {number} {\bibinfo  {number} { (9)},\ \bibinfo {pages} {026}},\ }\Eprint {https://arxiv.org/abs/1312.3947} {arXiv:1312.3947 [astro-ph.HE]} \BibitemShut {NoStop}%
\bibitem [{\citenamefont {{Kohri}}\ and\ \citenamefont {{Kodama}}(2017)}]{2017PhRvD..96e1701K}%
  \BibitemOpen
  \bibfield  {author} {\bibinfo {author} {\bibfnamefont {K.}~\bibnamefont {{Kohri}}}\ and\ \bibinfo {author} {\bibfnamefont {H.}~\bibnamefont {{Kodama}}},\ }\bibfield  {title} {\bibinfo {title} {{Axion-like particles and recent observations of the cosmic infrared background radiation}},\ }\href {https://doi.org/10.1103/PhysRevD.96.051701} {\bibfield  {journal} {\bibinfo  {journal} {\prd}\ }\textbf {\bibinfo {volume} {96}},\ \bibinfo {eid} {051701} (\bibinfo {year} {2017})},\ \Eprint {https://arxiv.org/abs/1704.05189} {arXiv:1704.05189 [hep-ph]} \BibitemShut {NoStop}%
\bibitem [{\citenamefont {{Obata}}\ \emph {et~al.}(2018)\citenamefont {{Obata}}, \citenamefont {{Fujita}},\ and\ \citenamefont {{Michimura}}}]{2018PhRvL.121p1301O}%
  \BibitemOpen
  \bibfield  {author} {\bibinfo {author} {\bibfnamefont {I.}~\bibnamefont {{Obata}}}, \bibinfo {author} {\bibfnamefont {T.}~\bibnamefont {{Fujita}}},\ and\ \bibinfo {author} {\bibfnamefont {Y.}~\bibnamefont {{Michimura}}},\ }\bibfield  {title} {\bibinfo {title} {{Optical Ring Cavity Search for Axion Dark Matter}},\ }\href {https://doi.org/10.1103/PhysRevLett.121.161301} {\bibfield  {journal} {\bibinfo  {journal} {\prl}\ }\textbf {\bibinfo {volume} {121}},\ \bibinfo {eid} {161301} (\bibinfo {year} {2018})},\ \Eprint {https://arxiv.org/abs/1805.11753} {arXiv:1805.11753 [astro-ph.CO]} \BibitemShut {NoStop}%
\bibitem [{\citenamefont {{Sikivie}}\ \emph {et~al.}(1994)\citenamefont {{Sikivie}}, \citenamefont {{Tanner}},\ and\ \citenamefont {{Wang}}}]{1994PhRvD..50.4744S}%
  \BibitemOpen
  \bibfield  {author} {\bibinfo {author} {\bibfnamefont {P.}~\bibnamefont {{Sikivie}}}, \bibinfo {author} {\bibfnamefont {D.~B.}\ \bibnamefont {{Tanner}}},\ and\ \bibinfo {author} {\bibfnamefont {Y.}~\bibnamefont {{Wang}}},\ }\bibfield  {title} {\bibinfo {title} {{Axion detection in the 10$^{-4}$ eV mass range}},\ }\href {https://doi.org/10.1103/PhysRevD.50.4744} {\bibfield  {journal} {\bibinfo  {journal} {\prd}\ }\textbf {\bibinfo {volume} {50}},\ \bibinfo {pages} {4744} (\bibinfo {year} {1994})},\ \Eprint {https://arxiv.org/abs/hep-ph/9305264} {arXiv:hep-ph/9305264 [hep-ph]} \BibitemShut {NoStop}%
\bibitem [{\citenamefont {{Horns}}\ \emph {et~al.}(2013)\citenamefont {{Horns}}, \citenamefont {{Jaeckel}}, \citenamefont {{Lindner}}, \citenamefont {{Lobanov}}, \citenamefont {{Redondo}},\ and\ \citenamefont {{Ringwald}}}]{2013JCAP...04..016H}%
  \BibitemOpen
  \bibfield  {author} {\bibinfo {author} {\bibfnamefont {D.}~\bibnamefont {{Horns}}}, \bibinfo {author} {\bibfnamefont {J.}~\bibnamefont {{Jaeckel}}}, \bibinfo {author} {\bibfnamefont {A.}~\bibnamefont {{Lindner}}}, \bibinfo {author} {\bibfnamefont {A.}~\bibnamefont {{Lobanov}}}, \bibinfo {author} {\bibfnamefont {J.}~\bibnamefont {{Redondo}}},\ and\ \bibinfo {author} {\bibfnamefont {A.}~\bibnamefont {{Ringwald}}},\ }\bibfield  {title} {\bibinfo {title} {{Searching for WISPy cold dark matter with a dish antenna}},\ }\href {https://doi.org/10.1088/1475-7516/2013/04/016} {\bibfield  {journal} {\bibinfo  {journal} {Journal of Cosmology and Astroparticle Physics}\ }\textbf {\bibinfo {volume} {2013}}\bibfield  {number} {\bibinfo  {number} { (4)},\ \bibinfo {eid} {016}},\ }\Eprint {https://arxiv.org/abs/1212.2970} {arXiv:1212.2970 [hep-ph]} \BibitemShut {NoStop}%
\bibitem [{\citenamefont {{Brun}}\ \emph {et~al.}(2019)\citenamefont {{Brun}}, \citenamefont {{Caldwell}}, \citenamefont {{Chevalier}}, \citenamefont {{Dvali}}, \citenamefont {{Freire}}, \citenamefont {{Garutti}}, \citenamefont {{Heyminck}}, \citenamefont {{Jochum}}, \citenamefont {{Knirck}}, \citenamefont {{Kramer}}, \citenamefont {{Krieger}}, \citenamefont {{Lasserre}}, \citenamefont {{Lee}}, \citenamefont {{Li}}, \citenamefont {{Lindner}}, \citenamefont {{Majorovits}}, \citenamefont {{Martens}}, \citenamefont {{Matysek}}, \citenamefont {{Millar}}, \citenamefont {{Raffelt}}, \citenamefont {{Redondo}}, \citenamefont {{Reimann}}, \citenamefont {{Ringwald}}, \citenamefont {{Saikawa}}, \citenamefont {{Schaffran}}, \citenamefont {{Schmidt}}, \citenamefont {{Sch{\"u}tte-Engel}}, \citenamefont {{Steffen}}, \citenamefont {{Strandhagen}},\ and\ \citenamefont {{Wieching}}}]{2019EPJC...79..186B}%
  \BibitemOpen
  \bibfield  {author} {\bibinfo {author} {\bibfnamefont {P.}~\bibnamefont {{Brun}}}, \bibinfo {author} {\bibfnamefont {A.}~\bibnamefont {{Caldwell}}}, \bibinfo {author} {\bibfnamefont {L.}~\bibnamefont {{Chevalier}}}, \bibinfo {author} {\bibfnamefont {G.}~\bibnamefont {{Dvali}}}, \bibinfo {author} {\bibfnamefont {P.}~\bibnamefont {{Freire}}}, \bibinfo {author} {\bibfnamefont {E.}~\bibnamefont {{Garutti}}}, \bibinfo {author} {\bibfnamefont {S.}~\bibnamefont {{Heyminck}}}, \bibinfo {author} {\bibfnamefont {J.}~\bibnamefont {{Jochum}}}, \bibinfo {author} {\bibfnamefont {S.}~\bibnamefont {{Knirck}}}, \bibinfo {author} {\bibfnamefont {M.}~\bibnamefont {{Kramer}}}, \bibinfo {author} {\bibfnamefont {C.}~\bibnamefont {{Krieger}}}, \bibinfo {author} {\bibfnamefont {T.}~\bibnamefont {{Lasserre}}}, \bibinfo {author} {\bibfnamefont {C.}~\bibnamefont {{Lee}}}, \bibinfo {author} {\bibfnamefont {X.}~\bibnamefont {{Li}}}, \bibinfo {author} {\bibfnamefont {A.}~\bibnamefont {{Lindner}}}, \bibinfo {author} {\bibfnamefont
  {B.}~\bibnamefont {{Majorovits}}}, \bibinfo {author} {\bibfnamefont {S.}~\bibnamefont {{Martens}}}, \bibinfo {author} {\bibfnamefont {M.}~\bibnamefont {{Matysek}}}, \bibinfo {author} {\bibfnamefont {A.}~\bibnamefont {{Millar}}}, \bibinfo {author} {\bibfnamefont {G.}~\bibnamefont {{Raffelt}}}, \bibinfo {author} {\bibfnamefont {J.}~\bibnamefont {{Redondo}}}, \bibinfo {author} {\bibfnamefont {O.}~\bibnamefont {{Reimann}}}, \bibinfo {author} {\bibfnamefont {A.}~\bibnamefont {{Ringwald}}}, \bibinfo {author} {\bibfnamefont {K.}~\bibnamefont {{Saikawa}}}, \bibinfo {author} {\bibfnamefont {J.}~\bibnamefont {{Schaffran}}}, \bibinfo {author} {\bibfnamefont {A.}~\bibnamefont {{Schmidt}}}, \bibinfo {author} {\bibfnamefont {J.}~\bibnamefont {{Sch{\"u}tte-Engel}}}, \bibinfo {author} {\bibfnamefont {F.}~\bibnamefont {{Steffen}}}, \bibinfo {author} {\bibfnamefont {C.}~\bibnamefont {{Strandhagen}}},\ and\ \bibinfo {author} {\bibfnamefont {G.}~\bibnamefont {{Wieching}}},\ }\bibfield  {title} {\bibinfo {title} {{A new
  experimental approach to probe QCD axion dark matter in the mass range above \{ 40\} \{{\ensuremath{\mu}} \}\{eV\}}},\ }\href {https://doi.org/10.1140/epjc/s10052-019-6683-x} {\bibfield  {journal} {\bibinfo  {journal} {European Physical Journal C}\ }\textbf {\bibinfo {volume} {79}},\ \bibinfo {eid} {186} (\bibinfo {year} {2019})},\ \Eprint {https://arxiv.org/abs/1901.07401} {arXiv:1901.07401 [physics.ins-det]} \BibitemShut {NoStop}%
\bibitem [{\citenamefont {{Nagano}}\ \emph {et~al.}(2019)\citenamefont {{Nagano}}, \citenamefont {{Fujita}}, \citenamefont {{Michimura}},\ and\ \citenamefont {{Obata}}}]{2019PhRvL.123k1301N}%
  \BibitemOpen
  \bibfield  {author} {\bibinfo {author} {\bibfnamefont {K.}~\bibnamefont {{Nagano}}}, \bibinfo {author} {\bibfnamefont {T.}~\bibnamefont {{Fujita}}}, \bibinfo {author} {\bibfnamefont {Y.}~\bibnamefont {{Michimura}}},\ and\ \bibinfo {author} {\bibfnamefont {I.}~\bibnamefont {{Obata}}},\ }\bibfield  {title} {\bibinfo {title} {{Axion Dark Matter Search with Interferometric Gravitational Wave Detectors}},\ }\href {https://doi.org/10.1103/PhysRevLett.123.111301} {\bibfield  {journal} {\bibinfo  {journal} {\prl}\ }\textbf {\bibinfo {volume} {123}},\ \bibinfo {eid} {111301} (\bibinfo {year} {2019})},\ \Eprint {https://arxiv.org/abs/1903.02017} {arXiv:1903.02017 [hep-ph]} \BibitemShut {NoStop}%
\bibitem [{\citenamefont {{Sikivie}}(2021)}]{2021RvMP...93a5004S}%
  \BibitemOpen
  \bibfield  {author} {\bibinfo {author} {\bibfnamefont {P.}~\bibnamefont {{Sikivie}}},\ }\bibfield  {title} {\bibinfo {title} {{Invisible axion search methods}},\ }\href {https://doi.org/10.1103/RevModPhys.93.015004} {\bibfield  {journal} {\bibinfo  {journal} {Reviews of Modern Physics}\ }\textbf {\bibinfo {volume} {93}},\ \bibinfo {eid} {015004} (\bibinfo {year} {2021})},\ \Eprint {https://arxiv.org/abs/2003.02206} {arXiv:2003.02206 [hep-ph]} \BibitemShut {NoStop}%
\bibitem [{\citenamefont {{Adams}}\ \emph {et~al.}(2022)\citenamefont {{Adams}}, \citenamefont {{Aggarwal}}, \citenamefont {{Agrawal}}, \citenamefont {{Balafendiev}}, \citenamefont {{Bartram}}, \citenamefont {{Baryakhtar}}, \citenamefont {{Bekker}}, \citenamefont {{Belov}}, \citenamefont {{Berggren}}, \citenamefont {{Berlin}}, \citenamefont {{Boutan}}, \citenamefont {{Bowring}}, \citenamefont {{Budker}}, \citenamefont {{Caldwell}}, \citenamefont {{Carenza}}, \citenamefont {{Carosi}}, \citenamefont {{Cervantes}}, \citenamefont {{Chakrabarty}}, \citenamefont {{Chaudhuri}}, \citenamefont {{Chen}}, \citenamefont {{Cheong}}, \citenamefont {{Chou}}, \citenamefont {{Co}}, \citenamefont {{Conrad}}, \citenamefont {{Croon}}, \citenamefont {{D'Agnolo}}, \citenamefont {{Demarteau}}, \citenamefont {{DePorzio}}, \citenamefont {{Descalle}}, \citenamefont {{Desch}}, \citenamefont {{Di Luzio}}, \citenamefont {{Diaz-Morcillo}}, \citenamefont {{Dona}}, \citenamefont {{Drachnev}}, \citenamefont {{Droster}}, \citenamefont {{Du}},
  \citenamefont {{Dunne}}, \citenamefont {{D{\"o}brich}}, \citenamefont {{Ellis}}, \citenamefont {{Essig}}, \citenamefont {{Fan}}, \citenamefont {{Foster}}, \citenamefont {{Fry}}, \citenamefont {{Gallo Rosso}}, \citenamefont {{Garc{\'\i}a Barcel{\'o}}}, \citenamefont {{Irastorza}}, \citenamefont {{Gardner}}, \citenamefont {{Geraci}}, \citenamefont {{Ghosh}}, \citenamefont {{Giaccone}}, \citenamefont {{Giannotti}}, \citenamefont {{Gimeno}}, \citenamefont {{Grin}}, \citenamefont {{Grote}}, \citenamefont {{Guzzetti}}, \citenamefont {{Awida}}, \citenamefont {{Henning}}, \citenamefont {{Hoof}}, \citenamefont {{Hoshino}}, \citenamefont {{Irsic}}, \citenamefont {{Irwin}}, \citenamefont {{Jackson}}, \citenamefont {{Kimball}}, \citenamefont {{Jaeckel}}, \citenamefont {{Jakovcic}}, \citenamefont {{Jewell}}, \citenamefont {{Kagan}}, \citenamefont {{Kahn}}, \citenamefont {{Khatiwada}}, \citenamefont {{Knirck}}, \citenamefont {{Kovachy}}, \citenamefont {{Krueger}}, \citenamefont {{Kuenstner}}, \citenamefont {{Kurinsky}},
  \citenamefont {{Leane}}, \citenamefont {{Leder}}, \citenamefont {{Lee}}, \citenamefont {{Lehnert}}, \citenamefont {{Lentz}}, \citenamefont {{Lewis}}, \citenamefont {{Liu}}, \citenamefont {{Lynn}}, \citenamefont {{Majorovits}}, \citenamefont {{Marsh}}, \citenamefont {{Maruyama}}, \citenamefont {{McAllister}}, \citenamefont {{Millar}}, \citenamefont {{Miller}}, \citenamefont {{Mitchell}}, \citenamefont {{Morampudi}}, \citenamefont {{Mueller}}, \citenamefont {{Nagaitsev}}, \citenamefont {{Nardi}}, \citenamefont {{Noroozian}}, \citenamefont {{O'Hare}}, \citenamefont {{Oblath}}, \citenamefont {{Ouellet}}, \citenamefont {{Pappas}}, \citenamefont {{Peiris}}, \citenamefont {{Perez}}, \citenamefont {{Phipps}}, \citenamefont {{Pivovaroff}}, \citenamefont {{Qu{\'\i}lez}}, \citenamefont {{Rapidis}}, \citenamefont {{Robles}}, \citenamefont {{Rogers}}, \citenamefont {{Rudolph}}, \citenamefont {{Ruz}}, \citenamefont {{Rybka}}, \citenamefont {{Safdari}}, \citenamefont {{Safdi}}, \citenamefont {{Safronova}}, \citenamefont
  {{Salemi}}, \citenamefont {{Schuster}}, \citenamefont {{Schwartzman}}, \citenamefont {{Shu}}, \citenamefont {{Simanovskaia}}, \citenamefont {{Singh}}, \citenamefont {{Singh}}, \citenamefont {{Sinha}}, \citenamefont {{Sinnis}}, \citenamefont {{Siodlaczek}}, \citenamefont {{Smith}}, \citenamefont {{Snow}}, \citenamefont {{Sokolov}}, \citenamefont {{Sonnenschein}}, \citenamefont {{Speller}}, \citenamefont {{Stadnik}}, \citenamefont {{Sun}}, \citenamefont {{Sushkov}}, \citenamefont {{Tait}}, \citenamefont {{Takhistov}}, \citenamefont {{Tanner}}, \citenamefont {{Tavecchio}}, \citenamefont {{Temples}}, \citenamefont {{Thomas}}, \citenamefont {{Tobar}}, \citenamefont {{Toro}}, \citenamefont {{Tsai}}, \citenamefont {{van Assendelft}}, \citenamefont {{van Bibber}}, \citenamefont {{Vandegar}}, \citenamefont {{Visinelli}}, \citenamefont {{Vitagliano}}, \citenamefont {{Vogel}}, \citenamefont {{Wang}}, \citenamefont {{Wickenbrock}}, \citenamefont {{Winslow}}, \citenamefont {{Withington}}, \citenamefont {{Wooten}},
  \citenamefont {{Yang}}, \citenamefont {{Young}}, \citenamefont {{Yu}}, \citenamefont {{Zhou}},\ and\ \citenamefont {{Zhou}}}]{2022arXiv220314923A}%
  \BibitemOpen
  \bibfield  {author} {\bibinfo {author} {\bibfnamefont {C.~B.}\ \bibnamefont {{Adams}}}, \bibinfo {author} {\bibfnamefont {N.}~\bibnamefont {{Aggarwal}}}, \bibinfo {author} {\bibfnamefont {A.}~\bibnamefont {{Agrawal}}}, \bibinfo {author} {\bibfnamefont {R.}~\bibnamefont {{Balafendiev}}}, \bibinfo {author} {\bibfnamefont {C.}~\bibnamefont {{Bartram}}}, \bibinfo {author} {\bibfnamefont {M.}~\bibnamefont {{Baryakhtar}}}, \bibinfo {author} {\bibfnamefont {H.}~\bibnamefont {{Bekker}}}, \bibinfo {author} {\bibfnamefont {P.}~\bibnamefont {{Belov}}}, \bibinfo {author} {\bibfnamefont {K.~K.}\ \bibnamefont {{Berggren}}}, \bibinfo {author} {\bibfnamefont {A.}~\bibnamefont {{Berlin}}}, \bibinfo {author} {\bibfnamefont {C.}~\bibnamefont {{Boutan}}}, \bibinfo {author} {\bibfnamefont {D.}~\bibnamefont {{Bowring}}}, \bibinfo {author} {\bibfnamefont {D.}~\bibnamefont {{Budker}}}, \bibinfo {author} {\bibfnamefont {A.}~\bibnamefont {{Caldwell}}}, \bibinfo {author} {\bibfnamefont {P.}~\bibnamefont {{Carenza}}}, \bibinfo
  {author} {\bibfnamefont {G.}~\bibnamefont {{Carosi}}}, \bibinfo {author} {\bibfnamefont {R.}~\bibnamefont {{Cervantes}}}, \bibinfo {author} {\bibfnamefont {S.~S.}\ \bibnamefont {{Chakrabarty}}}, \bibinfo {author} {\bibfnamefont {S.}~\bibnamefont {{Chaudhuri}}}, \bibinfo {author} {\bibfnamefont {T.~Y.}\ \bibnamefont {{Chen}}}, \bibinfo {author} {\bibfnamefont {S.}~\bibnamefont {{Cheong}}}, \bibinfo {author} {\bibfnamefont {A.}~\bibnamefont {{Chou}}}, \bibinfo {author} {\bibfnamefont {R.~T.}\ \bibnamefont {{Co}}}, \bibinfo {author} {\bibfnamefont {J.}~\bibnamefont {{Conrad}}}, \bibinfo {author} {\bibfnamefont {D.}~\bibnamefont {{Croon}}}, \bibinfo {author} {\bibfnamefont {R.~T.}\ \bibnamefont {{D'Agnolo}}}, \bibinfo {author} {\bibfnamefont {M.}~\bibnamefont {{Demarteau}}}, \bibinfo {author} {\bibfnamefont {N.}~\bibnamefont {{DePorzio}}}, \bibinfo {author} {\bibfnamefont {M.}~\bibnamefont {{Descalle}}}, \bibinfo {author} {\bibfnamefont {K.}~\bibnamefont {{Desch}}}, \bibinfo {author} {\bibfnamefont
  {L.}~\bibnamefont {{Di Luzio}}}, \bibinfo {author} {\bibfnamefont {A.}~\bibnamefont {{Diaz-Morcillo}}}, \bibinfo {author} {\bibfnamefont {K.}~\bibnamefont {{Dona}}}, \bibinfo {author} {\bibfnamefont {I.~S.}\ \bibnamefont {{Drachnev}}}, \bibinfo {author} {\bibfnamefont {A.}~\bibnamefont {{Droster}}}, \bibinfo {author} {\bibfnamefont {N.}~\bibnamefont {{Du}}}, \bibinfo {author} {\bibfnamefont {K.}~\bibnamefont {{Dunne}}}, \bibinfo {author} {\bibfnamefont {B.}~\bibnamefont {{D{\"o}brich}}}, \bibinfo {author} {\bibfnamefont {S.~A.~R.}\ \bibnamefont {{Ellis}}}, \bibinfo {author} {\bibfnamefont {R.}~\bibnamefont {{Essig}}}, \bibinfo {author} {\bibfnamefont {J.}~\bibnamefont {{Fan}}}, \bibinfo {author} {\bibfnamefont {J.~W.}\ \bibnamefont {{Foster}}}, \bibinfo {author} {\bibfnamefont {J.~T.}\ \bibnamefont {{Fry}}}, \bibinfo {author} {\bibfnamefont {A.}~\bibnamefont {{Gallo Rosso}}}, \bibinfo {author} {\bibfnamefont {J.~M.}\ \bibnamefont {{Garc{\'\i}a Barcel{\'o}}}}, \bibinfo {author} {\bibfnamefont {I.~G.}\
  \bibnamefont {{Irastorza}}}, \bibinfo {author} {\bibfnamefont {S.}~\bibnamefont {{Gardner}}}, \bibinfo {author} {\bibfnamefont {A.~A.}\ \bibnamefont {{Geraci}}}, \bibinfo {author} {\bibfnamefont {S.}~\bibnamefont {{Ghosh}}}, \bibinfo {author} {\bibfnamefont {B.}~\bibnamefont {{Giaccone}}}, \bibinfo {author} {\bibfnamefont {M.}~\bibnamefont {{Giannotti}}}, \bibinfo {author} {\bibfnamefont {B.}~\bibnamefont {{Gimeno}}}, \bibinfo {author} {\bibfnamefont {D.}~\bibnamefont {{Grin}}}, \bibinfo {author} {\bibfnamefont {H.}~\bibnamefont {{Grote}}}, \bibinfo {author} {\bibfnamefont {M.}~\bibnamefont {{Guzzetti}}}, \bibinfo {author} {\bibfnamefont {M.~H.}\ \bibnamefont {{Awida}}}, \bibinfo {author} {\bibfnamefont {R.}~\bibnamefont {{Henning}}}, \bibinfo {author} {\bibfnamefont {S.}~\bibnamefont {{Hoof}}}, \bibinfo {author} {\bibfnamefont {G.}~\bibnamefont {{Hoshino}}}, \bibinfo {author} {\bibfnamefont {V.}~\bibnamefont {{Irsic}}}, \bibinfo {author} {\bibfnamefont {K.~D.}\ \bibnamefont {{Irwin}}}, \bibinfo {author}
  {\bibfnamefont {H.}~\bibnamefont {{Jackson}}}, \bibinfo {author} {\bibfnamefont {D.~F.~J.}\ \bibnamefont {{Kimball}}}, \bibinfo {author} {\bibfnamefont {J.}~\bibnamefont {{Jaeckel}}}, \bibinfo {author} {\bibfnamefont {K.}~\bibnamefont {{Jakovcic}}}, \bibinfo {author} {\bibfnamefont {M.~J.}\ \bibnamefont {{Jewell}}}, \bibinfo {author} {\bibfnamefont {M.}~\bibnamefont {{Kagan}}}, \bibinfo {author} {\bibfnamefont {Y.}~\bibnamefont {{Kahn}}}, \bibinfo {author} {\bibfnamefont {R.}~\bibnamefont {{Khatiwada}}}, \bibinfo {author} {\bibfnamefont {S.}~\bibnamefont {{Knirck}}}, \bibinfo {author} {\bibfnamefont {T.}~\bibnamefont {{Kovachy}}}, \bibinfo {author} {\bibfnamefont {P.}~\bibnamefont {{Krueger}}}, \bibinfo {author} {\bibfnamefont {S.~E.}\ \bibnamefont {{Kuenstner}}}, \bibinfo {author} {\bibfnamefont {N.~A.}\ \bibnamefont {{Kurinsky}}}, \bibinfo {author} {\bibfnamefont {R.~K.}\ \bibnamefont {{Leane}}}, \bibinfo {author} {\bibfnamefont {A.~F.}\ \bibnamefont {{Leder}}}, \bibinfo {author} {\bibfnamefont
  {C.}~\bibnamefont {{Lee}}}, \bibinfo {author} {\bibfnamefont {K.~W.}\ \bibnamefont {{Lehnert}}}, \bibinfo {author} {\bibfnamefont {E.~W.}\ \bibnamefont {{Lentz}}}, \bibinfo {author} {\bibfnamefont {S.~M.}\ \bibnamefont {{Lewis}}}, \bibinfo {author} {\bibfnamefont {J.}~\bibnamefont {{Liu}}}, \bibinfo {author} {\bibfnamefont {M.}~\bibnamefont {{Lynn}}}, \bibinfo {author} {\bibfnamefont {B.}~\bibnamefont {{Majorovits}}}, \bibinfo {author} {\bibfnamefont {D.~J.~E.}\ \bibnamefont {{Marsh}}}, \bibinfo {author} {\bibfnamefont {R.~H.}\ \bibnamefont {{Maruyama}}}, \bibinfo {author} {\bibfnamefont {B.~T.}\ \bibnamefont {{McAllister}}}, \bibinfo {author} {\bibfnamefont {A.~J.}\ \bibnamefont {{Millar}}}, \bibinfo {author} {\bibfnamefont {D.~W.}\ \bibnamefont {{Miller}}}, \bibinfo {author} {\bibfnamefont {J.}~\bibnamefont {{Mitchell}}}, \bibinfo {author} {\bibfnamefont {S.}~\bibnamefont {{Morampudi}}}, \bibinfo {author} {\bibfnamefont {G.}~\bibnamefont {{Mueller}}}, \bibinfo {author} {\bibfnamefont {S.}~\bibnamefont
  {{Nagaitsev}}}, \bibinfo {author} {\bibfnamefont {E.}~\bibnamefont {{Nardi}}}, \bibinfo {author} {\bibfnamefont {O.}~\bibnamefont {{Noroozian}}}, \bibinfo {author} {\bibfnamefont {C.~A.~J.}\ \bibnamefont {{O'Hare}}}, \bibinfo {author} {\bibfnamefont {N.~S.}\ \bibnamefont {{Oblath}}}, \bibinfo {author} {\bibfnamefont {J.~L.}\ \bibnamefont {{Ouellet}}}, \bibinfo {author} {\bibfnamefont {K.~M.~W.}\ \bibnamefont {{Pappas}}}, \bibinfo {author} {\bibfnamefont {H.~V.}\ \bibnamefont {{Peiris}}}, \bibinfo {author} {\bibfnamefont {K.}~\bibnamefont {{Perez}}}, \bibinfo {author} {\bibfnamefont {A.}~\bibnamefont {{Phipps}}}, \bibinfo {author} {\bibfnamefont {M.~J.}\ \bibnamefont {{Pivovaroff}}}, \bibinfo {author} {\bibfnamefont {P.}~\bibnamefont {{Qu{\'\i}lez}}}, \bibinfo {author} {\bibfnamefont {N.~M.}\ \bibnamefont {{Rapidis}}}, \bibinfo {author} {\bibfnamefont {V.~H.}\ \bibnamefont {{Robles}}}, \bibinfo {author} {\bibfnamefont {K.~K.}\ \bibnamefont {{Rogers}}}, \bibinfo {author} {\bibfnamefont {J.}~\bibnamefont
  {{Rudolph}}}, \bibinfo {author} {\bibfnamefont {J.}~\bibnamefont {{Ruz}}}, \bibinfo {author} {\bibfnamefont {G.}~\bibnamefont {{Rybka}}}, \bibinfo {author} {\bibfnamefont {M.}~\bibnamefont {{Safdari}}}, \bibinfo {author} {\bibfnamefont {B.~R.}\ \bibnamefont {{Safdi}}}, \bibinfo {author} {\bibfnamefont {M.~S.}\ \bibnamefont {{Safronova}}}, \bibinfo {author} {\bibfnamefont {C.~P.}\ \bibnamefont {{Salemi}}}, \bibinfo {author} {\bibfnamefont {P.}~\bibnamefont {{Schuster}}}, \bibinfo {author} {\bibfnamefont {A.}~\bibnamefont {{Schwartzman}}}, \bibinfo {author} {\bibfnamefont {J.}~\bibnamefont {{Shu}}}, \bibinfo {author} {\bibfnamefont {M.}~\bibnamefont {{Simanovskaia}}}, \bibinfo {author} {\bibfnamefont {J.}~\bibnamefont {{Singh}}}, \bibinfo {author} {\bibfnamefont {S.}~\bibnamefont {{Singh}}}, \bibinfo {author} {\bibfnamefont {K.}~\bibnamefont {{Sinha}}}, \bibinfo {author} {\bibfnamefont {J.~T.}\ \bibnamefont {{Sinnis}}}, \bibinfo {author} {\bibfnamefont {M.}~\bibnamefont {{Siodlaczek}}}, \bibinfo {author}
  {\bibfnamefont {M.~S.}\ \bibnamefont {{Smith}}}, \bibinfo {author} {\bibfnamefont {W.~M.}\ \bibnamefont {{Snow}}}, \bibinfo {author} {\bibfnamefont {A.~V.}\ \bibnamefont {{Sokolov}}}, \bibinfo {author} {\bibfnamefont {A.}~\bibnamefont {{Sonnenschein}}}, \bibinfo {author} {\bibfnamefont {D.~H.}\ \bibnamefont {{Speller}}}, \bibinfo {author} {\bibfnamefont {Y.~V.}\ \bibnamefont {{Stadnik}}}, \bibinfo {author} {\bibfnamefont {C.}~\bibnamefont {{Sun}}}, \bibinfo {author} {\bibfnamefont {A.~O.}\ \bibnamefont {{Sushkov}}}, \bibinfo {author} {\bibfnamefont {T.~M.~P.}\ \bibnamefont {{Tait}}}, \bibinfo {author} {\bibfnamefont {V.}~\bibnamefont {{Takhistov}}}, \bibinfo {author} {\bibfnamefont {D.~B.}\ \bibnamefont {{Tanner}}}, \bibinfo {author} {\bibfnamefont {F.}~\bibnamefont {{Tavecchio}}}, \bibinfo {author} {\bibfnamefont {D.~J.}\ \bibnamefont {{Temples}}}, \bibinfo {author} {\bibfnamefont {J.~H.}\ \bibnamefont {{Thomas}}}, \bibinfo {author} {\bibfnamefont {M.~E.}\ \bibnamefont {{Tobar}}}, \bibinfo {author}
  {\bibfnamefont {N.}~\bibnamefont {{Toro}}}, \bibinfo {author} {\bibfnamefont {Y.~D.}\ \bibnamefont {{Tsai}}}, \bibinfo {author} {\bibfnamefont {E.~C.}\ \bibnamefont {{van Assendelft}}}, \bibinfo {author} {\bibfnamefont {K.}~\bibnamefont {{van Bibber}}}, \bibinfo {author} {\bibfnamefont {M.}~\bibnamefont {{Vandegar}}}, \bibinfo {author} {\bibfnamefont {L.}~\bibnamefont {{Visinelli}}}, \bibinfo {author} {\bibfnamefont {E.}~\bibnamefont {{Vitagliano}}}, \bibinfo {author} {\bibfnamefont {J.~K.}\ \bibnamefont {{Vogel}}}, \bibinfo {author} {\bibfnamefont {Z.}~\bibnamefont {{Wang}}}, \bibinfo {author} {\bibfnamefont {A.}~\bibnamefont {{Wickenbrock}}}, \bibinfo {author} {\bibfnamefont {L.}~\bibnamefont {{Winslow}}}, \bibinfo {author} {\bibfnamefont {S.}~\bibnamefont {{Withington}}}, \bibinfo {author} {\bibfnamefont {M.}~\bibnamefont {{Wooten}}}, \bibinfo {author} {\bibfnamefont {J.}~\bibnamefont {{Yang}}}, \bibinfo {author} {\bibfnamefont {B.~A.}\ \bibnamefont {{Young}}}, \bibinfo {author} {\bibfnamefont
  {F.}~\bibnamefont {{Yu}}}, \bibinfo {author} {\bibfnamefont {K.}~\bibnamefont {{Zhou}}},\ and\ \bibinfo {author} {\bibfnamefont {T.}~\bibnamefont {{Zhou}}},\ }\bibfield  {title} {\bibinfo {title} {{Axion Dark Matter}},\ }\href {https://doi.org/10.48550/arXiv.2203.14923} {\bibfield  {journal} {\bibinfo  {journal} {arXiv e-prints}\ ,\ \bibinfo {eid} {arXiv:2203.14923}} (\bibinfo {year} {2022})},\ \Eprint {https://arxiv.org/abs/2203.14923} {arXiv:2203.14923 [hep-ex]} \BibitemShut {NoStop}%
\bibitem [{\citenamefont {{Carroll}}\ \emph {et~al.}(1990)\citenamefont {{Carroll}}, \citenamefont {{Field}},\ and\ \citenamefont {{Jackiw}}}]{1990PhRvD..41.1231C}%
  \BibitemOpen
  \bibfield  {author} {\bibinfo {author} {\bibfnamefont {S.~M.}\ \bibnamefont {{Carroll}}}, \bibinfo {author} {\bibfnamefont {G.~B.}\ \bibnamefont {{Field}}},\ and\ \bibinfo {author} {\bibfnamefont {R.}~\bibnamefont {{Jackiw}}},\ }\bibfield  {title} {\bibinfo {title} {{Limits on a Lorentz- and parity-violating modification of electrodynamics}},\ }\href {https://doi.org/10.1103/PhysRevD.41.1231} {\bibfield  {journal} {\bibinfo  {journal} {\prd}\ }\textbf {\bibinfo {volume} {41}},\ \bibinfo {pages} {1231} (\bibinfo {year} {1990})}\BibitemShut {NoStop}%
\bibitem [{\citenamefont {{Harari}}\ and\ \citenamefont {{Sikivie}}(1992)}]{1992PhLB..289...67H}%
  \BibitemOpen
  \bibfield  {author} {\bibinfo {author} {\bibfnamefont {D.}~\bibnamefont {{Harari}}}\ and\ \bibinfo {author} {\bibfnamefont {P.}~\bibnamefont {{Sikivie}}},\ }\bibfield  {title} {\bibinfo {title} {{Effects of a Nambu-Goldstone boson on the polarization of radio galaxies and the cosmic microwave background}},\ }\href {https://doi.org/10.1016/0370-2693(92)91363-E} {\bibfield  {journal} {\bibinfo  {journal} {Physics Letters B}\ }\textbf {\bibinfo {volume} {289}},\ \bibinfo {pages} {67} (\bibinfo {year} {1992})}\BibitemShut {NoStop}%
\bibitem [{\citenamefont {{Carroll}}(1998)}]{1998PhRvL..81.3067C}%
  \BibitemOpen
  \bibfield  {author} {\bibinfo {author} {\bibfnamefont {S.~M.}\ \bibnamefont {{Carroll}}},\ }\bibfield  {title} {\bibinfo {title} {{Quintessence and the Rest of the World: Suppressing Long-Range Interactions}},\ }\href {https://doi.org/10.1103/PhysRevLett.81.3067} {\bibfield  {journal} {\bibinfo  {journal} {\prl}\ }\textbf {\bibinfo {volume} {81}},\ \bibinfo {pages} {3067} (\bibinfo {year} {1998})},\ \Eprint {https://arxiv.org/abs/astro-ph/9806099} {arXiv:astro-ph/9806099 [astro-ph]} \BibitemShut {NoStop}%
\bibitem [{\citenamefont {{Lue}}\ \emph {et~al.}(1999)\citenamefont {{Lue}}, \citenamefont {{Wang}},\ and\ \citenamefont {{Kamionkowski}}}]{1999PhRvL..83.1506L}%
  \BibitemOpen
  \bibfield  {author} {\bibinfo {author} {\bibfnamefont {A.}~\bibnamefont {{Lue}}}, \bibinfo {author} {\bibfnamefont {L.}~\bibnamefont {{Wang}}},\ and\ \bibinfo {author} {\bibfnamefont {M.}~\bibnamefont {{Kamionkowski}}},\ }\bibfield  {title} {\bibinfo {title} {{Cosmological Signature of New Parity-Violating Interactions}},\ }\href {https://doi.org/10.1103/PhysRevLett.83.1506} {\bibfield  {journal} {\bibinfo  {journal} {\prl}\ }\textbf {\bibinfo {volume} {83}},\ \bibinfo {pages} {1506} (\bibinfo {year} {1999})},\ \Eprint {https://arxiv.org/abs/astro-ph/9812088} {arXiv:astro-ph/9812088 [astro-ph]} \BibitemShut {NoStop}%
\bibitem [{\citenamefont {{DeRocco}}\ and\ \citenamefont {{Hook}}(2018)}]{2018PhRvD..98c5021D}%
  \BibitemOpen
  \bibfield  {author} {\bibinfo {author} {\bibfnamefont {W.}~\bibnamefont {{DeRocco}}}\ and\ \bibinfo {author} {\bibfnamefont {A.}~\bibnamefont {{Hook}}},\ }\bibfield  {title} {\bibinfo {title} {{Axion interferometry}},\ }\href {https://doi.org/10.1103/PhysRevD.98.035021} {\bibfield  {journal} {\bibinfo  {journal} {\prd}\ }\textbf {\bibinfo {volume} {98}},\ \bibinfo {eid} {035021} (\bibinfo {year} {2018})},\ \Eprint {https://arxiv.org/abs/1802.07273} {arXiv:1802.07273 [hep-ph]} \BibitemShut {NoStop}%
\bibitem [{\citenamefont {{Sigl}}\ and\ \citenamefont {{Trivedi}}(2018)}]{2018arXiv181107873S}%
  \BibitemOpen
  \bibfield  {author} {\bibinfo {author} {\bibfnamefont {G.}~\bibnamefont {{Sigl}}}\ and\ \bibinfo {author} {\bibfnamefont {P.}~\bibnamefont {{Trivedi}}},\ }\bibfield  {title} {\bibinfo {title} {{Axion-like Dark Matter Constraints from CMB Birefringence}},\ }\href {https://doi.org/10.48550/arXiv.1811.07873} {\bibfield  {journal} {\bibinfo  {journal} {arXiv e-prints}\ ,\ \bibinfo {eid} {arXiv:1811.07873}} (\bibinfo {year} {2018})},\ \Eprint {https://arxiv.org/abs/1811.07873} {arXiv:1811.07873 [astro-ph.CO]} \BibitemShut {NoStop}%
\bibitem [{\citenamefont {{Fedderke}}\ \emph {et~al.}(2019)\citenamefont {{Fedderke}}, \citenamefont {{Graham}},\ and\ \citenamefont {{Rajendran}}}]{2019PhRvD.100a5040F}%
  \BibitemOpen
  \bibfield  {author} {\bibinfo {author} {\bibfnamefont {M.~A.}\ \bibnamefont {{Fedderke}}}, \bibinfo {author} {\bibfnamefont {P.~W.}\ \bibnamefont {{Graham}}},\ and\ \bibinfo {author} {\bibfnamefont {S.}~\bibnamefont {{Rajendran}}},\ }\bibfield  {title} {\bibinfo {title} {{Axion dark matter detection with CMB polarization}},\ }\href {https://doi.org/10.1103/PhysRevD.100.015040} {\bibfield  {journal} {\bibinfo  {journal} {\prd}\ }\textbf {\bibinfo {volume} {100}},\ \bibinfo {eid} {015040} (\bibinfo {year} {2019})},\ \Eprint {https://arxiv.org/abs/1903.02666} {arXiv:1903.02666 [astro-ph.CO]} \BibitemShut {NoStop}%
\bibitem [{\citenamefont {{Ivanov}}\ \emph {et~al.}(2019)\citenamefont {{Ivanov}}, \citenamefont {{Kovalev}}, \citenamefont {{Lister}}, \citenamefont {{Panin}}, \citenamefont {{Pushkarev}}, \citenamefont {{Savolainen}},\ and\ \citenamefont {{Troitsky}}}]{2019JCAP...02..059I}%
  \BibitemOpen
  \bibfield  {author} {\bibinfo {author} {\bibfnamefont {M.~M.}\ \bibnamefont {{Ivanov}}}, \bibinfo {author} {\bibfnamefont {Y.~Y.}\ \bibnamefont {{Kovalev}}}, \bibinfo {author} {\bibfnamefont {M.~L.}\ \bibnamefont {{Lister}}}, \bibinfo {author} {\bibfnamefont {A.~G.}\ \bibnamefont {{Panin}}}, \bibinfo {author} {\bibfnamefont {A.~B.}\ \bibnamefont {{Pushkarev}}}, \bibinfo {author} {\bibfnamefont {T.}~\bibnamefont {{Savolainen}}},\ and\ \bibinfo {author} {\bibfnamefont {S.~V.}\ \bibnamefont {{Troitsky}}},\ }\bibfield  {title} {\bibinfo {title} {{Constraining the photon coupling of ultra-light dark-matter axion-like particles by polarization variations of parsec-scale jets in active galaxies}},\ }\href {https://doi.org/10.1088/1475-7516/2019/02/059} {\bibfield  {journal} {\bibinfo  {journal} {Journal of Cosmology and Astroparticle Physics}\ }\textbf {\bibinfo {volume} {2019}}\bibfield  {number} {\bibinfo  {number} { (2)},\ \bibinfo {eid} {059}},\ }\Eprint {https://arxiv.org/abs/1811.10997} {arXiv:1811.10997
  [astro-ph.CO]} \BibitemShut {NoStop}%
\bibitem [{\citenamefont {{Liu}}\ \emph {et~al.}(2020)\citenamefont {{Liu}}, \citenamefont {{Smoot}},\ and\ \citenamefont {{Zhao}}}]{2020PhRvD.101f3012L}%
  \BibitemOpen
  \bibfield  {author} {\bibinfo {author} {\bibfnamefont {T.}~\bibnamefont {{Liu}}}, \bibinfo {author} {\bibfnamefont {G.}~\bibnamefont {{Smoot}}},\ and\ \bibinfo {author} {\bibfnamefont {Y.}~\bibnamefont {{Zhao}}},\ }\bibfield  {title} {\bibinfo {title} {{Detecting axionlike dark matter with linearly polarized pulsar light}},\ }\href {https://doi.org/10.1103/PhysRevD.101.063012} {\bibfield  {journal} {\bibinfo  {journal} {\prd}\ }\textbf {\bibinfo {volume} {101}},\ \bibinfo {eid} {063012} (\bibinfo {year} {2020})},\ \Eprint {https://arxiv.org/abs/1901.10981} {arXiv:1901.10981 [astro-ph.CO]} \BibitemShut {NoStop}%
\bibitem [{\citenamefont {{Liu}}\ \emph {et~al.}(2023)\citenamefont {{Liu}}, \citenamefont {{Lou}},\ and\ \citenamefont {{Ren}}}]{2023PhRvL.130l1401L}%
  \BibitemOpen
  \bibfield  {author} {\bibinfo {author} {\bibfnamefont {T.}~\bibnamefont {{Liu}}}, \bibinfo {author} {\bibfnamefont {X.}~\bibnamefont {{Lou}}},\ and\ \bibinfo {author} {\bibfnamefont {J.}~\bibnamefont {{Ren}}},\ }\bibfield  {title} {\bibinfo {title} {{Pulsar Polarization Arrays}},\ }\href {https://doi.org/10.1103/PhysRevLett.130.121401} {\bibfield  {journal} {\bibinfo  {journal} {\prl}\ }\textbf {\bibinfo {volume} {130}},\ \bibinfo {eid} {121401} (\bibinfo {year} {2023})}\BibitemShut {NoStop}%
\bibitem [{\citenamefont {{Fujita}}\ \emph {et~al.}(2019)\citenamefont {{Fujita}}, \citenamefont {{Tazaki}},\ and\ \citenamefont {{Toma}}}]{2019PhRvL.122s1101F}%
  \BibitemOpen
  \bibfield  {author} {\bibinfo {author} {\bibfnamefont {T.}~\bibnamefont {{Fujita}}}, \bibinfo {author} {\bibfnamefont {R.}~\bibnamefont {{Tazaki}}},\ and\ \bibinfo {author} {\bibfnamefont {K.}~\bibnamefont {{Toma}}},\ }\bibfield  {title} {\bibinfo {title} {{Hunting Axion Dark Matter with Protoplanetary Disk Polarimetry}},\ }\href {https://doi.org/10.1103/PhysRevLett.122.191101} {\bibfield  {journal} {\bibinfo  {journal} {\prl}\ }\textbf {\bibinfo {volume} {122}},\ \bibinfo {eid} {191101} (\bibinfo {year} {2019})},\ \Eprint {https://arxiv.org/abs/1811.03525} {arXiv:1811.03525 [astro-ph.CO]} \BibitemShut {NoStop}%
\bibitem [{\citenamefont {{Basu}}\ \emph {et~al.}(2021)\citenamefont {{Basu}}, \citenamefont {{Goswami}}, \citenamefont {{Schwarz}},\ and\ \citenamefont {{Urakawa}}}]{2021PhRvL.126s1102B}%
  \BibitemOpen
  \bibfield  {author} {\bibinfo {author} {\bibfnamefont {A.}~\bibnamefont {{Basu}}}, \bibinfo {author} {\bibfnamefont {J.}~\bibnamefont {{Goswami}}}, \bibinfo {author} {\bibfnamefont {D.~J.}\ \bibnamefont {{Schwarz}}},\ and\ \bibinfo {author} {\bibfnamefont {Y.}~\bibnamefont {{Urakawa}}},\ }\bibfield  {title} {\bibinfo {title} {{Searching for Axionlike Particles under Strong Gravitational Lenses}},\ }\href {https://doi.org/10.1103/PhysRevLett.126.191102} {\bibfield  {journal} {\bibinfo  {journal} {\prl}\ }\textbf {\bibinfo {volume} {126}},\ \bibinfo {eid} {191102} (\bibinfo {year} {2021})},\ \Eprint {https://arxiv.org/abs/2007.01440} {arXiv:2007.01440 [astro-ph.CO]} \BibitemShut {NoStop}%
\bibitem [{\citenamefont {{Lorimer}}\ \emph {et~al.}(2007)\citenamefont {{Lorimer}}, \citenamefont {{Bailes}}, \citenamefont {{McLaughlin}}, \citenamefont {{Narkevic}},\ and\ \citenamefont {{Crawford}}}]{2007Sci...318..777L}%
  \BibitemOpen
  \bibfield  {author} {\bibinfo {author} {\bibfnamefont {D.~R.}\ \bibnamefont {{Lorimer}}}, \bibinfo {author} {\bibfnamefont {M.}~\bibnamefont {{Bailes}}}, \bibinfo {author} {\bibfnamefont {M.~A.}\ \bibnamefont {{McLaughlin}}}, \bibinfo {author} {\bibfnamefont {D.~J.}\ \bibnamefont {{Narkevic}}},\ and\ \bibinfo {author} {\bibfnamefont {F.}~\bibnamefont {{Crawford}}},\ }\bibfield  {title} {\bibinfo {title} {{A Bright Millisecond Radio Burst of Extragalactic Origin}},\ }\href {https://doi.org/10.1126/science.1147532} {\bibfield  {journal} {\bibinfo  {journal} {Science}\ }\textbf {\bibinfo {volume} {318}},\ \bibinfo {pages} {777} (\bibinfo {year} {2007})},\ \Eprint {https://arxiv.org/abs/0709.4301} {arXiv:0709.4301 [astro-ph]} \BibitemShut {NoStop}%
\bibitem [{\citenamefont {{Chatterjee}}\ \emph {et~al.}(2017)\citenamefont {{Chatterjee}}, \citenamefont {{Law}}, \citenamefont {{Wharton}}, \citenamefont {{Burke-Spolaor}}, \citenamefont {{Hessels}}, \citenamefont {{Bower}}, \citenamefont {{Cordes}}, \citenamefont {{Tendulkar}}, \citenamefont {{Bassa}}, \citenamefont {{Demorest}}, \citenamefont {{Butler}}, \citenamefont {{Seymour}}, \citenamefont {{Scholz}}, \citenamefont {{Abruzzo}}, \citenamefont {{Bogdanov}}, \citenamefont {{Kaspi}}, \citenamefont {{Keimpema}}, \citenamefont {{Lazio}}, \citenamefont {{Marcote}}, \citenamefont {{McLaughlin}}, \citenamefont {{Paragi}}, \citenamefont {{Ransom}}, \citenamefont {{Rupen}}, \citenamefont {{Spitler}},\ and\ \citenamefont {{van Langevelde}}}]{2017Natur.541...58C}%
  \BibitemOpen
  \bibfield  {author} {\bibinfo {author} {\bibfnamefont {S.}~\bibnamefont {{Chatterjee}}}, \bibinfo {author} {\bibfnamefont {C.~J.}\ \bibnamefont {{Law}}}, \bibinfo {author} {\bibfnamefont {R.~S.}\ \bibnamefont {{Wharton}}}, \bibinfo {author} {\bibfnamefont {S.}~\bibnamefont {{Burke-Spolaor}}}, \bibinfo {author} {\bibfnamefont {J.~W.~T.}\ \bibnamefont {{Hessels}}}, \bibinfo {author} {\bibfnamefont {G.~C.}\ \bibnamefont {{Bower}}}, \bibinfo {author} {\bibfnamefont {J.~M.}\ \bibnamefont {{Cordes}}}, \bibinfo {author} {\bibfnamefont {S.~P.}\ \bibnamefont {{Tendulkar}}}, \bibinfo {author} {\bibfnamefont {C.~G.}\ \bibnamefont {{Bassa}}}, \bibinfo {author} {\bibfnamefont {P.}~\bibnamefont {{Demorest}}}, \bibinfo {author} {\bibfnamefont {B.~J.}\ \bibnamefont {{Butler}}}, \bibinfo {author} {\bibfnamefont {A.}~\bibnamefont {{Seymour}}}, \bibinfo {author} {\bibfnamefont {P.}~\bibnamefont {{Scholz}}}, \bibinfo {author} {\bibfnamefont {M.~W.}\ \bibnamefont {{Abruzzo}}}, \bibinfo {author} {\bibfnamefont {S.}~\bibnamefont
  {{Bogdanov}}}, \bibinfo {author} {\bibfnamefont {V.~M.}\ \bibnamefont {{Kaspi}}}, \bibinfo {author} {\bibfnamefont {A.}~\bibnamefont {{Keimpema}}}, \bibinfo {author} {\bibfnamefont {T.~J.~W.}\ \bibnamefont {{Lazio}}}, \bibinfo {author} {\bibfnamefont {B.}~\bibnamefont {{Marcote}}}, \bibinfo {author} {\bibfnamefont {M.~A.}\ \bibnamefont {{McLaughlin}}}, \bibinfo {author} {\bibfnamefont {Z.}~\bibnamefont {{Paragi}}}, \bibinfo {author} {\bibfnamefont {S.~M.}\ \bibnamefont {{Ransom}}}, \bibinfo {author} {\bibfnamefont {M.}~\bibnamefont {{Rupen}}}, \bibinfo {author} {\bibfnamefont {L.~G.}\ \bibnamefont {{Spitler}}},\ and\ \bibinfo {author} {\bibfnamefont {H.~J.}\ \bibnamefont {{van Langevelde}}},\ }\bibfield  {title} {\bibinfo {title} {{A direct localization of a fast radio burst and its host}},\ }\href {https://doi.org/10.1038/nature20797} {\bibfield  {journal} {\bibinfo  {journal} {\nat}\ }\textbf {\bibinfo {volume} {541}},\ \bibinfo {pages} {58} (\bibinfo {year} {2017})},\ \Eprint
  {https://arxiv.org/abs/1701.01098} {arXiv:1701.01098 [astro-ph.HE]} \BibitemShut {NoStop}%
\bibitem [{\citenamefont {{Thornton}}\ \emph {et~al.}(2013)\citenamefont {{Thornton}}, \citenamefont {{Stappers}}, \citenamefont {{Bailes}}, \citenamefont {{Barsdell}}, \citenamefont {{Bates}}, \citenamefont {{Bhat}}, \citenamefont {{Burgay}}, \citenamefont {{Burke-Spolaor}}, \citenamefont {{Champion}}, \citenamefont {{Coster}}, \citenamefont {{D'Amico}}, \citenamefont {{Jameson}}, \citenamefont {{Johnston}}, \citenamefont {{Keith}}, \citenamefont {{Kramer}}, \citenamefont {{Levin}}, \citenamefont {{Milia}}, \citenamefont {{Ng}}, \citenamefont {{Possenti}},\ and\ \citenamefont {{van Straten}}}]{2013Sci...341...53T}%
  \BibitemOpen
  \bibfield  {author} {\bibinfo {author} {\bibfnamefont {D.}~\bibnamefont {{Thornton}}}, \bibinfo {author} {\bibfnamefont {B.}~\bibnamefont {{Stappers}}}, \bibinfo {author} {\bibfnamefont {M.}~\bibnamefont {{Bailes}}}, \bibinfo {author} {\bibfnamefont {B.}~\bibnamefont {{Barsdell}}}, \bibinfo {author} {\bibfnamefont {S.}~\bibnamefont {{Bates}}}, \bibinfo {author} {\bibfnamefont {N.~D.~R.}\ \bibnamefont {{Bhat}}}, \bibinfo {author} {\bibfnamefont {M.}~\bibnamefont {{Burgay}}}, \bibinfo {author} {\bibfnamefont {S.}~\bibnamefont {{Burke-Spolaor}}}, \bibinfo {author} {\bibfnamefont {D.~J.}\ \bibnamefont {{Champion}}}, \bibinfo {author} {\bibfnamefont {P.}~\bibnamefont {{Coster}}}, \bibinfo {author} {\bibfnamefont {N.}~\bibnamefont {{D'Amico}}}, \bibinfo {author} {\bibfnamefont {A.}~\bibnamefont {{Jameson}}}, \bibinfo {author} {\bibfnamefont {S.}~\bibnamefont {{Johnston}}}, \bibinfo {author} {\bibfnamefont {M.}~\bibnamefont {{Keith}}}, \bibinfo {author} {\bibfnamefont {M.}~\bibnamefont {{Kramer}}}, \bibinfo
  {author} {\bibfnamefont {L.}~\bibnamefont {{Levin}}}, \bibinfo {author} {\bibfnamefont {S.}~\bibnamefont {{Milia}}}, \bibinfo {author} {\bibfnamefont {C.}~\bibnamefont {{Ng}}}, \bibinfo {author} {\bibfnamefont {A.}~\bibnamefont {{Possenti}}},\ and\ \bibinfo {author} {\bibfnamefont {W.}~\bibnamefont {{van Straten}}},\ }\bibfield  {title} {\bibinfo {title} {{A Population of Fast Radio Bursts at Cosmological Distances}},\ }\href {https://doi.org/10.1126/science.1236789} {\bibfield  {journal} {\bibinfo  {journal} {Science}\ }\textbf {\bibinfo {volume} {341}},\ \bibinfo {pages} {53} (\bibinfo {year} {2013})},\ \Eprint {https://arxiv.org/abs/1307.1628} {arXiv:1307.1628 [astro-ph.HE]} \BibitemShut {NoStop}%
\bibitem [{\citenamefont {{Champion}}\ \emph {et~al.}(2016)\citenamefont {{Champion}}, \citenamefont {{Petroff}}, \citenamefont {{Kramer}}, \citenamefont {{Keith}}, \citenamefont {{Bailes}}, \citenamefont {{Barr}}, \citenamefont {{Bates}}, \citenamefont {{Bhat}}, \citenamefont {{Burgay}}, \citenamefont {{Burke-Spolaor}}, \citenamefont {{Flynn}}, \citenamefont {{Jameson}}, \citenamefont {{Johnston}}, \citenamefont {{Ng}}, \citenamefont {{Levin}}, \citenamefont {{Possenti}}, \citenamefont {{Stappers}}, \citenamefont {{van Straten}}, \citenamefont {{Thornton}}, \citenamefont {{Tiburzi}},\ and\ \citenamefont {{Lyne}}}]{2016MNRAS.460L..30C}%
  \BibitemOpen
  \bibfield  {author} {\bibinfo {author} {\bibfnamefont {D.~J.}\ \bibnamefont {{Champion}}}, \bibinfo {author} {\bibfnamefont {E.}~\bibnamefont {{Petroff}}}, \bibinfo {author} {\bibfnamefont {M.}~\bibnamefont {{Kramer}}}, \bibinfo {author} {\bibfnamefont {M.~J.}\ \bibnamefont {{Keith}}}, \bibinfo {author} {\bibfnamefont {M.}~\bibnamefont {{Bailes}}}, \bibinfo {author} {\bibfnamefont {E.~D.}\ \bibnamefont {{Barr}}}, \bibinfo {author} {\bibfnamefont {S.~D.}\ \bibnamefont {{Bates}}}, \bibinfo {author} {\bibfnamefont {N.~D.~R.}\ \bibnamefont {{Bhat}}}, \bibinfo {author} {\bibfnamefont {M.}~\bibnamefont {{Burgay}}}, \bibinfo {author} {\bibfnamefont {S.}~\bibnamefont {{Burke-Spolaor}}}, \bibinfo {author} {\bibfnamefont {C.~M.~L.}\ \bibnamefont {{Flynn}}}, \bibinfo {author} {\bibfnamefont {A.}~\bibnamefont {{Jameson}}}, \bibinfo {author} {\bibfnamefont {S.}~\bibnamefont {{Johnston}}}, \bibinfo {author} {\bibfnamefont {C.}~\bibnamefont {{Ng}}}, \bibinfo {author} {\bibfnamefont {L.}~\bibnamefont {{Levin}}}, \bibinfo
  {author} {\bibfnamefont {A.}~\bibnamefont {{Possenti}}}, \bibinfo {author} {\bibfnamefont {B.~W.}\ \bibnamefont {{Stappers}}}, \bibinfo {author} {\bibfnamefont {W.}~\bibnamefont {{van Straten}}}, \bibinfo {author} {\bibfnamefont {D.}~\bibnamefont {{Thornton}}}, \bibinfo {author} {\bibfnamefont {C.}~\bibnamefont {{Tiburzi}}},\ and\ \bibinfo {author} {\bibfnamefont {A.~G.}\ \bibnamefont {{Lyne}}},\ }\bibfield  {title} {\bibinfo {title} {{Five new fast radio bursts from the HTRU high-latitude survey at Parkes: first evidence for two-component bursts}},\ }\href {https://doi.org/10.1093/mnrasl/slw069} {\bibfield  {journal} {\bibinfo  {journal} {Monthly Notices of the Royal Astronomical Society}\ }\textbf {\bibinfo {volume} {460}},\ \bibinfo {pages} {L30} (\bibinfo {year} {2016})},\ \Eprint {https://arxiv.org/abs/1511.07746} {arXiv:1511.07746 [astro-ph.HE]} \BibitemShut {NoStop}%
\bibitem [{\citenamefont {{Niu}}\ \emph {et~al.}(2021)\citenamefont {{Niu}}, \citenamefont {{Li}}, \citenamefont {{Luo}}, \citenamefont {{Wang}}, \citenamefont {{Yao}}, \citenamefont {{Zhang}}, \citenamefont {{Zhu}}, \citenamefont {{Wang}}, \citenamefont {{Ye}}, \citenamefont {{Zhang}}, \citenamefont {{Niu}}, \citenamefont {{Tang}}, \citenamefont {{Duan}}, \citenamefont {{Krco}}, \citenamefont {{Dai}}, \citenamefont {{Feng}}, \citenamefont {{Miao}}, \citenamefont {{Pan}}, \citenamefont {{Qian}}, \citenamefont {{Xue}}, \citenamefont {{Yuan}}, \citenamefont {{Yue}}, \citenamefont {{Zhang}},\ and\ \citenamefont {{Zhang}}}]{2021ApJ...909L...8N}%
  \BibitemOpen
  \bibfield  {author} {\bibinfo {author} {\bibfnamefont {C.-H.}\ \bibnamefont {{Niu}}}, \bibinfo {author} {\bibfnamefont {D.}~\bibnamefont {{Li}}}, \bibinfo {author} {\bibfnamefont {R.}~\bibnamefont {{Luo}}}, \bibinfo {author} {\bibfnamefont {W.-Y.}\ \bibnamefont {{Wang}}}, \bibinfo {author} {\bibfnamefont {J.}~\bibnamefont {{Yao}}}, \bibinfo {author} {\bibfnamefont {B.}~\bibnamefont {{Zhang}}}, \bibinfo {author} {\bibfnamefont {W.-W.}\ \bibnamefont {{Zhu}}}, \bibinfo {author} {\bibfnamefont {P.}~\bibnamefont {{Wang}}}, \bibinfo {author} {\bibfnamefont {H.}~\bibnamefont {{Ye}}}, \bibinfo {author} {\bibfnamefont {Y.-K.}\ \bibnamefont {{Zhang}}}, \bibinfo {author} {\bibfnamefont {J.-r.}\ \bibnamefont {{Niu}}}, \bibinfo {author} {\bibfnamefont {N.-y.}\ \bibnamefont {{Tang}}}, \bibinfo {author} {\bibfnamefont {R.}~\bibnamefont {{Duan}}}, \bibinfo {author} {\bibfnamefont {M.}~\bibnamefont {{Krco}}}, \bibinfo {author} {\bibfnamefont {S.}~\bibnamefont {{Dai}}}, \bibinfo {author} {\bibfnamefont {Y.}~\bibnamefont
  {{Feng}}}, \bibinfo {author} {\bibfnamefont {C.}~\bibnamefont {{Miao}}}, \bibinfo {author} {\bibfnamefont {Z.}~\bibnamefont {{Pan}}}, \bibinfo {author} {\bibfnamefont {L.}~\bibnamefont {{Qian}}}, \bibinfo {author} {\bibfnamefont {M.}~\bibnamefont {{Xue}}}, \bibinfo {author} {\bibfnamefont {M.}~\bibnamefont {{Yuan}}}, \bibinfo {author} {\bibfnamefont {Y.}~\bibnamefont {{Yue}}}, \bibinfo {author} {\bibfnamefont {L.}~\bibnamefont {{Zhang}}},\ and\ \bibinfo {author} {\bibfnamefont {X.}~\bibnamefont {{Zhang}}},\ }\bibfield  {title} {\bibinfo {title} {{CRAFTS for Fast Radio Bursts: Extending the Dispersion-Fluence Relation with New FRBs Detected by FAST}},\ }\href {https://doi.org/10.3847/2041-8213/abe7f0} {\bibfield  {journal} {\bibinfo  {journal} {The Astrophysical Journal Letters}\ }\textbf {\bibinfo {volume} {909}},\ \bibinfo {eid} {L8} (\bibinfo {year} {2021})},\ \Eprint {https://arxiv.org/abs/2102.10546} {arXiv:2102.10546 [astro-ph.HE]} \BibitemShut {NoStop}%
\bibitem [{\citenamefont {{Cordes}}\ and\ \citenamefont {{Chatterjee}}(2019)}]{2019ARA&A..57..417C}%
  \BibitemOpen
  \bibfield  {author} {\bibinfo {author} {\bibfnamefont {J.~M.}\ \bibnamefont {{Cordes}}}\ and\ \bibinfo {author} {\bibfnamefont {S.}~\bibnamefont {{Chatterjee}}},\ }\bibfield  {title} {\bibinfo {title} {{Fast Radio Bursts: An Extragalactic Enigma}},\ }\href {https://doi.org/10.1146/annurev-astro-091918-104501} {\bibfield  {journal} {\bibinfo  {journal} {Annual Review of Astronomy and Astrophysics}\ }\textbf {\bibinfo {volume} {57}},\ \bibinfo {pages} {417} (\bibinfo {year} {2019})},\ \Eprint {https://arxiv.org/abs/1906.05878} {arXiv:1906.05878 [astro-ph.HE]} \BibitemShut {NoStop}%
\bibitem [{\citenamefont {{Zhang}}(2022)}]{2022arXiv221203972Z}%
  \BibitemOpen
  \bibfield  {author} {\bibinfo {author} {\bibfnamefont {B.}~\bibnamefont {{Zhang}}},\ }\bibfield  {title} {\bibinfo {title} {{The Physics of Fast Radio Bursts}},\ }\href {https://doi.org/10.48550/arXiv.2212.03972} {\bibfield  {journal} {\bibinfo  {journal} {arXiv e-prints}\ ,\ \bibinfo {eid} {arXiv:2212.03972}} (\bibinfo {year} {2022})},\ \Eprint {https://arxiv.org/abs/2212.03972} {arXiv:2212.03972 [astro-ph.HE]} \BibitemShut {NoStop}%
\bibitem [{\citenamefont {{Mu{\~n}oz}}\ \emph {et~al.}(2016)\citenamefont {{Mu{\~n}oz}}, \citenamefont {{Kovetz}}, \citenamefont {{Dai}},\ and\ \citenamefont {{Kamionkowski}}}]{2016PhRvL.117i1301M}%
  \BibitemOpen
  \bibfield  {author} {\bibinfo {author} {\bibfnamefont {J.~B.}\ \bibnamefont {{Mu{\~n}oz}}}, \bibinfo {author} {\bibfnamefont {E.~D.}\ \bibnamefont {{Kovetz}}}, \bibinfo {author} {\bibfnamefont {L.}~\bibnamefont {{Dai}}},\ and\ \bibinfo {author} {\bibfnamefont {M.}~\bibnamefont {{Kamionkowski}}},\ }\bibfield  {title} {\bibinfo {title} {{Lensing of Fast Radio Bursts as a Probe of Compact Dark Matter}},\ }\href {https://doi.org/10.1103/PhysRevLett.117.091301} {\bibfield  {journal} {\bibinfo  {journal} {\prl}\ }\textbf {\bibinfo {volume} {117}},\ \bibinfo {eid} {091301} (\bibinfo {year} {2016})},\ \Eprint {https://arxiv.org/abs/1605.00008} {arXiv:1605.00008 [astro-ph.CO]} \BibitemShut {NoStop}%
\bibitem [{\citenamefont {{Wang}}\ and\ \citenamefont {{Wang}}(2018)}]{2018A&A...614A..50W}%
  \BibitemOpen
  \bibfield  {author} {\bibinfo {author} {\bibfnamefont {Y.~K.}\ \bibnamefont {{Wang}}}\ and\ \bibinfo {author} {\bibfnamefont {F.~Y.}\ \bibnamefont {{Wang}}},\ }\bibfield  {title} {\bibinfo {title} {{Lensing of fast radio bursts by binaries to probe compact dark matter}},\ }\href {https://doi.org/10.1051/0004-6361/201731160} {\bibfield  {journal} {\bibinfo  {journal} {Astronomy \& Astrophysics}\ }\textbf {\bibinfo {volume} {614}},\ \bibinfo {eid} {A50} (\bibinfo {year} {2018})},\ \Eprint {https://arxiv.org/abs/1801.07360} {arXiv:1801.07360 [astro-ph.CO]} \BibitemShut {NoStop}%
\bibitem [{\citenamefont {{Katz}}\ \emph {et~al.}(2020)\citenamefont {{Katz}}, \citenamefont {{Kopp}}, \citenamefont {{Sibiryakov}},\ and\ \citenamefont {{Xue}}}]{2020MNRAS.496..564K}%
  \BibitemOpen
  \bibfield  {author} {\bibinfo {author} {\bibfnamefont {A.}~\bibnamefont {{Katz}}}, \bibinfo {author} {\bibfnamefont {J.}~\bibnamefont {{Kopp}}}, \bibinfo {author} {\bibfnamefont {S.}~\bibnamefont {{Sibiryakov}}},\ and\ \bibinfo {author} {\bibfnamefont {W.}~\bibnamefont {{Xue}}},\ }\bibfield  {title} {\bibinfo {title} {{Looking for MACHOs in the spectra of fast radio bursts}},\ }\href {https://doi.org/10.1093/mnras/staa1497} {\bibfield  {journal} {\bibinfo  {journal} {Monthly Notices of the Royal Astronomical Society}\ }\textbf {\bibinfo {volume} {496}},\ \bibinfo {pages} {564} (\bibinfo {year} {2020})},\ \Eprint {https://arxiv.org/abs/1912.07620} {arXiv:1912.07620 [astro-ph.CO]} \BibitemShut {NoStop}%
\bibitem [{\citenamefont {{Liao}}\ \emph {et~al.}(2020)\citenamefont {{Liao}}, \citenamefont {{Zhang}}, \citenamefont {{Li}},\ and\ \citenamefont {{Gao}}}]{2020ApJ...896L..11L}%
  \BibitemOpen
  \bibfield  {author} {\bibinfo {author} {\bibfnamefont {K.}~\bibnamefont {{Liao}}}, \bibinfo {author} {\bibfnamefont {S.~B.}\ \bibnamefont {{Zhang}}}, \bibinfo {author} {\bibfnamefont {Z.}~\bibnamefont {{Li}}},\ and\ \bibinfo {author} {\bibfnamefont {H.}~\bibnamefont {{Gao}}},\ }\bibfield  {title} {\bibinfo {title} {{Constraints on Compact Dark Matter with Fast Radio Burst Observations}},\ }\href {https://doi.org/10.3847/2041-8213/ab963e} {\bibfield  {journal} {\bibinfo  {journal} {The Astrophysical Journal Letters}\ }\textbf {\bibinfo {volume} {896}},\ \bibinfo {eid} {L11} (\bibinfo {year} {2020})},\ \Eprint {https://arxiv.org/abs/2003.13349} {arXiv:2003.13349 [astro-ph.CO]} \BibitemShut {NoStop}%
\bibitem [{\citenamefont {{Zhou}}\ \emph {et~al.}(2022)\citenamefont {{Zhou}}, \citenamefont {{Li}}, \citenamefont {{Liao}}, \citenamefont {{Niu}}, \citenamefont {{Gao}}, \citenamefont {{Huang}}, \citenamefont {{Huang}},\ and\ \citenamefont {{Zhang}}}]{2022ApJ...928..124Z}%
  \BibitemOpen
  \bibfield  {author} {\bibinfo {author} {\bibfnamefont {H.}~\bibnamefont {{Zhou}}}, \bibinfo {author} {\bibfnamefont {Z.}~\bibnamefont {{Li}}}, \bibinfo {author} {\bibfnamefont {K.}~\bibnamefont {{Liao}}}, \bibinfo {author} {\bibfnamefont {C.}~\bibnamefont {{Niu}}}, \bibinfo {author} {\bibfnamefont {H.}~\bibnamefont {{Gao}}}, \bibinfo {author} {\bibfnamefont {Z.}~\bibnamefont {{Huang}}}, \bibinfo {author} {\bibfnamefont {L.}~\bibnamefont {{Huang}}},\ and\ \bibinfo {author} {\bibfnamefont {B.}~\bibnamefont {{Zhang}}},\ }\bibfield  {title} {\bibinfo {title} {{Search for Lensing Signatures from the Latest Fast Radio Burst Observations and Constraints on the Abundance of Primordial Black Holes}},\ }\href {https://doi.org/10.3847/1538-4357/ac510d} {\bibfield  {journal} {\bibinfo  {journal} {\apj}\ }\textbf {\bibinfo {volume} {928}},\ \bibinfo {eid} {124} (\bibinfo {year} {2022})},\ \Eprint {https://arxiv.org/abs/2109.09251} {arXiv:2109.09251 [astro-ph.CO]} \BibitemShut {NoStop}%
\bibitem [{\citenamefont {{Krochek}}\ and\ \citenamefont {{Kovetz}}(2022)}]{2022PhRvD.105j3528K}%
  \BibitemOpen
  \bibfield  {author} {\bibinfo {author} {\bibfnamefont {K.}~\bibnamefont {{Krochek}}}\ and\ \bibinfo {author} {\bibfnamefont {E.~D.}\ \bibnamefont {{Kovetz}}},\ }\bibfield  {title} {\bibinfo {title} {{Constraining primordial black hole dark matter with CHIME fast radio bursts}},\ }\href {https://doi.org/10.1103/PhysRevD.105.103528} {\bibfield  {journal} {\bibinfo  {journal} {\prd}\ }\textbf {\bibinfo {volume} {105}},\ \bibinfo {eid} {103528} (\bibinfo {year} {2022})},\ \Eprint {https://arxiv.org/abs/2112.03721} {arXiv:2112.03721 [astro-ph.CO]} \BibitemShut {NoStop}%
\bibitem [{\citenamefont {{Dai}}\ and\ \citenamefont {{Lu}}(2017)}]{2017ApJ...847...19D}%
  \BibitemOpen
  \bibfield  {author} {\bibinfo {author} {\bibfnamefont {L.}~\bibnamefont {{Dai}}}\ and\ \bibinfo {author} {\bibfnamefont {W.}~\bibnamefont {{Lu}}},\ }\bibfield  {title} {\bibinfo {title} {{Probing Motion of Fast Radio Burst Sources by Timing Strongly Lensed Repeaters}},\ }\href {https://doi.org/10.3847/1538-4357/aa8873} {\bibfield  {journal} {\bibinfo  {journal} {\apj}\ }\textbf {\bibinfo {volume} {847}},\ \bibinfo {eid} {19} (\bibinfo {year} {2017})},\ \Eprint {https://arxiv.org/abs/1706.06103} {arXiv:1706.06103 [astro-ph.HE]} \BibitemShut {NoStop}%
\bibitem [{\citenamefont {{Gao}}\ \emph {et~al.}(2022)\citenamefont {{Gao}}, \citenamefont {{Li}},\ and\ \citenamefont {{Gao}}}]{2022MNRAS.516.1977G}%
  \BibitemOpen
  \bibfield  {author} {\bibinfo {author} {\bibfnamefont {R.}~\bibnamefont {{Gao}}}, \bibinfo {author} {\bibfnamefont {Z.}~\bibnamefont {{Li}}},\ and\ \bibinfo {author} {\bibfnamefont {H.}~\bibnamefont {{Gao}}},\ }\bibfield  {title} {\bibinfo {title} {{Prospects of strongly lensed fast radio bursts: simultaneous measurement of post-Newtonian parameter and Hubble constant}},\ }\href {https://doi.org/10.1093/mnras/stac2270} {\bibfield  {journal} {\bibinfo  {journal} {Monthly Notices of the Royal Astronomical Society}\ }\textbf {\bibinfo {volume} {516}},\ \bibinfo {pages} {1977} (\bibinfo {year} {2022})},\ \Eprint {https://arxiv.org/abs/2208.10175} {arXiv:2208.10175 [astro-ph.CO]} \BibitemShut {NoStop}%
\bibitem [{\citenamefont {{Li}}\ \emph {et~al.}(2018)\citenamefont {{Li}}, \citenamefont {{Gao}}, \citenamefont {{Ding}}, \citenamefont {{Wang}},\ and\ \citenamefont {{Zhang}}}]{2018NatCo...9.3833L}%
  \BibitemOpen
  \bibfield  {author} {\bibinfo {author} {\bibfnamefont {Z.-X.}\ \bibnamefont {{Li}}}, \bibinfo {author} {\bibfnamefont {H.}~\bibnamefont {{Gao}}}, \bibinfo {author} {\bibfnamefont {X.-H.}\ \bibnamefont {{Ding}}}, \bibinfo {author} {\bibfnamefont {G.-J.}\ \bibnamefont {{Wang}}},\ and\ \bibinfo {author} {\bibfnamefont {B.}~\bibnamefont {{Zhang}}},\ }\bibfield  {title} {\bibinfo {title} {{Strongly lensed repeating fast radio bursts as precision probes of the universe}},\ }\href {https://doi.org/10.1038/s41467-018-06303-0} {\bibfield  {journal} {\bibinfo  {journal} {Nature Communications}\ }\textbf {\bibinfo {volume} {9}},\ \bibinfo {eid} {3833} (\bibinfo {year} {2018})},\ \Eprint {https://arxiv.org/abs/1708.06357} {arXiv:1708.06357 [astro-ph.CO]} \BibitemShut {NoStop}%
\bibitem [{\citenamefont {{Wucknitz}}\ \emph {et~al.}(2021)\citenamefont {{Wucknitz}}, \citenamefont {{Spitler}},\ and\ \citenamefont {{Pen}}}]{2021A&A...645A..44W}%
  \BibitemOpen
  \bibfield  {author} {\bibinfo {author} {\bibfnamefont {O.}~\bibnamefont {{Wucknitz}}}, \bibinfo {author} {\bibfnamefont {L.~G.}\ \bibnamefont {{Spitler}}},\ and\ \bibinfo {author} {\bibfnamefont {U.~L.}\ \bibnamefont {{Pen}}},\ }\bibfield  {title} {\bibinfo {title} {{Cosmology with gravitationally lensed repeating fast radio bursts}},\ }\href {https://doi.org/10.1051/0004-6361/202038248} {\bibfield  {journal} {\bibinfo  {journal} {Astronomy \& Astrophysics}\ }\textbf {\bibinfo {volume} {645}},\ \bibinfo {eid} {A44} (\bibinfo {year} {2021})},\ \Eprint {https://arxiv.org/abs/2004.11643} {arXiv:2004.11643 [astro-ph.CO]} \BibitemShut {NoStop}%
\bibitem [{\citenamefont {{Oguri}}(2019)}]{2019RPPh...82l6901O}%
  \BibitemOpen
  \bibfield  {author} {\bibinfo {author} {\bibfnamefont {M.}~\bibnamefont {{Oguri}}},\ }\bibfield  {title} {\bibinfo {title} {{Strong gravitational lensing of explosive transients}},\ }\href {https://doi.org/10.1088/1361-6633/ab4fc5} {\bibfield  {journal} {\bibinfo  {journal} {Reports on Progress in Physics}\ }\textbf {\bibinfo {volume} {82}},\ \bibinfo {eid} {126901} (\bibinfo {year} {2019})},\ \Eprint {https://arxiv.org/abs/1907.06830} {arXiv:1907.06830 [astro-ph.CO]} \BibitemShut {NoStop}%
\bibitem [{\citenamefont {{Liao}}\ \emph {et~al.}(2022)\citenamefont {{Liao}}, \citenamefont {{Biesiada}},\ and\ \citenamefont {{Zhu}}}]{2022ChPhL..39k9801L}%
  \BibitemOpen
  \bibfield  {author} {\bibinfo {author} {\bibfnamefont {K.}~\bibnamefont {{Liao}}}, \bibinfo {author} {\bibfnamefont {M.}~\bibnamefont {{Biesiada}}},\ and\ \bibinfo {author} {\bibfnamefont {Z.-H.}\ \bibnamefont {{Zhu}}},\ }\bibfield  {title} {\bibinfo {title} {{Strongly Lensed Transient Sources: A Review}},\ }\href {https://doi.org/10.1088/0256-307X/39/11/119801} {\bibfield  {journal} {\bibinfo  {journal} {Chinese Physics Letters}\ }\textbf {\bibinfo {volume} {39}},\ \bibinfo {eid} {119801} (\bibinfo {year} {2022})},\ \Eprint {https://arxiv.org/abs/2207.13489} {arXiv:2207.13489 [astro-ph.HE]} \BibitemShut {NoStop}%
\bibitem [{\citenamefont {{Wilczek}}(1987)}]{1987PhRvL..58.1799W}%
  \BibitemOpen
  \bibfield  {author} {\bibinfo {author} {\bibfnamefont {F.}~\bibnamefont {{Wilczek}}},\ }\bibfield  {title} {\bibinfo {title} {{Two applications of axion electrodynamics}},\ }\href {https://doi.org/10.1103/PhysRevLett.58.1799} {\bibfield  {journal} {\bibinfo  {journal} {\prl}\ }\textbf {\bibinfo {volume} {58}},\ \bibinfo {pages} {1799} (\bibinfo {year} {1987})}\BibitemShut {NoStop}%
\bibitem [{\citenamefont {{Schwarz}}\ \emph {et~al.}(2021)\citenamefont {{Schwarz}}, \citenamefont {{Goswami}},\ and\ \citenamefont {{Basu}}}]{2021PhRvD.103h1306S}%
  \BibitemOpen
  \bibfield  {author} {\bibinfo {author} {\bibfnamefont {D.~J.}\ \bibnamefont {{Schwarz}}}, \bibinfo {author} {\bibfnamefont {J.}~\bibnamefont {{Goswami}}},\ and\ \bibinfo {author} {\bibfnamefont {A.}~\bibnamefont {{Basu}}},\ }\bibfield  {title} {\bibinfo {title} {{Geometric optics in the presence of axionlike particles in curved spacetime}},\ }\href {https://doi.org/10.1103/PhysRevD.103.L081306} {\bibfield  {journal} {\bibinfo  {journal} {\prd}\ }\textbf {\bibinfo {volume} {103}},\ \bibinfo {eid} {L081306} (\bibinfo {year} {2021})},\ \Eprint {https://arxiv.org/abs/2003.10205} {arXiv:2003.10205 [hep-ph]} \BibitemShut {NoStop}%
\bibitem [{\citenamefont {{O'Sullivan}}\ \emph {et~al.}(2012)\citenamefont {{O'Sullivan}}, \citenamefont {{Brown}}, \citenamefont {{Robishaw}}, \citenamefont {{Schnitzeler}}, \citenamefont {{McClure-Griffiths}}, \citenamefont {{Feain}}, \citenamefont {{Taylor}}, \citenamefont {{Gaensler}}, \citenamefont {{Landecker}}, \citenamefont {{Harvey-Smith}},\ and\ \citenamefont {{Carretti}}}]{2012MNRAS.421.3300O}%
  \BibitemOpen
  \bibfield  {author} {\bibinfo {author} {\bibfnamefont {S.~P.}\ \bibnamefont {{O'Sullivan}}}, \bibinfo {author} {\bibfnamefont {S.}~\bibnamefont {{Brown}}}, \bibinfo {author} {\bibfnamefont {T.}~\bibnamefont {{Robishaw}}}, \bibinfo {author} {\bibfnamefont {D.~H.~F.~M.}\ \bibnamefont {{Schnitzeler}}}, \bibinfo {author} {\bibfnamefont {N.~M.}\ \bibnamefont {{McClure-Griffiths}}}, \bibinfo {author} {\bibfnamefont {I.~J.}\ \bibnamefont {{Feain}}}, \bibinfo {author} {\bibfnamefont {A.~R.}\ \bibnamefont {{Taylor}}}, \bibinfo {author} {\bibfnamefont {B.~M.}\ \bibnamefont {{Gaensler}}}, \bibinfo {author} {\bibfnamefont {T.~L.}\ \bibnamefont {{Landecker}}}, \bibinfo {author} {\bibfnamefont {L.}~\bibnamefont {{Harvey-Smith}}},\ and\ \bibinfo {author} {\bibfnamefont {E.}~\bibnamefont {{Carretti}}},\ }\bibfield  {title} {\bibinfo {title} {{Complex Faraday depth structure of active galactic nuclei as revealed by broad-band radio polarimetry}},\ }\href {https://doi.org/10.1111/j.1365-2966.2012.20554.x} {\bibfield
  {journal} {\bibinfo  {journal} {Monthly Notices of the Royal Astronomical Society}\ }\textbf {\bibinfo {volume} {421}},\ \bibinfo {pages} {3300} (\bibinfo {year} {2012})},\ \Eprint {https://arxiv.org/abs/1201.3161} {arXiv:1201.3161 [astro-ph.CO]} \BibitemShut {NoStop}%
\bibitem [{\citenamefont {{O'Sullivan}}\ \emph {et~al.}(2017)\citenamefont {{O'Sullivan}}, \citenamefont {{Purcell}}, \citenamefont {{Anderson}}, \citenamefont {{Farnes}}, \citenamefont {{Sun}},\ and\ \citenamefont {{Gaensler}}}]{2017MNRAS.469.4034O}%
  \BibitemOpen
  \bibfield  {author} {\bibinfo {author} {\bibfnamefont {S.~P.}\ \bibnamefont {{O'Sullivan}}}, \bibinfo {author} {\bibfnamefont {C.~R.}\ \bibnamefont {{Purcell}}}, \bibinfo {author} {\bibfnamefont {C.~S.}\ \bibnamefont {{Anderson}}}, \bibinfo {author} {\bibfnamefont {J.~S.}\ \bibnamefont {{Farnes}}}, \bibinfo {author} {\bibfnamefont {X.~H.}\ \bibnamefont {{Sun}}},\ and\ \bibinfo {author} {\bibfnamefont {B.~M.}\ \bibnamefont {{Gaensler}}},\ }\bibfield  {title} {\bibinfo {title} {{Broad-band, radio spectro-polarimetric study of 100 radiative-mode and jet-mode AGN}},\ }\href {https://doi.org/10.1093/mnras/stx1133} {\bibfield  {journal} {\bibinfo  {journal} {Monthly Notices of the Royal Astronomical Society}\ }\textbf {\bibinfo {volume} {469}},\ \bibinfo {pages} {4034} (\bibinfo {year} {2017})},\ \Eprint {https://arxiv.org/abs/1705.00102} {arXiv:1705.00102 [astro-ph.GA]} \BibitemShut {NoStop}%
\bibitem [{\citenamefont {{Jiang}}\ \emph {et~al.}(2022)\citenamefont {{Jiang}}, \citenamefont {{Wang}}, \citenamefont {{Xu}}, \citenamefont {{Xu}}, \citenamefont {{Zhang}}, \citenamefont {{Wang}}, \citenamefont {{Zhou}}, \citenamefont {{Zhang}}, \citenamefont {{Niu}}, \citenamefont {{Lee}}, \citenamefont {{Zhang}}, \citenamefont {{Han}}, \citenamefont {{Li}}, \citenamefont {{Zhu}}, \citenamefont {{Dai}}, \citenamefont {{Feng}}, \citenamefont {{Jing}}, \citenamefont {{Li}}, \citenamefont {{Luo}}, \citenamefont {{Miao}}, \citenamefont {{Niu}}, \citenamefont {{Tsai}}, \citenamefont {{Wang}}, \citenamefont {{Wang}}, \citenamefont {{Xu}}, \citenamefont {{Yang}}, \citenamefont {{Yang}}, \citenamefont {{Yao}},\ and\ \citenamefont {{Yuan}}}]{2022RAA....22l4003J}%
  \BibitemOpen
  \bibfield  {author} {\bibinfo {author} {\bibfnamefont {J.-C.}\ \bibnamefont {{Jiang}}}, \bibinfo {author} {\bibfnamefont {W.-Y.}\ \bibnamefont {{Wang}}}, \bibinfo {author} {\bibfnamefont {H.}~\bibnamefont {{Xu}}}, \bibinfo {author} {\bibfnamefont {J.-W.}\ \bibnamefont {{Xu}}}, \bibinfo {author} {\bibfnamefont {C.-F.}\ \bibnamefont {{Zhang}}}, \bibinfo {author} {\bibfnamefont {B.-J.}\ \bibnamefont {{Wang}}}, \bibinfo {author} {\bibfnamefont {D.-J.}\ \bibnamefont {{Zhou}}}, \bibinfo {author} {\bibfnamefont {Y.-K.}\ \bibnamefont {{Zhang}}}, \bibinfo {author} {\bibfnamefont {J.-R.}\ \bibnamefont {{Niu}}}, \bibinfo {author} {\bibfnamefont {K.-J.}\ \bibnamefont {{Lee}}}, \bibinfo {author} {\bibfnamefont {B.}~\bibnamefont {{Zhang}}}, \bibinfo {author} {\bibfnamefont {J.-L.}\ \bibnamefont {{Han}}}, \bibinfo {author} {\bibfnamefont {D.}~\bibnamefont {{Li}}}, \bibinfo {author} {\bibfnamefont {W.-W.}\ \bibnamefont {{Zhu}}}, \bibinfo {author} {\bibfnamefont {Z.-G.}\ \bibnamefont {{Dai}}}, \bibinfo {author}
  {\bibfnamefont {Y.}~\bibnamefont {{Feng}}}, \bibinfo {author} {\bibfnamefont {W.-C.}\ \bibnamefont {{Jing}}}, \bibinfo {author} {\bibfnamefont {D.-Z.}\ \bibnamefont {{Li}}}, \bibinfo {author} {\bibfnamefont {R.}~\bibnamefont {{Luo}}}, \bibinfo {author} {\bibfnamefont {C.-C.}\ \bibnamefont {{Miao}}}, \bibinfo {author} {\bibfnamefont {C.-H.}\ \bibnamefont {{Niu}}}, \bibinfo {author} {\bibfnamefont {C.-W.}\ \bibnamefont {{Tsai}}}, \bibinfo {author} {\bibfnamefont {F.-Y.}\ \bibnamefont {{Wang}}}, \bibinfo {author} {\bibfnamefont {P.}~\bibnamefont {{Wang}}}, \bibinfo {author} {\bibfnamefont {R.-X.}\ \bibnamefont {{Xu}}}, \bibinfo {author} {\bibfnamefont {Y.-P.}\ \bibnamefont {{Yang}}}, \bibinfo {author} {\bibfnamefont {Z.-L.}\ \bibnamefont {{Yang}}}, \bibinfo {author} {\bibfnamefont {J.-M.}\ \bibnamefont {{Yao}}},\ and\ \bibinfo {author} {\bibfnamefont {M.}~\bibnamefont {{Yuan}}},\ }\bibfield  {title} {\bibinfo {title} {{FAST Observations of an Extremely Active Episode of FRB 20201124A. III. Polarimetry}},\
  }\href {https://doi.org/10.1088/1674-4527/ac98f6} {\bibfield  {journal} {\bibinfo  {journal} {Research in Astronomy and Astrophysics}\ }\textbf {\bibinfo {volume} {22}},\ \bibinfo {eid} {124003} (\bibinfo {year} {2022})},\ \Eprint {https://arxiv.org/abs/2210.03609} {arXiv:2210.03609 [astro-ph.HE]} \BibitemShut {NoStop}%
\bibitem [{\citenamefont {{Ching}}\ \emph {et~al.}(2022)\citenamefont {{Ching}}, \citenamefont {{Li}}, \citenamefont {{Heiles}}, \citenamefont {{Li}}, \citenamefont {{Qian}}, \citenamefont {{Yue}}, \citenamefont {{Tang}},\ and\ \citenamefont {{Jiao}}}]{2022Natur.601...49C}%
  \BibitemOpen
  \bibfield  {author} {\bibinfo {author} {\bibfnamefont {T.~C.}\ \bibnamefont {{Ching}}}, \bibinfo {author} {\bibfnamefont {D.}~\bibnamefont {{Li}}}, \bibinfo {author} {\bibfnamefont {C.}~\bibnamefont {{Heiles}}}, \bibinfo {author} {\bibfnamefont {Z.~Y.}\ \bibnamefont {{Li}}}, \bibinfo {author} {\bibfnamefont {L.}~\bibnamefont {{Qian}}}, \bibinfo {author} {\bibfnamefont {Y.~L.}\ \bibnamefont {{Yue}}}, \bibinfo {author} {\bibfnamefont {J.}~\bibnamefont {{Tang}}},\ and\ \bibinfo {author} {\bibfnamefont {S.~H.}\ \bibnamefont {{Jiao}}},\ }\bibfield  {title} {\bibinfo {title} {{An early transition to magnetic supercriticality in star formation}},\ }\href {https://doi.org/10.1038/s41586-021-04159-x} {\bibfield  {journal} {\bibinfo  {journal} {\nat}\ }\textbf {\bibinfo {volume} {601}},\ \bibinfo {pages} {49} (\bibinfo {year} {2022})},\ \Eprint {https://arxiv.org/abs/2112.12644} {arXiv:2112.12644 [astro-ph.GA]} \BibitemShut {NoStop}%
\bibitem [{\citenamefont {{Turner}}(1986)}]{1986PhRvD..33..889T}%
  \BibitemOpen
  \bibfield  {author} {\bibinfo {author} {\bibfnamefont {M.~S.}\ \bibnamefont {{Turner}}},\ }\bibfield  {title} {\bibinfo {title} {{Cosmic and local mass density of ``invisible'' axions}},\ }\href {https://doi.org/10.1103/PhysRevD.33.889} {\bibfield  {journal} {\bibinfo  {journal} {\prd}\ }\textbf {\bibinfo {volume} {33}},\ \bibinfo {pages} {889} (\bibinfo {year} {1986})}\BibitemShut {NoStop}%
\bibitem [{\citenamefont {{Liu}}\ \emph {et~al.}(2021)\citenamefont {{Liu}}, \citenamefont {{Lou}},\ and\ \citenamefont {{Ren}}}]{2021arXiv211110615L}%
  \BibitemOpen
  \bibfield  {author} {\bibinfo {author} {\bibfnamefont {T.}~\bibnamefont {{Liu}}}, \bibinfo {author} {\bibfnamefont {X.}~\bibnamefont {{Lou}}},\ and\ \bibinfo {author} {\bibfnamefont {J.}~\bibnamefont {{Ren}}},\ }\bibfield  {title} {\bibinfo {title} {{Pulsar Polarization Arrays}},\ }\href {https://doi.org/10.48550/arXiv.2111.10615} {\bibfield  {journal} {\bibinfo  {journal} {arXiv e-prints}\ ,\ \bibinfo {eid} {arXiv:2111.10615}} (\bibinfo {year} {2021})},\ \Eprint {https://arxiv.org/abs/2111.10615} {arXiv:2111.10615 [astro-ph.HE]} \BibitemShut {NoStop}%
\bibitem [{\citenamefont {{Castillo}}\ \emph {et~al.}(2022)\citenamefont {{Castillo}}, \citenamefont {{Martin-Camalich}}, \citenamefont {{Terol-Calvo}}, \citenamefont {{Blas}}, \citenamefont {{Caputo}}, \citenamefont {{G{\'e}nova Santos}}, \citenamefont {{Sberna}}, \citenamefont {{Peel}},\ and\ \citenamefont {{Rubi{\~n}o-Mart{\'\i}n}}}]{2022JCAP...06..014C}%
  \BibitemOpen
  \bibfield  {author} {\bibinfo {author} {\bibfnamefont {A.}~\bibnamefont {{Castillo}}}, \bibinfo {author} {\bibfnamefont {J.}~\bibnamefont {{Martin-Camalich}}}, \bibinfo {author} {\bibfnamefont {J.}~\bibnamefont {{Terol-Calvo}}}, \bibinfo {author} {\bibfnamefont {D.}~\bibnamefont {{Blas}}}, \bibinfo {author} {\bibfnamefont {A.}~\bibnamefont {{Caputo}}}, \bibinfo {author} {\bibfnamefont {R.~T.}\ \bibnamefont {{G{\'e}nova Santos}}}, \bibinfo {author} {\bibfnamefont {L.}~\bibnamefont {{Sberna}}}, \bibinfo {author} {\bibfnamefont {M.}~\bibnamefont {{Peel}}},\ and\ \bibinfo {author} {\bibfnamefont {J.~A.}\ \bibnamefont {{Rubi{\~n}o-Mart{\'\i}n}}},\ }\bibfield  {title} {\bibinfo {title} {{Searching for dark-matter waves with PPTA and QUIJOTE pulsar polarimetry}},\ }\href {https://doi.org/10.1088/1475-7516/2022/06/014} {\bibfield  {journal} {\bibinfo  {journal} {Journal of Cosmology and Astroparticle Physics}\ }\textbf {\bibinfo {volume} {2022}}\bibfield  {number} {\bibinfo  {number} { (6)},\ \bibinfo {eid}
  {014}},\ }\Eprint {https://arxiv.org/abs/2201.03422} {arXiv:2201.03422 [astro-ph.CO]} \BibitemShut {NoStop}%
\bibitem [{\citenamefont {{Ferguson}}\ \emph {et~al.}(2022)\citenamefont {{Ferguson}}, \citenamefont {{Anderson}}, \citenamefont {{Whitehorn}}, \citenamefont {{Ade}}, \citenamefont {{Archipley}}, \citenamefont {{Avva}}, \citenamefont {{Balkenhol}}, \citenamefont {{Benabed}}, \citenamefont {{Bender}}, \citenamefont {{Benson}}, \citenamefont {{Bianchini}}, \citenamefont {{Bleem}}, \citenamefont {{Bouchet}}, \citenamefont {{Bryant}}, \citenamefont {{Camphuis}}, \citenamefont {{Carlstrom}}, \citenamefont {{Cecil}}, \citenamefont {{Chang}}, \citenamefont {{Chaubal}}, \citenamefont {{Chichura}}, \citenamefont {{Chou}}, \citenamefont {{Crawford}}, \citenamefont {{Cukierman}}, \citenamefont {{Daley}}, \citenamefont {{de Haan}}, \citenamefont {{Dibert}}, \citenamefont {{Dobbs}}, \citenamefont {{Doussot}}, \citenamefont {{Dutcher}}, \citenamefont {{Everett}}, \citenamefont {{Feng}}, \citenamefont {{Foster}}, \citenamefont {{Galli}}, \citenamefont {{Gambrel}}, \citenamefont {{Gardner}}, \citenamefont {{Goeckner-Wald}},
  \citenamefont {{Gualtieri}}, \citenamefont {{Guidi}}, \citenamefont {{Guns}}, \citenamefont {{Halverson}}, \citenamefont {{Hivon}}, \citenamefont {{Holder}}, \citenamefont {{Holzapfel}}, \citenamefont {{Hood}}, \citenamefont {{Huang}}, \citenamefont {{Knox}}, \citenamefont {{Korman}}, \citenamefont {{Kuo}}, \citenamefont {{Lee}}, \citenamefont {{Lowitz}}, \citenamefont {{Lu}}, \citenamefont {{Millea}}, \citenamefont {{Montgomery}}, \citenamefont {{Natoli}}, \citenamefont {{Noble}}, \citenamefont {{Novosad}}, \citenamefont {{Omori}}, \citenamefont {{Padin}}, \citenamefont {{Pan}}, \citenamefont {{Paschos}}, \citenamefont {{Prabhu}}, \citenamefont {{Quan}}, \citenamefont {{Rahlin}}, \citenamefont {{Reichardt}}, \citenamefont {{Rouble}}, \citenamefont {{Ruhl}}, \citenamefont {{Schiappucci}}, \citenamefont {{Smecher}}, \citenamefont {{Sobrin}}, \citenamefont {{Stephen}}, \citenamefont {{Suzuki}}, \citenamefont {{Tandoi}}, \citenamefont {{Thompson}}, \citenamefont {{Thorne}}, \citenamefont {{Tucker}},
  \citenamefont {{Umilta}}, \citenamefont {{Vieira}}, \citenamefont {{Wang}}, \citenamefont {{Wu}}, \citenamefont {{Yefremenko}}, \citenamefont {{Young}},\ and\ \citenamefont {{SPT-3G Collaboration}}}]{2022PhRvD.106d2011F}%
  \BibitemOpen
  \bibfield  {author} {\bibinfo {author} {\bibfnamefont {K.~R.}\ \bibnamefont {{Ferguson}}}, \bibinfo {author} {\bibfnamefont {A.~J.}\ \bibnamefont {{Anderson}}}, \bibinfo {author} {\bibfnamefont {N.}~\bibnamefont {{Whitehorn}}}, \bibinfo {author} {\bibfnamefont {P.~A.~R.}\ \bibnamefont {{Ade}}}, \bibinfo {author} {\bibfnamefont {M.}~\bibnamefont {{Archipley}}}, \bibinfo {author} {\bibfnamefont {J.~S.}\ \bibnamefont {{Avva}}}, \bibinfo {author} {\bibfnamefont {L.}~\bibnamefont {{Balkenhol}}}, \bibinfo {author} {\bibfnamefont {K.}~\bibnamefont {{Benabed}}}, \bibinfo {author} {\bibfnamefont {A.~N.}\ \bibnamefont {{Bender}}}, \bibinfo {author} {\bibfnamefont {B.~A.}\ \bibnamefont {{Benson}}}, \bibinfo {author} {\bibfnamefont {F.}~\bibnamefont {{Bianchini}}}, \bibinfo {author} {\bibfnamefont {L.~E.}\ \bibnamefont {{Bleem}}}, \bibinfo {author} {\bibfnamefont {F.~R.}\ \bibnamefont {{Bouchet}}}, \bibinfo {author} {\bibfnamefont {L.}~\bibnamefont {{Bryant}}}, \bibinfo {author} {\bibfnamefont {E.}~\bibnamefont
  {{Camphuis}}}, \bibinfo {author} {\bibfnamefont {J.~E.}\ \bibnamefont {{Carlstrom}}}, \bibinfo {author} {\bibfnamefont {T.~W.}\ \bibnamefont {{Cecil}}}, \bibinfo {author} {\bibfnamefont {C.~L.}\ \bibnamefont {{Chang}}}, \bibinfo {author} {\bibfnamefont {P.}~\bibnamefont {{Chaubal}}}, \bibinfo {author} {\bibfnamefont {P.~M.}\ \bibnamefont {{Chichura}}}, \bibinfo {author} {\bibfnamefont {T.~L.}\ \bibnamefont {{Chou}}}, \bibinfo {author} {\bibfnamefont {T.~M.}\ \bibnamefont {{Crawford}}}, \bibinfo {author} {\bibfnamefont {A.}~\bibnamefont {{Cukierman}}}, \bibinfo {author} {\bibfnamefont {C.}~\bibnamefont {{Daley}}}, \bibinfo {author} {\bibfnamefont {T.}~\bibnamefont {{de Haan}}}, \bibinfo {author} {\bibfnamefont {K.~R.}\ \bibnamefont {{Dibert}}}, \bibinfo {author} {\bibfnamefont {M.~A.}\ \bibnamefont {{Dobbs}}}, \bibinfo {author} {\bibfnamefont {A.}~\bibnamefont {{Doussot}}}, \bibinfo {author} {\bibfnamefont {D.}~\bibnamefont {{Dutcher}}}, \bibinfo {author} {\bibfnamefont {W.}~\bibnamefont {{Everett}}},
  \bibinfo {author} {\bibfnamefont {C.}~\bibnamefont {{Feng}}}, \bibinfo {author} {\bibfnamefont {A.}~\bibnamefont {{Foster}}}, \bibinfo {author} {\bibfnamefont {S.}~\bibnamefont {{Galli}}}, \bibinfo {author} {\bibfnamefont {A.~E.}\ \bibnamefont {{Gambrel}}}, \bibinfo {author} {\bibfnamefont {R.~W.}\ \bibnamefont {{Gardner}}}, \bibinfo {author} {\bibfnamefont {N.}~\bibnamefont {{Goeckner-Wald}}}, \bibinfo {author} {\bibfnamefont {R.}~\bibnamefont {{Gualtieri}}}, \bibinfo {author} {\bibfnamefont {F.}~\bibnamefont {{Guidi}}}, \bibinfo {author} {\bibfnamefont {S.}~\bibnamefont {{Guns}}}, \bibinfo {author} {\bibfnamefont {N.~W.}\ \bibnamefont {{Halverson}}}, \bibinfo {author} {\bibfnamefont {E.}~\bibnamefont {{Hivon}}}, \bibinfo {author} {\bibfnamefont {G.~P.}\ \bibnamefont {{Holder}}}, \bibinfo {author} {\bibfnamefont {W.~L.}\ \bibnamefont {{Holzapfel}}}, \bibinfo {author} {\bibfnamefont {J.~C.}\ \bibnamefont {{Hood}}}, \bibinfo {author} {\bibfnamefont {N.}~\bibnamefont {{Huang}}}, \bibinfo {author}
  {\bibfnamefont {L.}~\bibnamefont {{Knox}}}, \bibinfo {author} {\bibfnamefont {M.}~\bibnamefont {{Korman}}}, \bibinfo {author} {\bibfnamefont {C.~L.}\ \bibnamefont {{Kuo}}}, \bibinfo {author} {\bibfnamefont {A.~T.}\ \bibnamefont {{Lee}}}, \bibinfo {author} {\bibfnamefont {A.~E.}\ \bibnamefont {{Lowitz}}}, \bibinfo {author} {\bibfnamefont {C.}~\bibnamefont {{Lu}}}, \bibinfo {author} {\bibfnamefont {M.}~\bibnamefont {{Millea}}}, \bibinfo {author} {\bibfnamefont {J.}~\bibnamefont {{Montgomery}}}, \bibinfo {author} {\bibfnamefont {T.}~\bibnamefont {{Natoli}}}, \bibinfo {author} {\bibfnamefont {G.~I.}\ \bibnamefont {{Noble}}}, \bibinfo {author} {\bibfnamefont {V.}~\bibnamefont {{Novosad}}}, \bibinfo {author} {\bibfnamefont {Y.}~\bibnamefont {{Omori}}}, \bibinfo {author} {\bibfnamefont {S.}~\bibnamefont {{Padin}}}, \bibinfo {author} {\bibfnamefont {Z.}~\bibnamefont {{Pan}}}, \bibinfo {author} {\bibfnamefont {P.}~\bibnamefont {{Paschos}}}, \bibinfo {author} {\bibfnamefont {K.}~\bibnamefont {{Prabhu}}}, \bibinfo
  {author} {\bibfnamefont {W.}~\bibnamefont {{Quan}}}, \bibinfo {author} {\bibfnamefont {A.}~\bibnamefont {{Rahlin}}}, \bibinfo {author} {\bibfnamefont {C.~L.}\ \bibnamefont {{Reichardt}}}, \bibinfo {author} {\bibfnamefont {M.}~\bibnamefont {{Rouble}}}, \bibinfo {author} {\bibfnamefont {J.~E.}\ \bibnamefont {{Ruhl}}}, \bibinfo {author} {\bibfnamefont {E.}~\bibnamefont {{Schiappucci}}}, \bibinfo {author} {\bibfnamefont {G.}~\bibnamefont {{Smecher}}}, \bibinfo {author} {\bibfnamefont {J.~A.}\ \bibnamefont {{Sobrin}}}, \bibinfo {author} {\bibfnamefont {J.}~\bibnamefont {{Stephen}}}, \bibinfo {author} {\bibfnamefont {A.}~\bibnamefont {{Suzuki}}}, \bibinfo {author} {\bibfnamefont {C.}~\bibnamefont {{Tandoi}}}, \bibinfo {author} {\bibfnamefont {K.~L.}\ \bibnamefont {{Thompson}}}, \bibinfo {author} {\bibfnamefont {B.}~\bibnamefont {{Thorne}}}, \bibinfo {author} {\bibfnamefont {C.}~\bibnamefont {{Tucker}}}, \bibinfo {author} {\bibfnamefont {C.}~\bibnamefont {{Umilta}}}, \bibinfo {author} {\bibfnamefont {J.~D.}\
  \bibnamefont {{Vieira}}}, \bibinfo {author} {\bibfnamefont {G.}~\bibnamefont {{Wang}}}, \bibinfo {author} {\bibfnamefont {W.~L.~K.}\ \bibnamefont {{Wu}}}, \bibinfo {author} {\bibfnamefont {V.}~\bibnamefont {{Yefremenko}}}, \bibinfo {author} {\bibfnamefont {M.~R.}\ \bibnamefont {{Young}}},\ and\ \bibinfo {author} {\bibnamefont {{SPT-3G Collaboration}}},\ }\bibfield  {title} {\bibinfo {title} {{Searching for axionlike time-dependent cosmic birefringence with data from SPT-3G}},\ }\href {https://doi.org/10.1103/PhysRevD.106.042011} {\bibfield  {journal} {\bibinfo  {journal} {\prd}\ }\textbf {\bibinfo {volume} {106}},\ \bibinfo {eid} {042011} (\bibinfo {year} {2022})},\ \Eprint {https://arxiv.org/abs/2203.16567} {arXiv:2203.16567 [astro-ph.CO]} \BibitemShut {NoStop}%
\bibitem [{\citenamefont {{Planck Collaboration}}\ \emph {et~al.}(2020)\citenamefont {{Planck Collaboration}}, \citenamefont {{Aghanim}}, \citenamefont {{Akrami}}, \citenamefont {{Ashdown}}, \citenamefont {{Aumont}}, \citenamefont {{Baccigalupi}}, \citenamefont {{Ballardini}}, \citenamefont {{Banday}}, \citenamefont {{Barreiro}}, \citenamefont {{Bartolo}}, \citenamefont {{Basak}}, \citenamefont {{Battye}}, \citenamefont {{Benabed}}, \citenamefont {{Bernard}}, \citenamefont {{Bersanelli}}, \citenamefont {{Bielewicz}}, \citenamefont {{Bock}}, \citenamefont {{Bond}}, \citenamefont {{Borrill}}, \citenamefont {{Bouchet}}, \citenamefont {{Boulanger}}, \citenamefont {{Bucher}}, \citenamefont {{Burigana}}, \citenamefont {{Butler}}, \citenamefont {{Calabrese}}, \citenamefont {{Cardoso}}, \citenamefont {{Carron}}, \citenamefont {{Challinor}}, \citenamefont {{Chiang}}, \citenamefont {{Chluba}}, \citenamefont {{Colombo}}, \citenamefont {{Combet}}, \citenamefont {{Contreras}}, \citenamefont {{Crill}}, \citenamefont
  {{Cuttaia}}, \citenamefont {{de Bernardis}}, \citenamefont {{de Zotti}}, \citenamefont {{Delabrouille}}, \citenamefont {{Delouis}}, \citenamefont {{Di Valentino}}, \citenamefont {{Diego}}, \citenamefont {{Dor{\'e}}}, \citenamefont {{Douspis}}, \citenamefont {{Ducout}}, \citenamefont {{Dupac}}, \citenamefont {{Dusini}}, \citenamefont {{Efstathiou}}, \citenamefont {{Elsner}}, \citenamefont {{En{\ss}lin}}, \citenamefont {{Eriksen}}, \citenamefont {{Fantaye}}, \citenamefont {{Farhang}}, \citenamefont {{Fergusson}}, \citenamefont {{Fernandez-Cobos}}, \citenamefont {{Finelli}}, \citenamefont {{Forastieri}}, \citenamefont {{Frailis}}, \citenamefont {{Fraisse}}, \citenamefont {{Franceschi}}, \citenamefont {{Frolov}}, \citenamefont {{Galeotta}}, \citenamefont {{Galli}}, \citenamefont {{Ganga}}, \citenamefont {{G{\'e}nova-Santos}}, \citenamefont {{Gerbino}}, \citenamefont {{Ghosh}}, \citenamefont {{Gonz{\'a}lez-Nuevo}}, \citenamefont {{G{\'o}rski}}, \citenamefont {{Gratton}}, \citenamefont {{Gruppuso}}, \citenamefont
  {{Gudmundsson}}, \citenamefont {{Hamann}}, \citenamefont {{Handley}}, \citenamefont {{Hansen}}, \citenamefont {{Herranz}}, \citenamefont {{Hildebrandt}}, \citenamefont {{Hivon}}, \citenamefont {{Huang}}, \citenamefont {{Jaffe}}, \citenamefont {{Jones}}, \citenamefont {{Karakci}}, \citenamefont {{Keih{\"a}nen}}, \citenamefont {{Keskitalo}}, \citenamefont {{Kiiveri}}, \citenamefont {{Kim}}, \citenamefont {{Kisner}}, \citenamefont {{Knox}}, \citenamefont {{Krachmalnicoff}}, \citenamefont {{Kunz}}, \citenamefont {{Kurki-Suonio}}, \citenamefont {{Lagache}}, \citenamefont {{Lamarre}}, \citenamefont {{Lasenby}}, \citenamefont {{Lattanzi}}, \citenamefont {{Lawrence}}, \citenamefont {{Le Jeune}}, \citenamefont {{Lemos}}, \citenamefont {{Lesgourgues}}, \citenamefont {{Levrier}}, \citenamefont {{Lewis}}, \citenamefont {{Liguori}}, \citenamefont {{Lilje}}, \citenamefont {{Lilley}}, \citenamefont {{Lindholm}}, \citenamefont {{L{\'o}pez-Caniego}}, \citenamefont {{Lubin}}, \citenamefont {{Ma}}, \citenamefont
  {{Mac{\'\i}as-P{\'e}rez}}, \citenamefont {{Maggio}}, \citenamefont {{Maino}}, \citenamefont {{Mandolesi}}, \citenamefont {{Mangilli}}, \citenamefont {{Marcos-Caballero}}, \citenamefont {{Maris}}, \citenamefont {{Martin}}, \citenamefont {{Martinelli}}, \citenamefont {{Mart{\'\i}nez-Gonz{\'a}lez}}, \citenamefont {{Matarrese}}, \citenamefont {{Mauri}}, \citenamefont {{McEwen}}, \citenamefont {{Meinhold}}, \citenamefont {{Melchiorri}}, \citenamefont {{Mennella}}, \citenamefont {{Migliaccio}}, \citenamefont {{Millea}}, \citenamefont {{Mitra}}, \citenamefont {{Miville-Desch{\^e}nes}}, \citenamefont {{Molinari}}, \citenamefont {{Montier}}, \citenamefont {{Morgante}}, \citenamefont {{Moss}}, \citenamefont {{Natoli}}, \citenamefont {{N{\o}rgaard-Nielsen}}, \citenamefont {{Pagano}}, \citenamefont {{Paoletti}}, \citenamefont {{Partridge}}, \citenamefont {{Patanchon}}, \citenamefont {{Peiris}}, \citenamefont {{Perrotta}}, \citenamefont {{Pettorino}}, \citenamefont {{Piacentini}}, \citenamefont {{Polastri}},
  \citenamefont {{Polenta}}, \citenamefont {{Puget}}, \citenamefont {{Rachen}}, \citenamefont {{Reinecke}}, \citenamefont {{Remazeilles}}, \citenamefont {{Renzi}}, \citenamefont {{Rocha}}, \citenamefont {{Rosset}}, \citenamefont {{Roudier}}, \citenamefont {{Rubi{\~n}o-Mart{\'\i}n}}, \citenamefont {{Ruiz-Granados}}, \citenamefont {{Salvati}}, \citenamefont {{Sandri}}, \citenamefont {{Savelainen}}, \citenamefont {{Scott}}, \citenamefont {{Shellard}}, \citenamefont {{Sirignano}}, \citenamefont {{Sirri}}, \citenamefont {{Spencer}}, \citenamefont {{Sunyaev}}, \citenamefont {{Suur-Uski}}, \citenamefont {{Tauber}}, \citenamefont {{Tavagnacco}}, \citenamefont {{Tenti}}, \citenamefont {{Toffolatti}}, \citenamefont {{Tomasi}}, \citenamefont {{Trombetti}}, \citenamefont {{Valenziano}}, \citenamefont {{Valiviita}}, \citenamefont {{Van Tent}}, \citenamefont {{Vibert}}, \citenamefont {{Vielva}}, \citenamefont {{Villa}}, \citenamefont {{Vittorio}}, \citenamefont {{Wandelt}}, \citenamefont {{Wehus}}, \citenamefont {{White}},
  \citenamefont {{White}}, \citenamefont {{Zacchei}},\ and\ \citenamefont {{Zonca}}}]{2020A&A...641A...6P}%
  \BibitemOpen
  \bibfield  {author} {\bibinfo {author} {\bibnamefont {{Planck Collaboration}}}, \bibinfo {author} {\bibfnamefont {N.}~\bibnamefont {{Aghanim}}}, \bibinfo {author} {\bibfnamefont {Y.}~\bibnamefont {{Akrami}}}, \bibinfo {author} {\bibfnamefont {M.}~\bibnamefont {{Ashdown}}}, \bibinfo {author} {\bibfnamefont {J.}~\bibnamefont {{Aumont}}}, \bibinfo {author} {\bibfnamefont {C.}~\bibnamefont {{Baccigalupi}}}, \bibinfo {author} {\bibfnamefont {M.}~\bibnamefont {{Ballardini}}}, \bibinfo {author} {\bibfnamefont {A.~J.}\ \bibnamefont {{Banday}}}, \bibinfo {author} {\bibfnamefont {R.~B.}\ \bibnamefont {{Barreiro}}}, \bibinfo {author} {\bibfnamefont {N.}~\bibnamefont {{Bartolo}}}, \bibinfo {author} {\bibfnamefont {S.}~\bibnamefont {{Basak}}}, \bibinfo {author} {\bibfnamefont {R.}~\bibnamefont {{Battye}}}, \bibinfo {author} {\bibfnamefont {K.}~\bibnamefont {{Benabed}}}, \bibinfo {author} {\bibfnamefont {J.~P.}\ \bibnamefont {{Bernard}}}, \bibinfo {author} {\bibfnamefont {M.}~\bibnamefont {{Bersanelli}}}, \bibinfo
  {author} {\bibfnamefont {P.}~\bibnamefont {{Bielewicz}}}, \bibinfo {author} {\bibfnamefont {J.~J.}\ \bibnamefont {{Bock}}}, \bibinfo {author} {\bibfnamefont {J.~R.}\ \bibnamefont {{Bond}}}, \bibinfo {author} {\bibfnamefont {J.}~\bibnamefont {{Borrill}}}, \bibinfo {author} {\bibfnamefont {F.~R.}\ \bibnamefont {{Bouchet}}}, \bibinfo {author} {\bibfnamefont {F.}~\bibnamefont {{Boulanger}}}, \bibinfo {author} {\bibfnamefont {M.}~\bibnamefont {{Bucher}}}, \bibinfo {author} {\bibfnamefont {C.}~\bibnamefont {{Burigana}}}, \bibinfo {author} {\bibfnamefont {R.~C.}\ \bibnamefont {{Butler}}}, \bibinfo {author} {\bibfnamefont {E.}~\bibnamefont {{Calabrese}}}, \bibinfo {author} {\bibfnamefont {J.~F.}\ \bibnamefont {{Cardoso}}}, \bibinfo {author} {\bibfnamefont {J.}~\bibnamefont {{Carron}}}, \bibinfo {author} {\bibfnamefont {A.}~\bibnamefont {{Challinor}}}, \bibinfo {author} {\bibfnamefont {H.~C.}\ \bibnamefont {{Chiang}}}, \bibinfo {author} {\bibfnamefont {J.}~\bibnamefont {{Chluba}}}, \bibinfo {author} {\bibfnamefont
  {L.~P.~L.}\ \bibnamefont {{Colombo}}}, \bibinfo {author} {\bibfnamefont {C.}~\bibnamefont {{Combet}}}, \bibinfo {author} {\bibfnamefont {D.}~\bibnamefont {{Contreras}}}, \bibinfo {author} {\bibfnamefont {B.~P.}\ \bibnamefont {{Crill}}}, \bibinfo {author} {\bibfnamefont {F.}~\bibnamefont {{Cuttaia}}}, \bibinfo {author} {\bibfnamefont {P.}~\bibnamefont {{de Bernardis}}}, \bibinfo {author} {\bibfnamefont {G.}~\bibnamefont {{de Zotti}}}, \bibinfo {author} {\bibfnamefont {J.}~\bibnamefont {{Delabrouille}}}, \bibinfo {author} {\bibfnamefont {J.~M.}\ \bibnamefont {{Delouis}}}, \bibinfo {author} {\bibfnamefont {E.}~\bibnamefont {{Di Valentino}}}, \bibinfo {author} {\bibfnamefont {J.~M.}\ \bibnamefont {{Diego}}}, \bibinfo {author} {\bibfnamefont {O.}~\bibnamefont {{Dor{\'e}}}}, \bibinfo {author} {\bibfnamefont {M.}~\bibnamefont {{Douspis}}}, \bibinfo {author} {\bibfnamefont {A.}~\bibnamefont {{Ducout}}}, \bibinfo {author} {\bibfnamefont {X.}~\bibnamefont {{Dupac}}}, \bibinfo {author} {\bibfnamefont {S.}~\bibnamefont
  {{Dusini}}}, \bibinfo {author} {\bibfnamefont {G.}~\bibnamefont {{Efstathiou}}}, \bibinfo {author} {\bibfnamefont {F.}~\bibnamefont {{Elsner}}}, \bibinfo {author} {\bibfnamefont {T.~A.}\ \bibnamefont {{En{\ss}lin}}}, \bibinfo {author} {\bibfnamefont {H.~K.}\ \bibnamefont {{Eriksen}}}, \bibinfo {author} {\bibfnamefont {Y.}~\bibnamefont {{Fantaye}}}, \bibinfo {author} {\bibfnamefont {M.}~\bibnamefont {{Farhang}}}, \bibinfo {author} {\bibfnamefont {J.}~\bibnamefont {{Fergusson}}}, \bibinfo {author} {\bibfnamefont {R.}~\bibnamefont {{Fernandez-Cobos}}}, \bibinfo {author} {\bibfnamefont {F.}~\bibnamefont {{Finelli}}}, \bibinfo {author} {\bibfnamefont {F.}~\bibnamefont {{Forastieri}}}, \bibinfo {author} {\bibfnamefont {M.}~\bibnamefont {{Frailis}}}, \bibinfo {author} {\bibfnamefont {A.~A.}\ \bibnamefont {{Fraisse}}}, \bibinfo {author} {\bibfnamefont {E.}~\bibnamefont {{Franceschi}}}, \bibinfo {author} {\bibfnamefont {A.}~\bibnamefont {{Frolov}}}, \bibinfo {author} {\bibfnamefont {S.}~\bibnamefont {{Galeotta}}},
  \bibinfo {author} {\bibfnamefont {S.}~\bibnamefont {{Galli}}}, \bibinfo {author} {\bibfnamefont {K.}~\bibnamefont {{Ganga}}}, \bibinfo {author} {\bibfnamefont {R.~T.}\ \bibnamefont {{G{\'e}nova-Santos}}}, \bibinfo {author} {\bibfnamefont {M.}~\bibnamefont {{Gerbino}}}, \bibinfo {author} {\bibfnamefont {T.}~\bibnamefont {{Ghosh}}}, \bibinfo {author} {\bibfnamefont {J.}~\bibnamefont {{Gonz{\'a}lez-Nuevo}}}, \bibinfo {author} {\bibfnamefont {K.~M.}\ \bibnamefont {{G{\'o}rski}}}, \bibinfo {author} {\bibfnamefont {S.}~\bibnamefont {{Gratton}}}, \bibinfo {author} {\bibfnamefont {A.}~\bibnamefont {{Gruppuso}}}, \bibinfo {author} {\bibfnamefont {J.~E.}\ \bibnamefont {{Gudmundsson}}}, \bibinfo {author} {\bibfnamefont {J.}~\bibnamefont {{Hamann}}}, \bibinfo {author} {\bibfnamefont {W.}~\bibnamefont {{Handley}}}, \bibinfo {author} {\bibfnamefont {F.~K.}\ \bibnamefont {{Hansen}}}, \bibinfo {author} {\bibfnamefont {D.}~\bibnamefont {{Herranz}}}, \bibinfo {author} {\bibfnamefont {S.~R.}\ \bibnamefont {{Hildebrandt}}},
  \bibinfo {author} {\bibfnamefont {E.}~\bibnamefont {{Hivon}}}, \bibinfo {author} {\bibfnamefont {Z.}~\bibnamefont {{Huang}}}, \bibinfo {author} {\bibfnamefont {A.~H.}\ \bibnamefont {{Jaffe}}}, \bibinfo {author} {\bibfnamefont {W.~C.}\ \bibnamefont {{Jones}}}, \bibinfo {author} {\bibfnamefont {A.}~\bibnamefont {{Karakci}}}, \bibinfo {author} {\bibfnamefont {E.}~\bibnamefont {{Keih{\"a}nen}}}, \bibinfo {author} {\bibfnamefont {R.}~\bibnamefont {{Keskitalo}}}, \bibinfo {author} {\bibfnamefont {K.}~\bibnamefont {{Kiiveri}}}, \bibinfo {author} {\bibfnamefont {J.}~\bibnamefont {{Kim}}}, \bibinfo {author} {\bibfnamefont {T.~S.}\ \bibnamefont {{Kisner}}}, \bibinfo {author} {\bibfnamefont {L.}~\bibnamefont {{Knox}}}, \bibinfo {author} {\bibfnamefont {N.}~\bibnamefont {{Krachmalnicoff}}}, \bibinfo {author} {\bibfnamefont {M.}~\bibnamefont {{Kunz}}}, \bibinfo {author} {\bibfnamefont {H.}~\bibnamefont {{Kurki-Suonio}}}, \bibinfo {author} {\bibfnamefont {G.}~\bibnamefont {{Lagache}}}, \bibinfo {author} {\bibfnamefont
  {J.~M.}\ \bibnamefont {{Lamarre}}}, \bibinfo {author} {\bibfnamefont {A.}~\bibnamefont {{Lasenby}}}, \bibinfo {author} {\bibfnamefont {M.}~\bibnamefont {{Lattanzi}}}, \bibinfo {author} {\bibfnamefont {C.~R.}\ \bibnamefont {{Lawrence}}}, \bibinfo {author} {\bibfnamefont {M.}~\bibnamefont {{Le Jeune}}}, \bibinfo {author} {\bibfnamefont {P.}~\bibnamefont {{Lemos}}}, \bibinfo {author} {\bibfnamefont {J.}~\bibnamefont {{Lesgourgues}}}, \bibinfo {author} {\bibfnamefont {F.}~\bibnamefont {{Levrier}}}, \bibinfo {author} {\bibfnamefont {A.}~\bibnamefont {{Lewis}}}, \bibinfo {author} {\bibfnamefont {M.}~\bibnamefont {{Liguori}}}, \bibinfo {author} {\bibfnamefont {P.~B.}\ \bibnamefont {{Lilje}}}, \bibinfo {author} {\bibfnamefont {M.}~\bibnamefont {{Lilley}}}, \bibinfo {author} {\bibfnamefont {V.}~\bibnamefont {{Lindholm}}}, \bibinfo {author} {\bibfnamefont {M.}~\bibnamefont {{L{\'o}pez-Caniego}}}, \bibinfo {author} {\bibfnamefont {P.~M.}\ \bibnamefont {{Lubin}}}, \bibinfo {author} {\bibfnamefont {Y.~Z.}\ \bibnamefont
  {{Ma}}}, \bibinfo {author} {\bibfnamefont {J.~F.}\ \bibnamefont {{Mac{\'\i}as-P{\'e}rez}}}, \bibinfo {author} {\bibfnamefont {G.}~\bibnamefont {{Maggio}}}, \bibinfo {author} {\bibfnamefont {D.}~\bibnamefont {{Maino}}}, \bibinfo {author} {\bibfnamefont {N.}~\bibnamefont {{Mandolesi}}}, \bibinfo {author} {\bibfnamefont {A.}~\bibnamefont {{Mangilli}}}, \bibinfo {author} {\bibfnamefont {A.}~\bibnamefont {{Marcos-Caballero}}}, \bibinfo {author} {\bibfnamefont {M.}~\bibnamefont {{Maris}}}, \bibinfo {author} {\bibfnamefont {P.~G.}\ \bibnamefont {{Martin}}}, \bibinfo {author} {\bibfnamefont {M.}~\bibnamefont {{Martinelli}}}, \bibinfo {author} {\bibfnamefont {E.}~\bibnamefont {{Mart{\'\i}nez-Gonz{\'a}lez}}}, \bibinfo {author} {\bibfnamefont {S.}~\bibnamefont {{Matarrese}}}, \bibinfo {author} {\bibfnamefont {N.}~\bibnamefont {{Mauri}}}, \bibinfo {author} {\bibfnamefont {J.~D.}\ \bibnamefont {{McEwen}}}, \bibinfo {author} {\bibfnamefont {P.~R.}\ \bibnamefont {{Meinhold}}}, \bibinfo {author} {\bibfnamefont
  {A.}~\bibnamefont {{Melchiorri}}}, \bibinfo {author} {\bibfnamefont {A.}~\bibnamefont {{Mennella}}}, \bibinfo {author} {\bibfnamefont {M.}~\bibnamefont {{Migliaccio}}}, \bibinfo {author} {\bibfnamefont {M.}~\bibnamefont {{Millea}}}, \bibinfo {author} {\bibfnamefont {S.}~\bibnamefont {{Mitra}}}, \bibinfo {author} {\bibfnamefont {M.~A.}\ \bibnamefont {{Miville-Desch{\^e}nes}}}, \bibinfo {author} {\bibfnamefont {D.}~\bibnamefont {{Molinari}}}, \bibinfo {author} {\bibfnamefont {L.}~\bibnamefont {{Montier}}}, \bibinfo {author} {\bibfnamefont {G.}~\bibnamefont {{Morgante}}}, \bibinfo {author} {\bibfnamefont {A.}~\bibnamefont {{Moss}}}, \bibinfo {author} {\bibfnamefont {P.}~\bibnamefont {{Natoli}}}, \bibinfo {author} {\bibfnamefont {H.~U.}\ \bibnamefont {{N{\o}rgaard-Nielsen}}}, \bibinfo {author} {\bibfnamefont {L.}~\bibnamefont {{Pagano}}}, \bibinfo {author} {\bibfnamefont {D.}~\bibnamefont {{Paoletti}}}, \bibinfo {author} {\bibfnamefont {B.}~\bibnamefont {{Partridge}}}, \bibinfo {author} {\bibfnamefont
  {G.}~\bibnamefont {{Patanchon}}}, \bibinfo {author} {\bibfnamefont {H.~V.}\ \bibnamefont {{Peiris}}}, \bibinfo {author} {\bibfnamefont {F.}~\bibnamefont {{Perrotta}}}, \bibinfo {author} {\bibfnamefont {V.}~\bibnamefont {{Pettorino}}}, \bibinfo {author} {\bibfnamefont {F.}~\bibnamefont {{Piacentini}}}, \bibinfo {author} {\bibfnamefont {L.}~\bibnamefont {{Polastri}}}, \bibinfo {author} {\bibfnamefont {G.}~\bibnamefont {{Polenta}}}, \bibinfo {author} {\bibfnamefont {J.~L.}\ \bibnamefont {{Puget}}}, \bibinfo {author} {\bibfnamefont {J.~P.}\ \bibnamefont {{Rachen}}}, \bibinfo {author} {\bibfnamefont {M.}~\bibnamefont {{Reinecke}}}, \bibinfo {author} {\bibfnamefont {M.}~\bibnamefont {{Remazeilles}}}, \bibinfo {author} {\bibfnamefont {A.}~\bibnamefont {{Renzi}}}, \bibinfo {author} {\bibfnamefont {G.}~\bibnamefont {{Rocha}}}, \bibinfo {author} {\bibfnamefont {C.}~\bibnamefont {{Rosset}}}, \bibinfo {author} {\bibfnamefont {G.}~\bibnamefont {{Roudier}}}, \bibinfo {author} {\bibfnamefont {J.~A.}\ \bibnamefont
  {{Rubi{\~n}o-Mart{\'\i}n}}}, \bibinfo {author} {\bibfnamefont {B.}~\bibnamefont {{Ruiz-Granados}}}, \bibinfo {author} {\bibfnamefont {L.}~\bibnamefont {{Salvati}}}, \bibinfo {author} {\bibfnamefont {M.}~\bibnamefont {{Sandri}}}, \bibinfo {author} {\bibfnamefont {M.}~\bibnamefont {{Savelainen}}}, \bibinfo {author} {\bibfnamefont {D.}~\bibnamefont {{Scott}}}, \bibinfo {author} {\bibfnamefont {E.~P.~S.}\ \bibnamefont {{Shellard}}}, \bibinfo {author} {\bibfnamefont {C.}~\bibnamefont {{Sirignano}}}, \bibinfo {author} {\bibfnamefont {G.}~\bibnamefont {{Sirri}}}, \bibinfo {author} {\bibfnamefont {L.~D.}\ \bibnamefont {{Spencer}}}, \bibinfo {author} {\bibfnamefont {R.}~\bibnamefont {{Sunyaev}}}, \bibinfo {author} {\bibfnamefont {A.~S.}\ \bibnamefont {{Suur-Uski}}}, \bibinfo {author} {\bibfnamefont {J.~A.}\ \bibnamefont {{Tauber}}}, \bibinfo {author} {\bibfnamefont {D.}~\bibnamefont {{Tavagnacco}}}, \bibinfo {author} {\bibfnamefont {M.}~\bibnamefont {{Tenti}}}, \bibinfo {author} {\bibfnamefont {L.}~\bibnamefont
  {{Toffolatti}}}, \bibinfo {author} {\bibfnamefont {M.}~\bibnamefont {{Tomasi}}}, \bibinfo {author} {\bibfnamefont {T.}~\bibnamefont {{Trombetti}}}, \bibinfo {author} {\bibfnamefont {L.}~\bibnamefont {{Valenziano}}}, \bibinfo {author} {\bibfnamefont {J.}~\bibnamefont {{Valiviita}}}, \bibinfo {author} {\bibfnamefont {B.}~\bibnamefont {{Van Tent}}}, \bibinfo {author} {\bibfnamefont {L.}~\bibnamefont {{Vibert}}}, \bibinfo {author} {\bibfnamefont {P.}~\bibnamefont {{Vielva}}}, \bibinfo {author} {\bibfnamefont {F.}~\bibnamefont {{Villa}}}, \bibinfo {author} {\bibfnamefont {N.}~\bibnamefont {{Vittorio}}}, \bibinfo {author} {\bibfnamefont {B.~D.}\ \bibnamefont {{Wandelt}}}, \bibinfo {author} {\bibfnamefont {I.~K.}\ \bibnamefont {{Wehus}}}, \bibinfo {author} {\bibfnamefont {M.}~\bibnamefont {{White}}}, \bibinfo {author} {\bibfnamefont {S.~D.~M.}\ \bibnamefont {{White}}}, \bibinfo {author} {\bibfnamefont {A.}~\bibnamefont {{Zacchei}}},\ and\ \bibinfo {author} {\bibfnamefont {A.}~\bibnamefont {{Zonca}}},\ }\bibfield
  {title} {\bibinfo {title} {{Planck 2018 results. VI. Cosmological parameters}},\ }\href {https://doi.org/10.1051/0004-6361/201833910} {\bibfield  {journal} {\bibinfo  {journal} {Astronomy \& Astrophysics}\ }\textbf {\bibinfo {volume} {641}},\ \bibinfo {eid} {A6} (\bibinfo {year} {2020})},\ \Eprint {https://arxiv.org/abs/1807.06209} {arXiv:1807.06209 [astro-ph.CO]} \BibitemShut {NoStop}%
\bibitem [{\citenamefont {{Calore}}\ \emph {et~al.}(2022)\citenamefont {{Calore}}, \citenamefont {{Carenza}}, \citenamefont {{Eckner}}, \citenamefont {{Fischer}}, \citenamefont {{Giannotti}}, \citenamefont {{Jaeckel}}, \citenamefont {{Kotake}}, \citenamefont {{Kuroda}}, \citenamefont {{Mirizzi}},\ and\ \citenamefont {{Sivo}}}]{2022PhRvD.105f3028C}%
  \BibitemOpen
  \bibfield  {author} {\bibinfo {author} {\bibfnamefont {F.}~\bibnamefont {{Calore}}}, \bibinfo {author} {\bibfnamefont {P.}~\bibnamefont {{Carenza}}}, \bibinfo {author} {\bibfnamefont {C.}~\bibnamefont {{Eckner}}}, \bibinfo {author} {\bibfnamefont {T.}~\bibnamefont {{Fischer}}}, \bibinfo {author} {\bibfnamefont {M.}~\bibnamefont {{Giannotti}}}, \bibinfo {author} {\bibfnamefont {J.}~\bibnamefont {{Jaeckel}}}, \bibinfo {author} {\bibfnamefont {K.}~\bibnamefont {{Kotake}}}, \bibinfo {author} {\bibfnamefont {T.}~\bibnamefont {{Kuroda}}}, \bibinfo {author} {\bibfnamefont {A.}~\bibnamefont {{Mirizzi}}},\ and\ \bibinfo {author} {\bibfnamefont {F.}~\bibnamefont {{Sivo}}},\ }\bibfield  {title} {\bibinfo {title} {{3D template-based Fermi-LAT constraints on the diffuse supernova axion-like particle background}},\ }\href {https://doi.org/10.1103/PhysRevD.105.063028} {\bibfield  {journal} {\bibinfo  {journal} {\prd}\ }\textbf {\bibinfo {volume} {105}},\ \bibinfo {eid} {063028} (\bibinfo {year} {2022})},\ \Eprint
  {https://arxiv.org/abs/2110.03679} {arXiv:2110.03679 [astro-ph.HE]} \BibitemShut {NoStop}%
\bibitem [{\citenamefont {{Wouters}}\ and\ \citenamefont {{Brun}}(2013)}]{2013ApJ...772...44W}%
  \BibitemOpen
  \bibfield  {author} {\bibinfo {author} {\bibfnamefont {D.}~\bibnamefont {{Wouters}}}\ and\ \bibinfo {author} {\bibfnamefont {P.}~\bibnamefont {{Brun}}},\ }\bibfield  {title} {\bibinfo {title} {{Constraints on Axion-like Particles from X-Ray Observations of the Hydra Galaxy Cluster}},\ }\href {https://doi.org/10.1088/0004-637X/772/1/44} {\bibfield  {journal} {\bibinfo  {journal} {\apj}\ }\textbf {\bibinfo {volume} {772}},\ \bibinfo {eid} {44} (\bibinfo {year} {2013})},\ \Eprint {https://arxiv.org/abs/1304.0989} {arXiv:1304.0989 [astro-ph.HE]} \BibitemShut {NoStop}%
\bibitem [{\citenamefont {{Dessert}}\ \emph {et~al.}(2020)\citenamefont {{Dessert}}, \citenamefont {{Foster}},\ and\ \citenamefont {{Safdi}}}]{2020PhRvL.125z1102D}%
  \BibitemOpen
  \bibfield  {author} {\bibinfo {author} {\bibfnamefont {C.}~\bibnamefont {{Dessert}}}, \bibinfo {author} {\bibfnamefont {J.~W.}\ \bibnamefont {{Foster}}},\ and\ \bibinfo {author} {\bibfnamefont {B.~R.}\ \bibnamefont {{Safdi}}},\ }\bibfield  {title} {\bibinfo {title} {{X-Ray Searches for Axions from Super Star Clusters}},\ }\href {https://doi.org/10.1103/PhysRevLett.125.261102} {\bibfield  {journal} {\bibinfo  {journal} {\prl}\ }\textbf {\bibinfo {volume} {125}},\ \bibinfo {eid} {261102} (\bibinfo {year} {2020})},\ \Eprint {https://arxiv.org/abs/2008.03305} {arXiv:2008.03305 [hep-ph]} \BibitemShut {NoStop}%
\bibitem [{\citenamefont {{Marsh}}\ \emph {et~al.}(2017)\citenamefont {{Marsh}}, \citenamefont {{Russell}}, \citenamefont {{Fabian}}, \citenamefont {{McNamara}}, \citenamefont {{Nulsen}},\ and\ \citenamefont {{Reynolds}}}]{2017JCAP...12..036M}%
  \BibitemOpen
  \bibfield  {author} {\bibinfo {author} {\bibfnamefont {M.~C.~D.}\ \bibnamefont {{Marsh}}}, \bibinfo {author} {\bibfnamefont {H.~R.}\ \bibnamefont {{Russell}}}, \bibinfo {author} {\bibfnamefont {A.~C.}\ \bibnamefont {{Fabian}}}, \bibinfo {author} {\bibfnamefont {B.~R.}\ \bibnamefont {{McNamara}}}, \bibinfo {author} {\bibfnamefont {P.}~\bibnamefont {{Nulsen}}},\ and\ \bibinfo {author} {\bibfnamefont {C.~S.}\ \bibnamefont {{Reynolds}}},\ }\bibfield  {title} {\bibinfo {title} {{A new bound on axion-like particles}},\ }\href {https://doi.org/10.1088/1475-7516/2017/12/036} {\bibfield  {journal} {\bibinfo  {journal} {Journal of Cosmology and Astroparticle Physics}\ }\textbf {\bibinfo {volume} {2017}}\bibfield  {number} {\bibinfo  {number} { (12)},\ \bibinfo {eid} {036}},\ }\Eprint {https://arxiv.org/abs/1703.07354} {arXiv:1703.07354 [hep-ph]} \BibitemShut {NoStop}%
\bibitem [{\citenamefont {{Sisk-Reyn{\'e}s}}\ \emph {et~al.}(2022)\citenamefont {{Sisk-Reyn{\'e}s}}, \citenamefont {{Matthews}}, \citenamefont {{Reynolds}}, \citenamefont {{Russell}}, \citenamefont {{Smith}},\ and\ \citenamefont {{Marsh}}}]{2022MNRAS.510.1264S}%
  \BibitemOpen
  \bibfield  {author} {\bibinfo {author} {\bibfnamefont {J.}~\bibnamefont {{Sisk-Reyn{\'e}s}}}, \bibinfo {author} {\bibfnamefont {J.~H.}\ \bibnamefont {{Matthews}}}, \bibinfo {author} {\bibfnamefont {C.~S.}\ \bibnamefont {{Reynolds}}}, \bibinfo {author} {\bibfnamefont {H.~R.}\ \bibnamefont {{Russell}}}, \bibinfo {author} {\bibfnamefont {R.~N.}\ \bibnamefont {{Smith}}},\ and\ \bibinfo {author} {\bibfnamefont {M.~C.~D.}\ \bibnamefont {{Marsh}}},\ }\bibfield  {title} {\bibinfo {title} {{New constraints on light axion-like particles using Chandra transmission grating spectroscopy of the powerful cluster-hosted quasar H1821+643}},\ }\href {https://doi.org/10.1093/mnras/stab3464} {\bibfield  {journal} {\bibinfo  {journal} {Monthly Notices of the Royal Astronomical Society}\ }\textbf {\bibinfo {volume} {510}},\ \bibinfo {pages} {1264} (\bibinfo {year} {2022})},\ \Eprint {https://arxiv.org/abs/2109.03261} {arXiv:2109.03261 [astro-ph.HE]} \BibitemShut {NoStop}%
\end{thebibliography}%

\end{document}